\documentclass[twocolumn,trackchanges]{aastex62}
\usepackage{hyperref}

\usepackage{mathrsfs}  
\usepackage{amssymb, amsmath, amsbsy}
\usepackage{longtable}
\usepackage{multirow}

\def\kms{\,km\,s$^{-1}$}
\def\fvar{$F\mathrm{_{var}}$}

\def\eddr{$L_{\rm bol}/L_{\rm Edd}$}
\def\mdot{$\dot{\mathcal{M}}$}
\def\ergs{erg\,s$^{-1}$}
\def\rfe{R$_\mathrm{FeII}$}
\def\mbh{$M\mathrm{_{BH}}$}
\def\oiii{[O {\sc iii}]~$\lambda\lambda5007,4959$}
\def\hb{H$\beta$}

\newenvironment{myfont}{\fontfamily{lmtt}\selectfont}{\par}
\DeclareTextFontCommand{\textmyfont}{\myfont}

\graphicspath{{./}{figures/}}

\received{\today}
\revised{\today}
\accepted{\today}
\submitjournal{ApJ}

\shorttitle{Radius-luminosity relations for Mg II sample}
\shortauthors{Martinez-Aldama et al.}

\begin{document}

\title{{Scatter Analysis Along the Multidimensional Radius-Luminosity Relations for Reverberation-Mapped Mg II Sources}}

\correspondingauthor{Mary Loli Mart\'inez--Aldama}
\email{mmary@cft.edu.pl}

\author[0000-0002-7843-7689]{Mary Loli Mart\'inez--Aldama}
\affiliation{Center for Theoretical Physics, Polish Academy of Sciences, Al. Lotnik\'ow 32/46, 02-668 Warsaw, Poland}

\author[0000-0001-6450-1187]{Michal Zaja\v{c}ek}
\affiliation{Center for Theoretical Physics, Polish Academy of Sciences, Al. Lotnik\'ow 32/46, 02-668 Warsaw, Poland}

\author[0000-0001-5848-4333]{Bo\.{z}ena Czerny}
\affiliation{Center for Theoretical Physics, Polish Academy of Sciences, Al. Lotnik\'ow 32/46, 02-668 Warsaw, Poland}

\author[0000-0002-5854-7426]{Swayamtrupta Panda}
\affiliation{Center for Theoretical Physics, Polish Academy of Sciences, Al. Lotnik\'ow 32/46, 02-668 Warsaw, Poland}
\affiliation{Nicolaus Copernicus Astronomical Center, Polish Academy of Sciences, ul. Bartycka 18, 00-716 Warsaw, Poland}



\begin{abstract}

The usage of the radius-luminosity (R-L) relation for the determination of black hole masses across the cosmic history as well as its application for cosmological studies motivates us to analyze its scatter, which has recently increased significantly both for the optical (H$\beta$) and UV (Mg II) lines. To this purpose, we determined the scatter along the R-L relation for an up-to-date reverberation-mapped Mg II sample. Studying linear combinations of the luminosity at 3000~\AA\ with independent parameters such as the Full Width at Half Maximum (FWHM), UV Fe~II strength (\rfe), and the fractional variability (\fvar) for the whole sample, we get only a small decrease in the scatter ($\sigma{_{\rm rms}}=0.29-0.30$ dex). Linear combinations with the dimensionless accretion rate (\mdot) and the Eddington ratio lead to significant reductions of the scatter ($\sigma_{\rm rms}\sim 0.1$ dex), albeit both suffering from the inter-dependency on the observed time-delay. After the division into two sub-samples considering the median value of the \mdot\ in the full sample, we find that the scatter decreases significantly for the highly accreting sub-sample. In particular, the smallest scatter of $\sigma{_{\rm rms}}=0.17$ dex is associated with the independent parameter  \rfe\,, followed by the combination with \fvar\, with $\sigma{_{\rm rms}}=0.19$ dex. Both of these independent observationally-inferred parameters are in turn correlated with \mdot\ and \eddr. These results suggest that the large scatter along the R-L relation is driven \textit{mainly} by the accretion rate intensity. 

\end{abstract}



\section{Introduction} 
\label{sec:intro}

Strong and broad emission lines are among the most characteristic features of the type 1 active galactic nuclei (AGN; for a review, see \citealt{netzer2013book}). The broad line region (BLR) where these lines originate was only recently marginally resolved in 3C 273 with the near-infrared interferometry instrument GRAVITY \citep{GRAVITY2018} but in all other objects the access to the BLR structure is only through variability studies. The reverberation mapping (RM) technique - the observation of the response of the BLR lines to the changing continuum \citep{1982ApJ...255..419B,2004AN....325..248P,2009NewAR..53..140G,2019OAst...28..200C} - has been applied now to over 100 objects, mostly using H$\beta$ lines but also for several other lines \citep[see e.g.][and the references therein]{netzer2020}.

The major discovery coming from this research field was the radius-luminosity (R-L) relation between the time delay of the H$\beta$ line and the continuum luminosity measured at 5100 \AA ~ \citep{kaspi2000,peterson2004,bentz2013}. Assuming that the line-emitting clouds are virialized, the R-L relation, with a relatively small scatter of $0.19$ dex \citep[only 0.13 dex with one source removed, see][]{bentz2013}, allowed for massive inference of black hole masses in AGN using just single-epoch spectra, with {the line width (Full Width at Half Maximum - FWHM, or $\sigma_\mathrm{line}$ - line dispersion) serving as a proxy for the BLR velocity \citep[e.g.][]{peterson2004, collin2006, vestergaard2006, denney2013, mejia2018, dallabonta2020, yu2020_sigmaline}, and the monochromatic luminosity serving as a proxy for the BLR radius}. It also started to be considered as a promising tool for measuring luminosity distances in cosmology, making AGN standardizable candles, as the time delay of the BLR lines could be used as a proxy for a redshift-independent measurement of the absolute luminosity \citep{watson2011,haas2011,czerny2013,martinez_aldama2019,2019arXiv190905572P}. The small scatter in the R-L relation is {potentially explained by the Failed Radiatively Accelerated Dusty Outflow (FRADO), a theoretical model of the BLR} which connected the inner radius of the BLR to the monochromatic flux through the request of the disk surface temperature to be equal to the dust sublimation temperature \citep{czhr2011,czerny2017}. 


{Reverberation mapping requires high cadence, long duration monitoring. Since it is very difficult to get a large number of time-critical observations on over-subscribed facilities for a technique that is still under development, early reverberation-mapping programs were carried out on smaller telescopes, which in turn restricted observations to apparently bright, relatively nearby AGNs, especially those already known to have variable broad emission lines. Furthermore, many reverberation programs used the \oiii\ narrow lines as an internal flux calibrator because they arise in a more extended, lower density environment and are expected to be constant in flux over reverberation time scales \citep{peterson1998}. These considerations conspired to bias the early reverberation samples toward lower Eddington ratio sources.} With a growing interest in the behaviour of different types of AGN \citep[e.g. Supermassive black holes with high accretion rates][]{dupu2014,dupu2018}, and with the start of {Sloan Digital Sky Survey Reverberation Mapping Project (SDSS-RM)} based on selection of a part of the sky instead of individual sources, the scatter in the R-L has increased considerably. The scatter was seen both in H$\beta$ studies \citep{grier2017} as well as in most recent studies based on Mg II line \citep{czerny2019,homayouni2020,zajacek2020}. {This poses two fundamental questions: what is the physical cause of this dispersion? and, is there a way to still use the  R-L relation reliably for black hole mass estimates as well as for the cosmology? }

{The scatter is apparently related to the accretion-rate intensities, where the sources with the largest accretion rate show the largest departures from the R-L relation \citep[e.g.][]{dupu2018}, which increases its overall scatter. In order to take into account the accretion rate effect, a correction based on the dimensionless accretion rate parameter as well as the Eddington ratio was proposed \citet{martinez_aldama2019} {and \citet{dallabonta2020}}. However, the inter-dependence between these parameters and the observed time delay ($\tau_\mathrm{{obs}}$) makes it less reliable \citep{fonseca-alvarez2019}. 
Since a few independent observationally inferred parameters are driven by the accretion rate intensity, they can also be considered to recover the low scatter. For instance, the optical Fe II strength is related to the accretion-rate intensity, then including this parameter in the R-L relation, the scatter decreases significantly  \citep[$\sigma\sim0.19$ dex;][]{du2019, Yu2020}.}

In the present paper we analyse the multidimensional view of the R-L relation using the measurements of the Mg II time delay. The sample used is relatively homogeneous, coming from a small number of research groups {(Section~\ref{sec:sample})}. We analyze how the scatter changes when the observed time delay is expressed as a linear combination of the logarithms of relevant quantities, including the monochromatic luminosity at 3000~\AA, the Full Width at Half Maximum (FWHM), the {inter-dependent} accretion rate parameters (dimensionless $\dot{\mathcal{M}}$ and the Eddington ratio, \eddr), {and other parameters correlated with the the accretion rate, such as} the strength of UV Fe~II line (\rfe), and the fractional variability, \fvar\ (Section~\ref{sec_results}). {We discuss certain relevant issues in Section~\ref{sec:discussion} and summarize our results in Section~\ref{sec:conclusions}.}


\section{Sample and measurements} \label{sec:sample}

\subsection{Sample description}
\label{sec:sampledes}

The full sample includes 68 objects with $42.8< \mathrm{log} L_{3000}<46.8$ at $0.003<z<1.89$. It includes all the objects with reverberation-mapped Mg II measurements reported till date. The sample includes the 57 Mg~II time lags {(henceforth SDSS-RM sample)} from the recent SDSS-RM monitoring \citep[][]{homayouni2020} and the 6 objects previously monitored by the same project \citep{shen2006}. Only one object from the previous SDSS-RM monitoring is included in the recent one; both measurements are considered. We also include the high-luminosity objects: CTS~252 monitored by \citet{lira2018}, CTS~C30.10 measured by \citet{czerny2019}, HE~0413-4031 monitored by \citet{zajacek2020}, and the 2 old IUE measurements reported for NGC~4151 by \citet{metzroth2006}.

The full sample considered is relatively homogeneous, since $\sim83\%$ of the objects come from the {SDSS-RM sample} and $\sim9\%$ from their previous program. The rest of the objects with the lowest (NGC~4151) and the highest (CTS~C30.10, HE~0413-4031, and CTS~252) luminosities are crucial for the detection of the trends in the R-L relation. For the recent SDSS-RM sample, the time delay is estimated using the JAVELIN method \citep{2011ApJ...735...80Z,2013ApJ...765..106Z,2016ApJ...819..122Z}, while for the rest of the sources other methods are applied, specifically interpolated cross-correlation function \citep[ICCF;][]{1987ApJS...65....1G,1998PASP..110..660P}, discrete correlation function \citep[DCF;][]{1988ApJ...333..646E}, $z$-transformed DCF \citep[zDCF;][]{1997ASSL..218..163A}, the light-curve similarity estimators \citep[Von Neumann or Bartels estimators;][]{2017ApJ...844..146C}, and the $\chi^2$ method \citep[][]{czerny2013,czerny2019,zajacek2020}, or an average of them. {Recent statistical analyses \citep{2019ApJ...884..119L,2020MNRAS.491.6045Y} points out that the JAVELIN method is more powerful than other traditional methods (ICCF and zDCF) to recover time delays. However, the analyses were performed on the mock sample where the continuum light curve is generated using the damped random-walk (DRW) process, and the JAVELIN makes use of the DRW to recover the time delay. Therefore, there may be a bias since the stochastic, red-noise part of the variable continuum is omitted in the generated light curves from their power-density spectra. In particular, the model-independent, discrete methods such as zDCF and von Neumann estimator may still be more suitable for the analysis of irregular and heterogeneous light curves of more distant quasars \citep{czerny2019,2019arXiv190703910Z,zajacek2020}. In particular, for the highly accreting source HE~0413-4031 at $z=1.39$, \citet{zajacek2020} saw a difference between JAVELIN and ICCF on one hand and discrete methods (DCF, zDCF, and von Neumann) on the other hand, in the determination of the primary time-delay peak, $\tau\sim 431$ and $\sim 303$ days, respectively.} 

In addition, each research group assumes different time delay significance criteria, which may not be satisfied for the objects of other samples. For example, the recent SDSS-RM {monitoring \citep{homayouni2020}} considers that a time delay is statistically significant if (1) a primary time-lag peak includes at least 60$\%$ of the weighted time-delay posterior samples, and (2) the time delay is well-detected at the 3$\sigma$ level different from the zero time-delay. Therefore, for completeness, we consider all the objects as {we analyze a mixed sample}, for which it is difficult to establish general statistically robust criteria.




\subsection{Measurements}
\label{sec:estimations}

For all the sources in the sample, we collected several measured parameters, which are summarized in  {Table~{\ref{tab:measurements}} in Appendix~\ref{appendix}}. For the SDSS-RM sample, the luminosity at 3000~\AA, Full Width at Half Maximum (FWHM), equivalent width (EW) of Mg II and Fe II (at 2250-2650~\AA) were taken from the {\citet{shen2019} catalog}, which assumes a flat $\Lambda$CDM cosmology with $\Omega_\Lambda$=0.7
and $H_0$= 70 \kms~Mpc$^{-1}$. The time delay is taken from \cite{homayouni2020}. They estimated time delays of Mg II by two methods: JAVELIN and CREAM. In this contribution we use the JAVELIN estimations since they are more reliable according to the authors. For rest of the sample, the measurements were taken from the compilation done by \citet[][see their Table~2]{zajacek2020}. 

Since in the optical range the strength of the Fe~II broad line shows a clear correlation with the accretion rate intensity \citep{boroson2002, negrete2018, 2019ApJ...882...79P, du2019}, {we explore a similar correlation considering the UV Fe~II. The relation between the UV Fe~II and the accretion parameters is scarcely studied \citep{dong2011}, but
since the UV and optical FeII are correlated, and in turn, both show similar anti-correlations with the FWHM of \hb\ \citep{dong2009,kovacevic15, sniegowska, dong2011}, it opens the possibility that the UV Fe~II is also related to the accretion rate intensity as is the optical Fe~II emission. In Appendix~\ref{sec:fe2uv}, we include a discussion about this issue considering the parameters estimated below}.

The UV Fe II strength is estimated using the parameter \rfe, which is defined as the ratio of the equivalent width of Fe II pseudo-continuum measured at 2250-2650~\AA\ to EW of Mg~II line:
\begin{equation}
\mathrm{R_{FeII}}=\frac{\mathrm{EW (UV Fe~II)}}{\mathrm{EW(Mg~II)}},
\label{equ:rfe}
\end{equation}

{The wavelength range selected for the EW of UV Fe II is defined by that one {reported in the \citet{shen2019} catalog}.} In the case of quasars CTS~C30.10 and HE~0413-4031, the UV Fe~II was directly measured only in the 2700-2900 \AA~range and for a different UV Fe~II template, but for consistency we refitted these spectra using the UV Fe~II template of \citet{vestergaard2001}, and rescaled the newly derived EW(UV Fe~II) from 2700-2900~\AA~to 2250-2650~\AA~ by the factor 2.32 appropriate for the \citet{vestergaard2001} UV Fe~II template. For NGC~4151 and CTS~252, the EW of Mg~II and UV Fe~II is not reported in the required wavelength range, and hence we do not consider these objects in the analysis where \rfe\ is used. 

We also estimated the level of variability using the \fvar\ parameter \citep{rodriguez-pascual1997} defined by:
\begin{equation}
F\mathrm{_{var}}=\frac{(\sigma^2-\Delta^2)^{1/2}}{\langle f \rangle},
\label{equ:fvar}
\end{equation}
where $\sigma^2$ is the variance of the flux, $\Delta$ is the mean square value of the uncertainties ($\Delta_{i}$) associated with each flux measurement ($f_{i}$), and $\langle f \rangle$ is the mean flux. \fvar\ or similar expressions of the variability level are also correlated with the accretion rate intensity \citep{wilhite2005, macleod2010,  sanchez-saenz2018, martinez_aldama2019}. For the SDSS-RM sample, we use the {fractional RMS variability} provided by the \citet{shen2019} catalog (see their Table 2) for estimating \fvar, while for the rest of the objects \fvar\ values were taken from their originals works.  

{Furthermore, we consider the secondary parameters which can be derived from the basic measurements and in turn from the black hole mass: dimensionless accretion rate (\mdot) and the Eddington ratio (\eddr). Several results indicate that line dispersion ($\sigma_{\rm line}$) is a better estimator of the velocity field and in turn a better entity for the estimation of the black hole mass, because it is less biased than the FWHM \citep[e.g.][]{peterson2004, denney2013, dallabonta2020, yu2020_sigmaline}. However since $\sigma_{\rm line}$ is not reported for the SDSS-RM sample which includes the majority  of the sources used in this paper, our estimations are done considering the FWHM taken from the \citet{shen2019} catalog.} Then, to get the black hole mass (\mbh) estimation we consider a virial factor anti-correlated with the FWHM of the emission line  defined as $f_{\rm c}=\left({{\mathrm{FWHM_{MgII}}}}\,/\,{3200 {\pm 800}}\right)^{ -1.21\pm0.24}$ \citep{mejia2018}, the time delay ($\tau_\mathrm{obs}$) of Mg~II reported in Table~{\ref{tab:measurements}} (Appendix~\ref{appendix}) and the virial relation, \mbh=$f_c\,c\, \tau{\mathrm{_{obs}}} \, \mathrm{FWHM}^2/{G}$, where $c$ is the speed of light and $G$ the gravitational constant.


The dimensionless accretion rate (\mdot) was introduced by \citet{wang2014_mdot} assuming a Shakura-Sunyaev (SS) disk. Originally, \mdot\ is adjusted for the continuum at 5100~\AA, and therefore it must be re-scaled for other wavelengths. Following the standard SS accretion disk we get:
\begin{equation} 
\label{equ:mdot_ss}
\dot{\mathcal{M}}= \alpha\  \left(\frac{\lambda}{5100}\right)^{-1/2}  L_{\lambda}^{3/2}\,M_{\mathrm{BH}}^{-2}, 
\end{equation}
where $\lambda$ is the wavelength of the continuum luminosity ($L_{\lambda}$). Comparing with the formula given by \citet{wang2014_mdot}, \mdot\ at 3000~\AA\ is given by:
\begin{equation} \label{equ:mdot_3000}
\dot{\mathcal{M}}=26.2\left(\frac{{L}\mathrm{_{44}}}{\mathrm{cos}\,\it{\theta}}\right)^{3/2} m_{7}^{-2},
\end{equation}
where ${L}\mathrm{_{44}}$ is the luminosity at 3000~\AA\ in units of 10$^{44}$ \ergs, $\theta$ is the inclination angle of disk to the line of sight, and $m_{7}$ is the black hole mass in units of 10$^7$ $M_\odot$. {We considered cos $\it{\theta}$ = 0.75, which is the mean disk inclination for type 1 AGN  and it is in agreement with the typical torus covering factor found \citep[e.g.][]{lawrenceandelvis2010, ichikawa2015}}.

Previously, in \citet{zajacek2020} we used the \mdot\ definition adjusted for 5100~\AA, therefore the \mdot\ values reported there must be re-scaled simply multiplying by a factor of 1.3038. The corrected \mdot\ values for \citet{zajacek2020} sample are already included in {Table~\ref{tab:measurements}}.

The Eddington ratio is defined as \eddr=$\mathrm{bc} \cdot L_{3000}/L_\mathrm{Edd}$, where $\mathrm{bc}=5.62\pm1.14$ \citep{richards2006} and  $L\mathrm{_{Edd}}=1.5\times10^{38}\left(\frac{M\mathrm{_{BH}}}{M_\odot}\right)$. In Table~\ref{tab:measurements}, we report the dimensionless accretion rates as well as the Eddington ratios.{We use both dimensionless accretion rate and the Eddington ratio since it is not clear which of these two dimensionless values is determined more accurately. Eddington ratio requires the knowledge of the broad band spectrum or use of the bolometric correction which neglects, for example, the spin issue in a quasar and is based on a specific SED shape. The $\dot{\mathcal{M}}$ determination does not require the knowledge of the bolometric correction but the dependence on the black hole mass determination is much stronger than for the Eddington ratio, and the spectral slope in the quasar might not be consistent with theoretical power law expected from the \citet{SS1973} accretion disk model.}

\section{Results}
\label{sec_results}

\subsection{Standard radius-luminosity relation}

First, we use the full reverberation-mapped Mg~II sample and study the relation between the measured time delay of the Mg II line and the continuum flux at 3000~\AA. In Fig.~\ref{fig_RL_scatter_linear_combination1} (upper left panel), a clear linear trend is visible, although the RMS scatter, defined using the sample size $N$, observed values $\tau_i$, and predicted values $\hat{{\tau}}_i$ as,
\begin{equation}
\sigma_{\rm rms}=\sqrt{\frac{1}{N}\sum_{i=1}^{N}(\tau_i-\hat{{\tau}}_i)^2}\,,
\label{eq_rms_scatter}
\end{equation}
is considerable with $\sigma_{\rm rms}= 0.3014$ dex. The best fit parameters are given in {Table~\ref{tab_multivariate_regression}}, where we also give the values of the RMS scatter and the Pearson correlation coefficient {($r$)}. The correlation is moderately strong but clearly visible. The scatter is visibly larger than in \citet{zajacek2020}, where only eleven sources were available and the R-L relation for all the available sources at that time had a scatter of $\sigma_{\rm rms}=0.22$ dex and, $\sigma_{\rm rms}=0.19$ dex when two sources with the largest offset were removed. The slope is now also much shallower than given by \citet{zajacek2020} ($0.42 \pm 0.05$ for all 11 sources and, $0.58 \pm 0.07$ with two outliers removed, see their Fig.~5, right panel), or obtained earlier by \citet{mclure2002} at the basis of the Mg II line shape. The R-L luminosity in the past was much better studied for H${\beta}$ time delay with respect to the continuum at 5100 \AA, and then the slope was close to 0.5, when the host contamination was carefully taken into account (e.g. \citealt{bentz2013}). The correlation {for Mg II line} is partially (but not entirely) driven by the extreme points representing the lowest {(NGC~4151)} and the highest {(CTS~C30.10, HE~0413-4031 and CTS~252)} luminosity sources. 

Since it is generally accepted now that the large dispersion is related to the range of dimensionless accretion rates \citep{du2016,Yu2020, zajacek2020}, we mark the points in the plot with the color, corresponding to \mdot ~ values.  A certain degree of dependence is visible but the trend is not at all clear - yellower points occupy mostly the lower part of the diagram but they also concentrate more towards higher values of the luminosity. \citet{zajacek2020} achieve the considerable reduction of the scatter when the observed time delay was corrected for the trend with  \mdot\ or \eddr\, particularly if the virial factor $f_c$ depending on FWHM \citep{mejia2018,martinez_aldama2019} instead of a constant value has been used. In their work, \citet{lusso2017} use the additional dependence on FWHM to reduce the scatter in their quasar relation between the X-ray, and the UV flux, and \citet{du2019} and \citet{Yu2020} applied the additional dependence on the \mdot\ and \rfe\ to reduce the scatter in R-L relation for H$\beta$.  Thus, we can expect that including more parameters in the fit will lead to the reduction in the scatter to a smaller or larger extent. 

\subsection{Scatter reduction using a linear combination of variables}
\label{sect:combi}

The need to reduce the large scatter along the R-L relation motivates us to search for extended, multidimensional radius--luminosity relations that involve linear combinations of the logarithms of monochromatic luminosity ($L_{44}$) with additional parameters, which can generally be written as $\log{\tau_{\rm obs}}=K_1\log{L_{44}}+\sum_{i=2}^{n} K_i\log{Q_i}+K_{n+1}$, where the parameters $Q_i$ are typically related to the accretion rate and we use \mdot, \eddr, \fvar\, and \rfe\ for this purpose. The total number of quantities is typically $n=2$, in a few cases we have $n=3$. FWHM is either included or in some cases omitted from the linear combination. An overview of all studied cases is in {Table~\ref{tab_multivariate_regression}} (third column). 

We use the python packages \textmyfont{sklearn} and \textmyfont{statsmodels} to perform a multivariate linear regression and to obtain regression coefficients $K_1$, $K_i$, and $K_{n+1}$, including their standard errors, and the correlation coefficients ($r$), which are listed in {Table~\ref{tab_multivariate_regression}} for each linear combination. In addition, we also include the {RMS} scatter calculated using Eq.~\ref{eq_rms_scatter}.

\begin{figure*}
    \centering
    \includegraphics[width=\columnwidth]{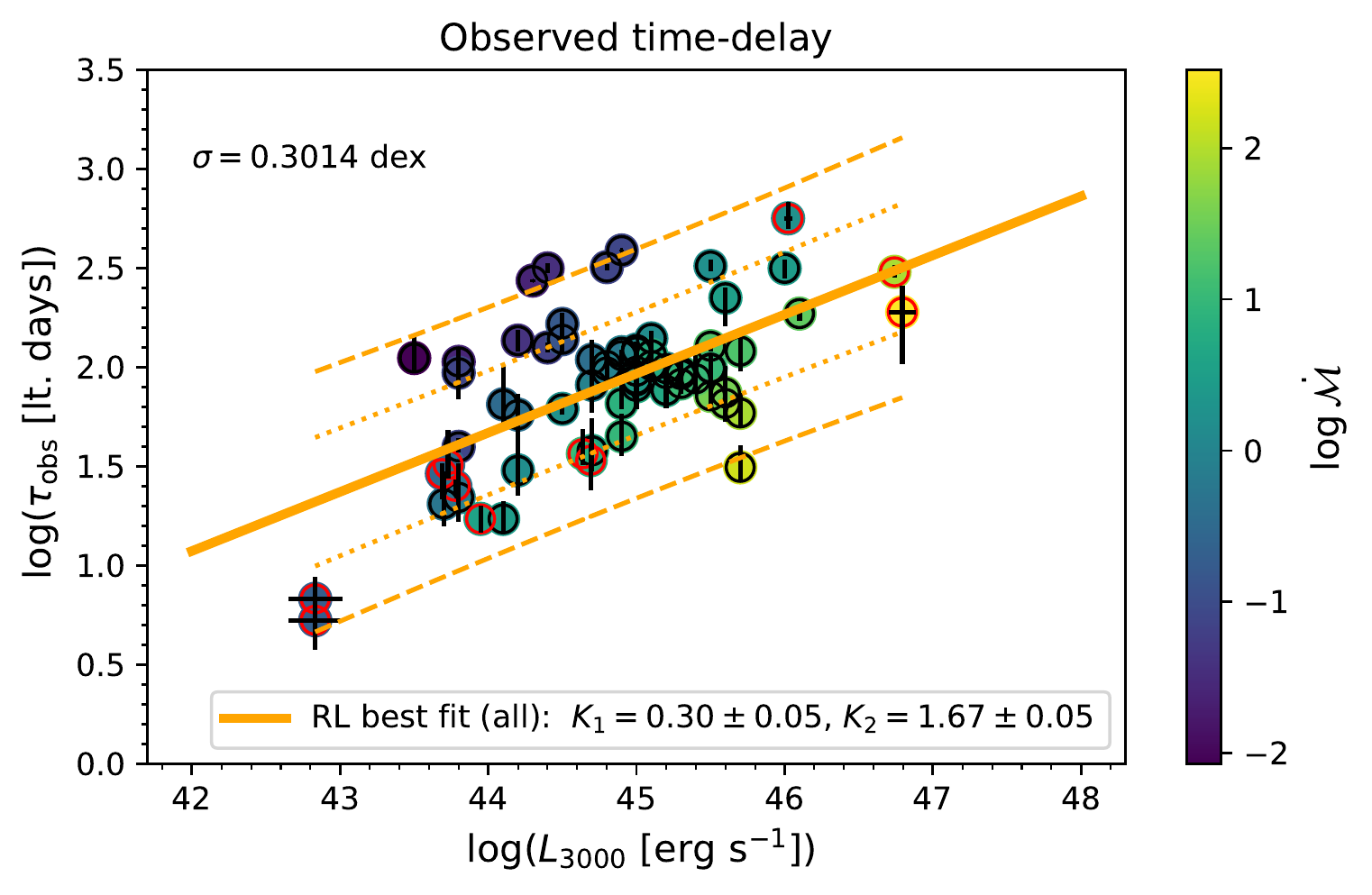}
    \includegraphics[width=\columnwidth]{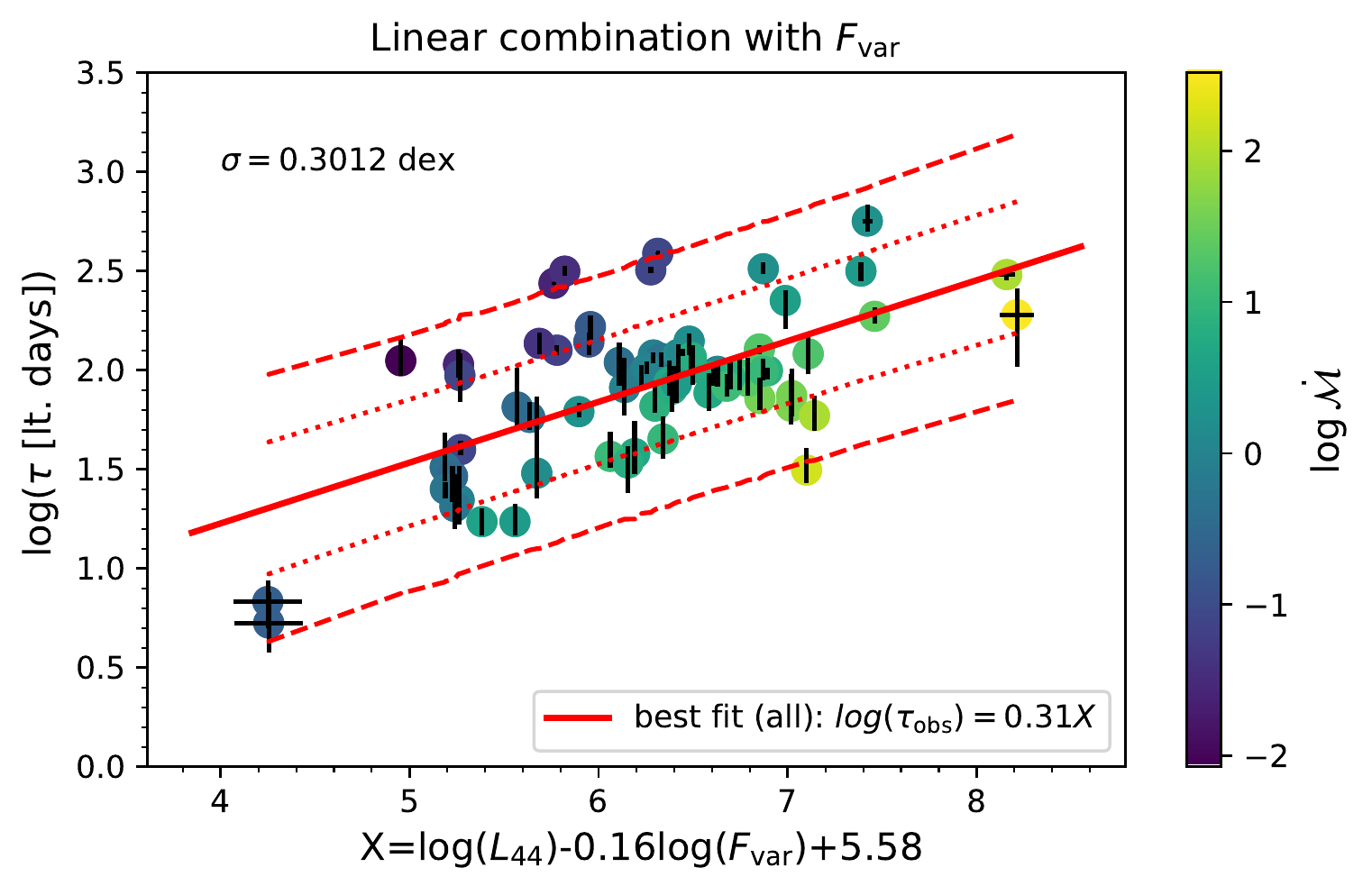}
    \includegraphics[width=\columnwidth]{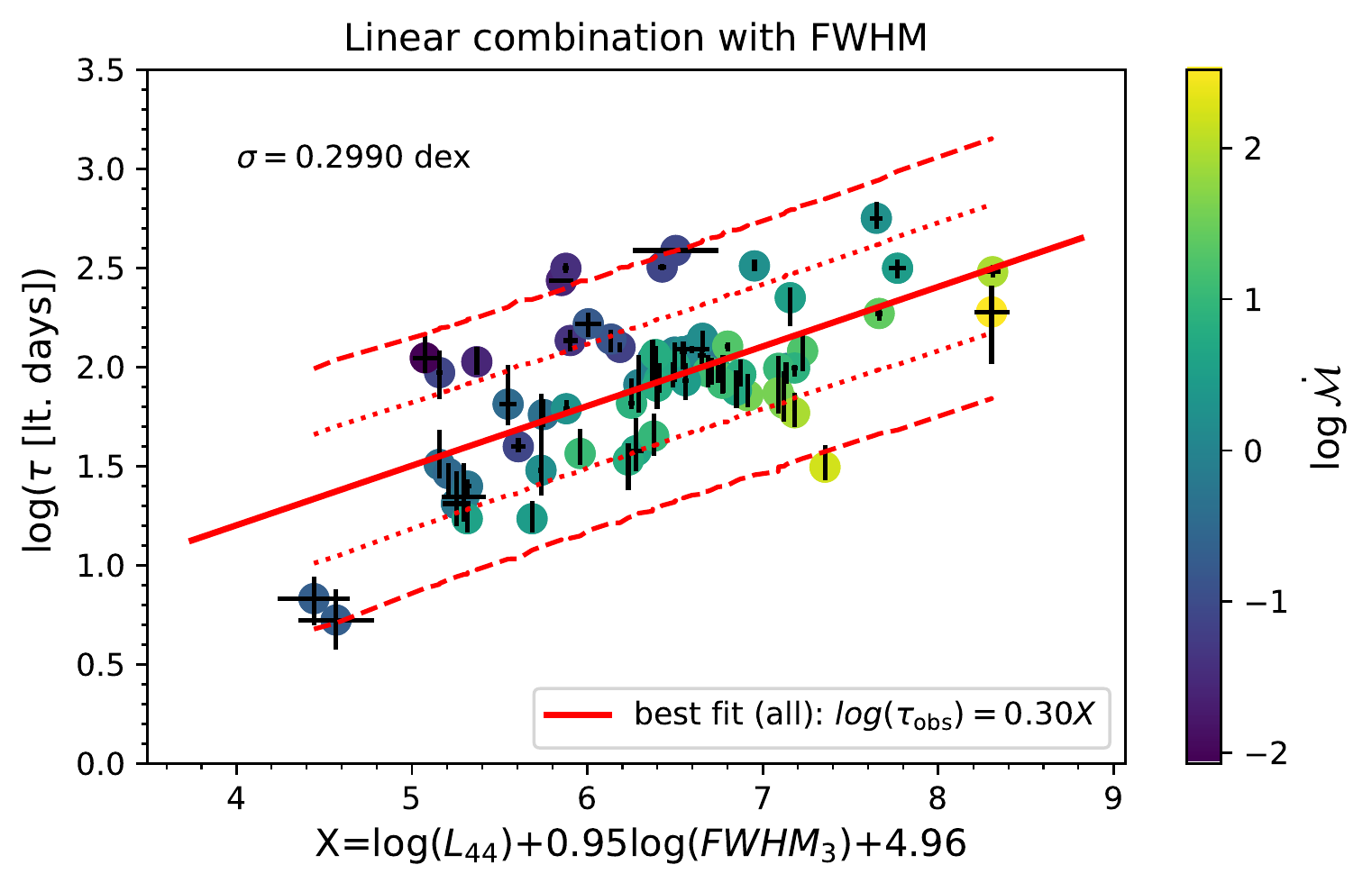}
    \includegraphics[width=\columnwidth]{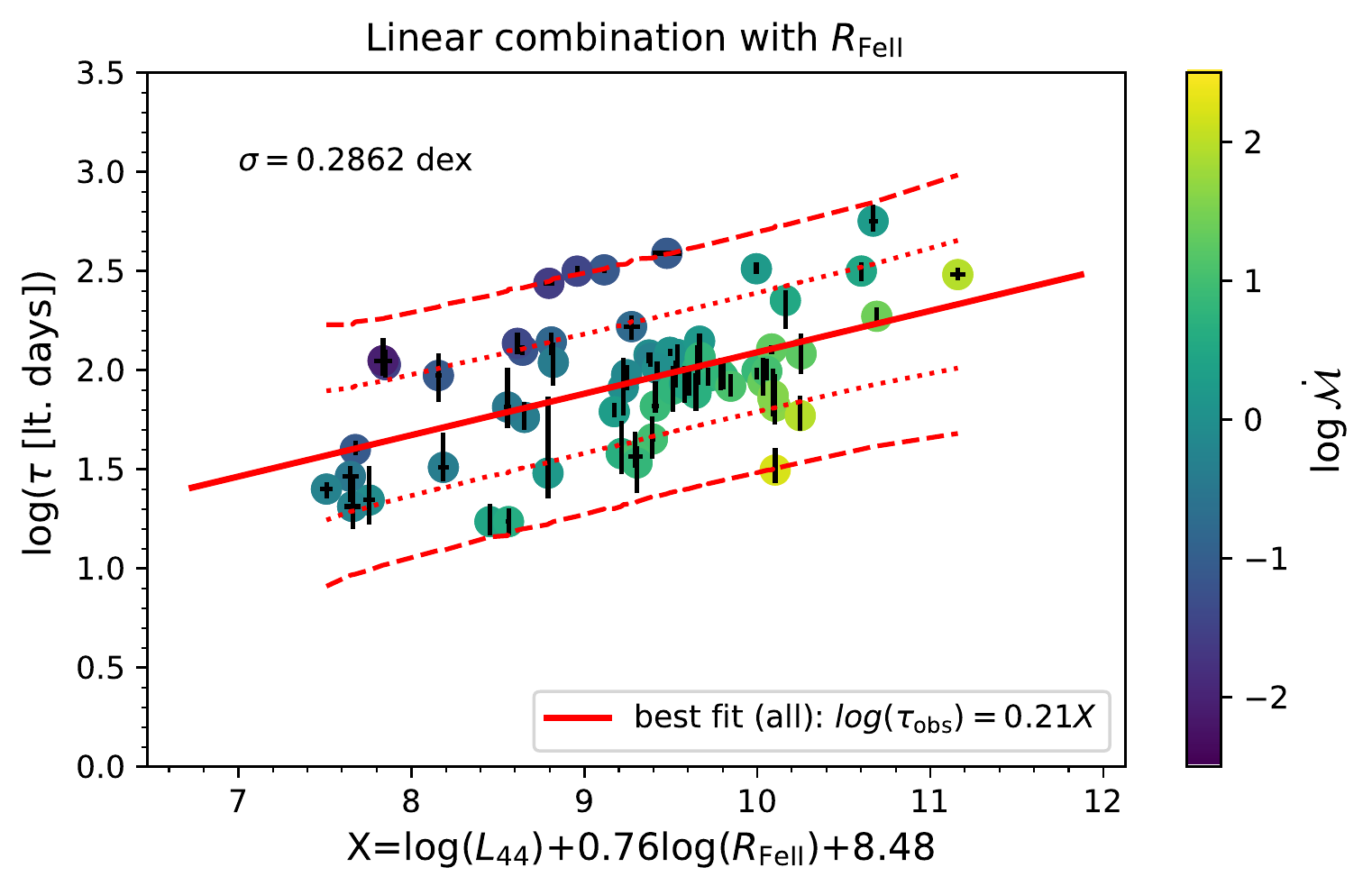}
     \includegraphics[width=\columnwidth]{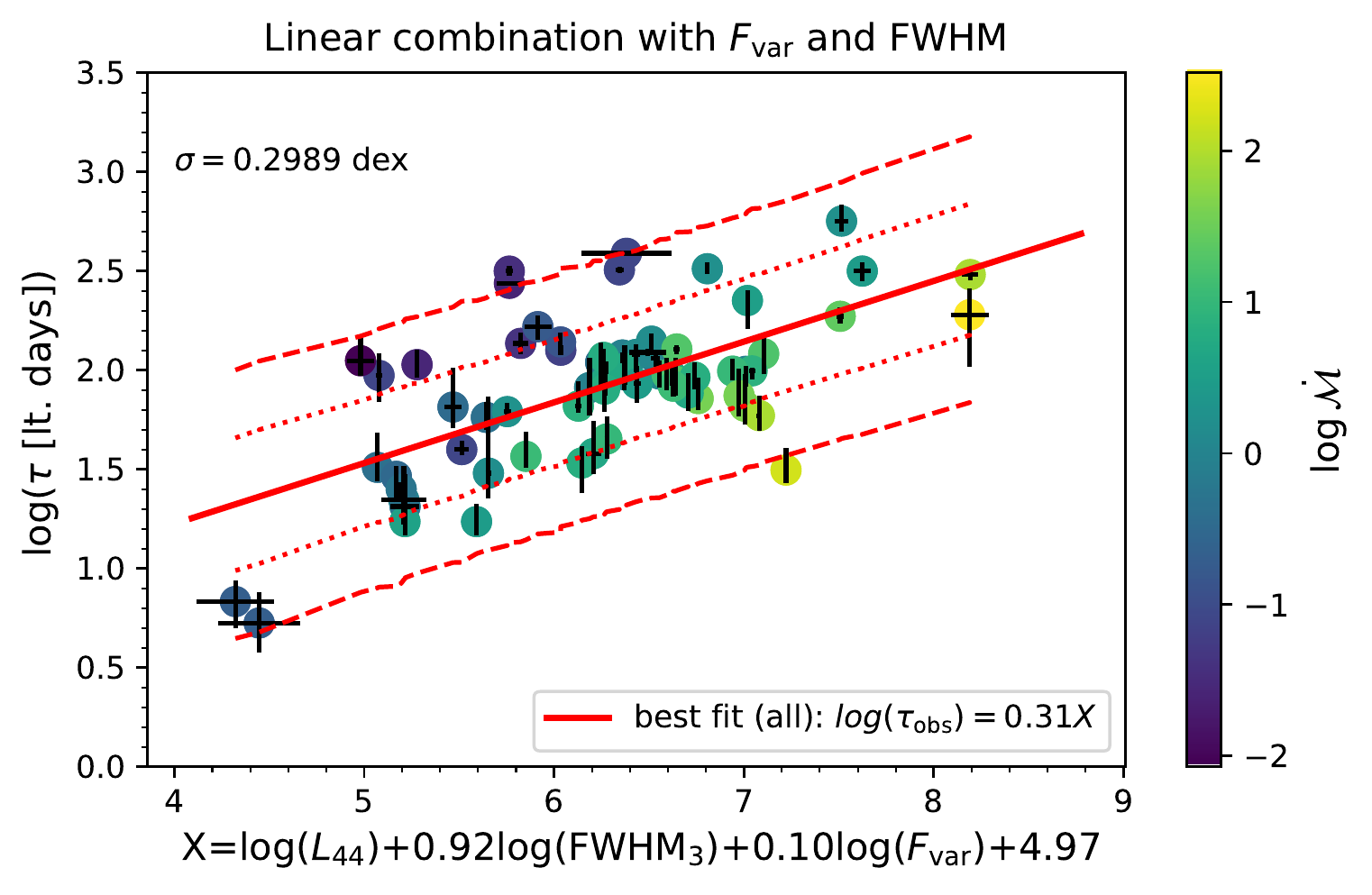} 
2    \caption{Observed time-delay ($\tau_{\rm obs}$) expressed as a linear combination of the monochromatic luminosity at 3000 \AA\ (top left panel), of luminosity $L_{44}$ and $F_{\rm var}$ parameter (top right panel), of $L_{44}$ and FWHM (middle left panel), {of $L_{44}$ and \rfe\ (middle right panel)} and of $L_{44}$, FWHM and $F_{\rm var}$ parameters (bottom panel). These combinations have a comparable scatter of $0.3\,{\rm dex}$. {Dotted and dashed lines denote the 68$\%$ and $95\%$ confidence intervals, respectively. Color code represents the intensity of the dimensionless accretion rate, \mdot.} }
    \label{fig_RL_scatter_linear_combination1}
\end{figure*}

\begin{figure*}
    \centering
    \includegraphics[width=\columnwidth]{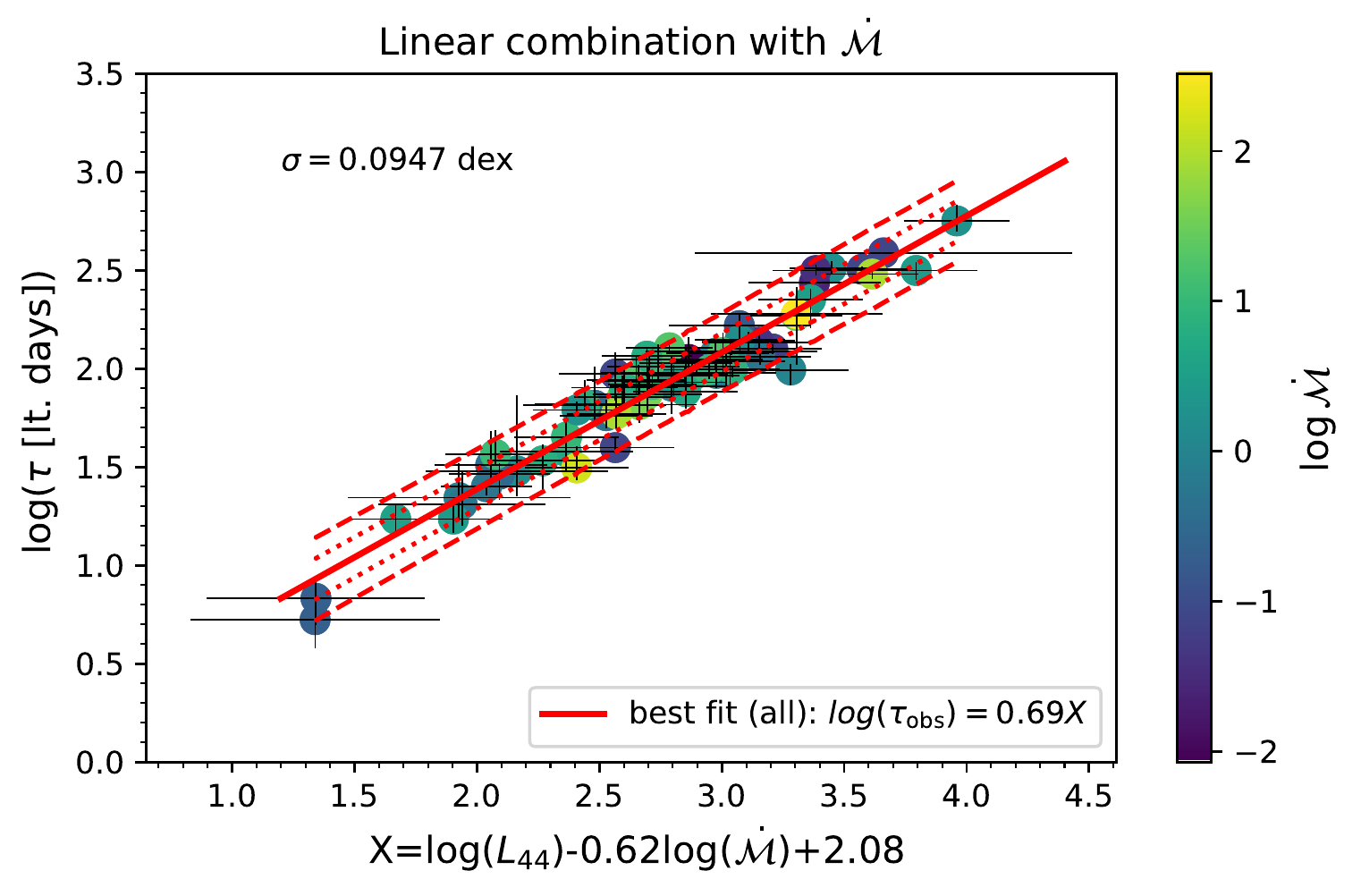}
    \includegraphics[width=\columnwidth]{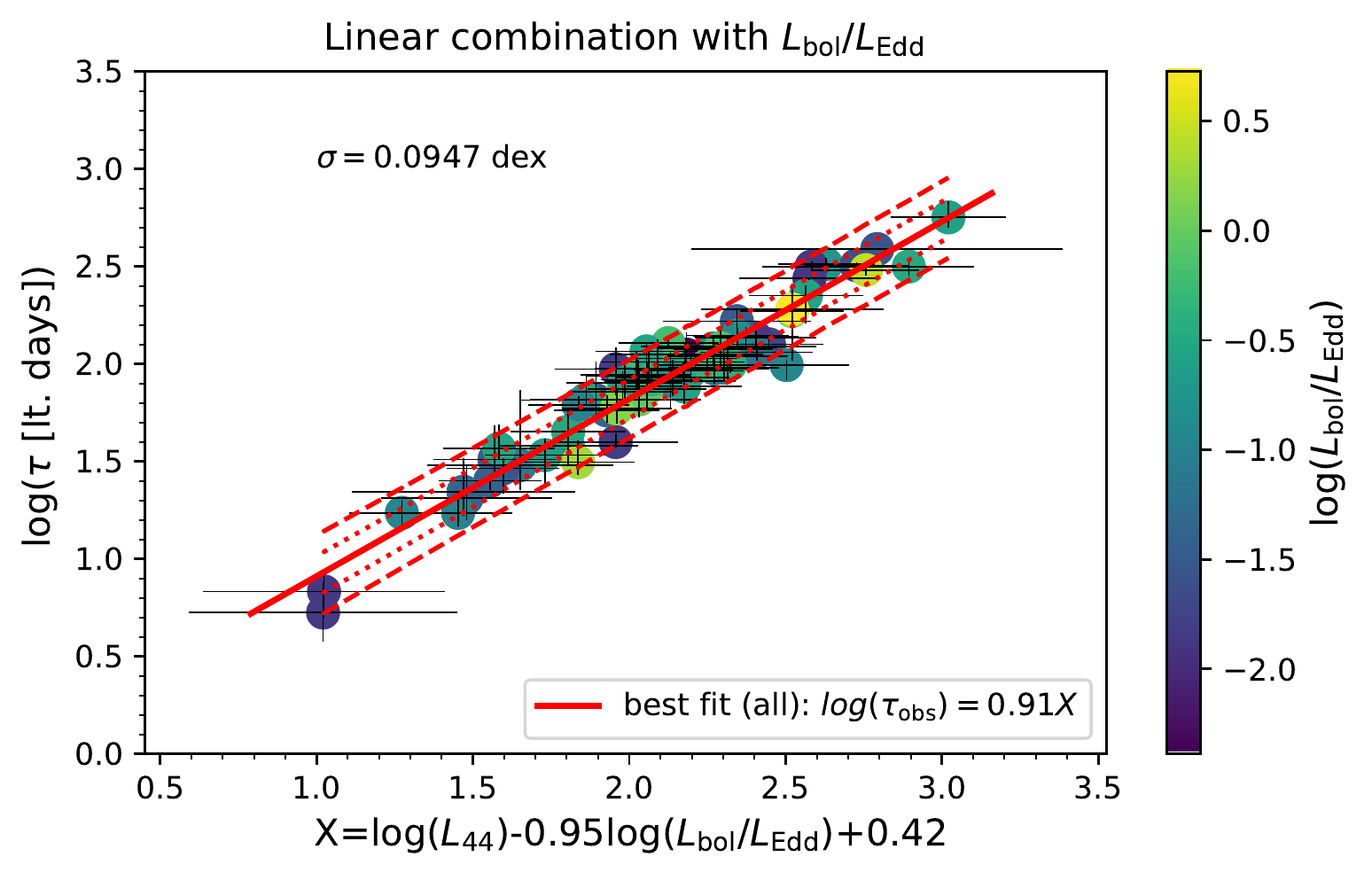}
    \caption{Observed time-delay (${\tau_{\rm obs}}$) as a function of the linear combination of the  of $L_{44}$ and $\dot{\mathcal{M}}$ (left panel) and of the  of $L_{44}$ and \eddr\ (right panel). The scatter along the linear relation is comparable in both cases, $\sigma_{\rm rms}\sim 0.1\,{\rm dex}$. {Dotted and dashed lines denote the 68$\%$ and $95\%$ confidence intervals, respectively. Color code represents the intensity of the dimensionless accretion rate, \mdot.} } 
    \label{fig_RL_scatter_linear_combination2}
\end{figure*}

We list all the relations in Table~\ref{tab_multivariate_regression} and graphically they are shown in {Figs.~\ref{fig_RL_scatter_linear_combination1} and \ref{fig_RL_scatter_linear_combination2}.} In these figures, for an easier comparison of the slopes $\tau_{\rm obs}\propto L_{44}^{\alpha}$, we plot $\log{\tau_{\rm obs}}$ vs. $\log{L_{44}}+\sum_{i=2}^{n} K_{i}/K_1\log{Q_i}+K_{n+1}/K_1$, where $K_1$ represents the slope $\alpha$. 

The smallest scatter of $\sigma_{\rm rms} \sim 0.1\,{\rm dex}$ is for combinations that include $L_{44}$ and either \mdot\ or \eddr\ (see Fig.~\ref{fig_RL_scatter_linear_combination2}). When in addition to \mdot\ or \eddr, FWHM is added to the combination, the scatter is of the order of $\sigma_{\rm rms} \sim 10^{-4}\,{\rm dex}$ only. However, in this case, the correlation is artificially enhanced (the correlation coefficient is essentially $1.00$) as both \mdot\ and \eddr\ depend on $\tau_{\rm obs}$ and FWHM via the black hole mass. {The use of such inter-dependent quantities is not welcome as that may easily create an apparent correlation, and the subsequent error determination when such a relation is used has to take that into account.} For this reason, we do not include these two cases in the overview of linear combinations in Table~\ref{tab_multivariate_regression}. 

{Considering independent quantities like FWHM, \rfe\, and \fvar\, decreases the scatter but the effect is relatively small and it is  comparable to the original radius--luminosity relation for Mg~II, $\sigma_{\rm rms}\sim 0.3\,{\rm dex}$ (see Fig.~\ref{fig_RL_scatter_linear_combination1}). The scatter drops by $0.8\%$ when FWHM is added to $L_{44}$ and only by $0.07\%$ when \fvar\ is added to $L_{\rm 44}$. 
This is not a significant improvement in terms of the scatter, but the relations define planes instead of a line and thus connect three or four independent observables, which can be relevant in terms of understanding mutual relations among them.}

An additional dependence on \rfe\, gives a better result, the scatter drops from 0.3 dex down to 0.286 dex (see Table~\ref{fig_RL_scatter_linear_combination1}) but the correlation coefficient also drops so apparently \rfe\, adds considerably to the scatter. Also, the slope of the relation is shallower. {Although \rfe\ and \fvar\ are physically related to the Eddington ratio, neither of these quantities leads to the considerable improvement when the whole sample is considered. }

Combining even more quantities does not provide an improvement either: a combination of $L_{44}$, FWHM, \fvar\ and \rfe\ still leads to almost the same scatter and the same value of the correlation coefficient as in the basic R-L relation. Other combinations also do not provide an improvement.

\begin{table*}[]
    \caption{Results of the parameter inference applied to the logarithm of the observed time-delay ($\tau\mathrm{_{obs}}$) expressed as a linear combination of the logarithm of the monochromatic luminosity {(expressed as $L_{3000}$ or $L_{44}$)} and other quantities. We analyze parameter values as well as the scatter for the whole Mg~II sample (denoted as `All') or its low- or high-accretion sub-samples (denoted as `Low' or `High', respectively), which is specified in the first column. The parameter inference type -- ordinary least squares (OLS) or Markov-Chain Monte Carlo (MCMC) -- is specified in the second column. The third column lists the analyzed parameter combination. The other columns contain coefficient values with standard errors {within 1$\sigma$}, RMS scatter ($\sigma{_{\rm rms}}$) along the linear relations in dex, and the {Pearson} correlation coefficient ($r$). For the two cases with the smallest RMS scatter, namely combinations with $\log{L_{44}}$, $\log{\dot{\mathcal{M}}}$ and $\log{L_{44}}$, $\log{R_{\rm FeII}}$, we perform Markov-Chain Monte Carlo (MCMC) fitting to cross-check our linear-regression results {with and without the underestimation factor $f$, which, if included, is listed in the same column as $K_4$ coefficient}.}
  \hspace{-3cm}
    \resizebox{2.5\columnwidth}{!}{
    \begin{tabular}{c|c|c|c|c|c|c|c|c}
    \hline
    \hline
    Sample & Inference & $\log{\tau_{\rm obs}}=$  & $K_1$ & $K_2$ & $K_3$ & $K_4$ or $f$ & $\sigma_{\rm rms}$ [dex] & $r$\\
    \hline
    All & OLS & $K_1\log{L_{44}}+K_2$ & $0.298 \pm 0.047$  & $1.670 \pm 0.053$  & - & - &  $0.3014$ & $0.62$    \\
    Low & OLS & $K_1\log{L_{44}}+K_2$ & $0.520 \pm 0.078$ & $1.732 \pm 0.056$ & - & - &  $0.2815$ & $0.76$    \\
    High & OLS & $K_1\log{L_{44}}+K_2$ & $0.414 \pm 0.058$  & $1.382 \pm 0.085$ & - & - &  $0.2012$ & $0.78$    \\
    \hline
    All & OLS & $K_1\log{L_{44}}+K_2\log{{\rm FWHM_3}}+K_3$ & $0.30\pm 0.05$ & $0.29 \pm 0.28$ & $1.49 \pm 0.18$ & - & $0.2990$ & $0.63$\\
    Low & OLS &  $K_1\log{L_{44}}+K_2\log{{\rm FWHM_3}}+K_3$ & $0.54 \pm 0.08$ & $-0.59 \pm 0.40$   & $2.12 \pm 0.27$  &  -   & $0.2722$ &  $0.78$\\
    High & OLS &  $K_1\log{L_{44}}+K_2\log{{\rm FWHM_3}}+K_3$ & $0.42 \pm 0.06$  & $-0.12 \pm 0.32$  & $1.44 \pm 0.17$  &  -   & $0.2007$ &  $0.78$\\
    \hline
    All & OLS & $K_1\log{L_{44}}+K_2\log{\dot{\mathcal{M}}}+K_3$ & $0.694\pm 0.022$ & $-0.432 \pm 0.018$ & $1.442 \pm 0.019$ & - & $0.0947$ & $0.97$\\
    All & MCMC & $K_1\log{L_{44}}+K_2\log{\dot{\mathcal{M}}}+K_3$ & $0.6965_{-0.0098}^{+0.0102}$ & $-0.4550_{-0.0054}^{+0.0054}$ & $1.4618_{-0.0128}^{+0.0124}$ & - & $0.0985$ & $0.97$\\
    All & MCMC (f) & $K_1\log{L_{44}}+K_2\log{\dot{\mathcal{M}}}+K_3$ & $0.6984_{-0.0234}^{+0.0236}$ & $-0.4558_{-0.0125}^{+0.0127}$  & $1.4600_{-0.0293}^{+0.0297}$ & $f=2.3730_{-0.1916}^{+0.2292}$ & $0.0985$ & $0.97$ \\
    \hline
     All & OLS & $K_1\log{L_{44}}+K_2\log{L/L\mathrm{_{Edd}}}+K_3$ & $0.910\pm 0.029$ & $-0.863 \pm 0.035$ & $0.380 \pm 0.056$ & - & $0.0947$ & $0.97$\\ 
     \hline
     All & OLS & $K_1\log{L_{44}}+K_2\log{R_{\rm FeII}}+K_3$ & $0.21 \pm 0.06$ & $0.16 \pm 0.16$ & $1.77 \pm 0.07$ & - & $0.2862$ & $0.51$\\
     Low & OLS &  $K_1\log{L_{44}}+K_2\log{R_{\rm FeII}}+K_3$ & $0.27\pm 0.11$ & $0.35 \pm 0.18$ & $1.92 \pm 0.08$ & - & $0.2521$ & $0.68$\\
     High & OLS &  $K_1\log{L_{44}}+K_2\log{R_{\rm FeII}}+K_3$ & $0.47\pm 0.06$ & $1.04 \pm 0.33$ & $1.25 \pm 0.08$ & - & $0.1718$ & $0.84$\\
     High & MCMC &  $K_1\log{L_{44}}+K_2\log{R_{\rm FeII}}+K_3$ & $0.4749_{-0.0178}^{+0.0177}$ & $1.0647_{-0.1146}^{+0.1120}$ & $1.2678_{-0.0328}^{+0.0327}$ & - & $0.1743$ & $0.85$\\ 
      High & MCMC (f) &  $K_1\log{L_{44}}+K_2\log{R_{\rm FeII}}+K_3$ & $0.4761_{-0.0397}^{+0.0404}$ & $1.0735_{-0.2725}^{+0.2663}$ & $1.2659_{-0.0751}^{+0.0718}$ & $f=2.3420_{-0.2671}^{+0.3458}$ & $0.1743$ & $0.85$\\
     \hline
     All & OLS & $K_1\log{L_{44}}+K_2\log{F_{\rm var}}+K_3$ & $0.307 \pm 0.056$ & $0.048 \pm 0.178$ & $1.712 \pm 0.165$ & - & $0.3012$ & $0.62$\\ 
     Low & OLS &   $K_1\log{L_{44}}+K_2\log{F_{\rm var}}+K_3$ & $0.550 \pm 0.087$ & $0.173 \pm 0.223$ & $1.879 \pm 0.198$ & - & $0.2788$ & $0.77$\\ 
     High & OLS &   $K_1\log{L_{44}}+K_2\log{F_{\rm var}}+K_3$ & $0.370 \pm 0.058$ & $-0.406 \pm 0.179$ & $0.980 \pm 0.194$ & - & $0.1863$ & $0.82$\\ 
     \hline
     All & OLS & $K_1\log{L_{44}}+K_2\log{{\rm FWHM}_3}+K_3\log{F_{\rm var}}+K_4$ & $0.306 \pm 0.056$ & $0.280 \pm 0.282$ & $0.032 \pm 0.178$ & $K_4=1.522 \pm 0.252$ & $0.2989$ & $0.63$\\
     \hline
     \hline
    \end{tabular}}
    \label{tab_multivariate_regression}
\end{table*}

\subsection{Sample division: low and high accretors}
\label{sec:sample_division}

\begin{figure}
    \centering
    \includegraphics[width=\columnwidth]{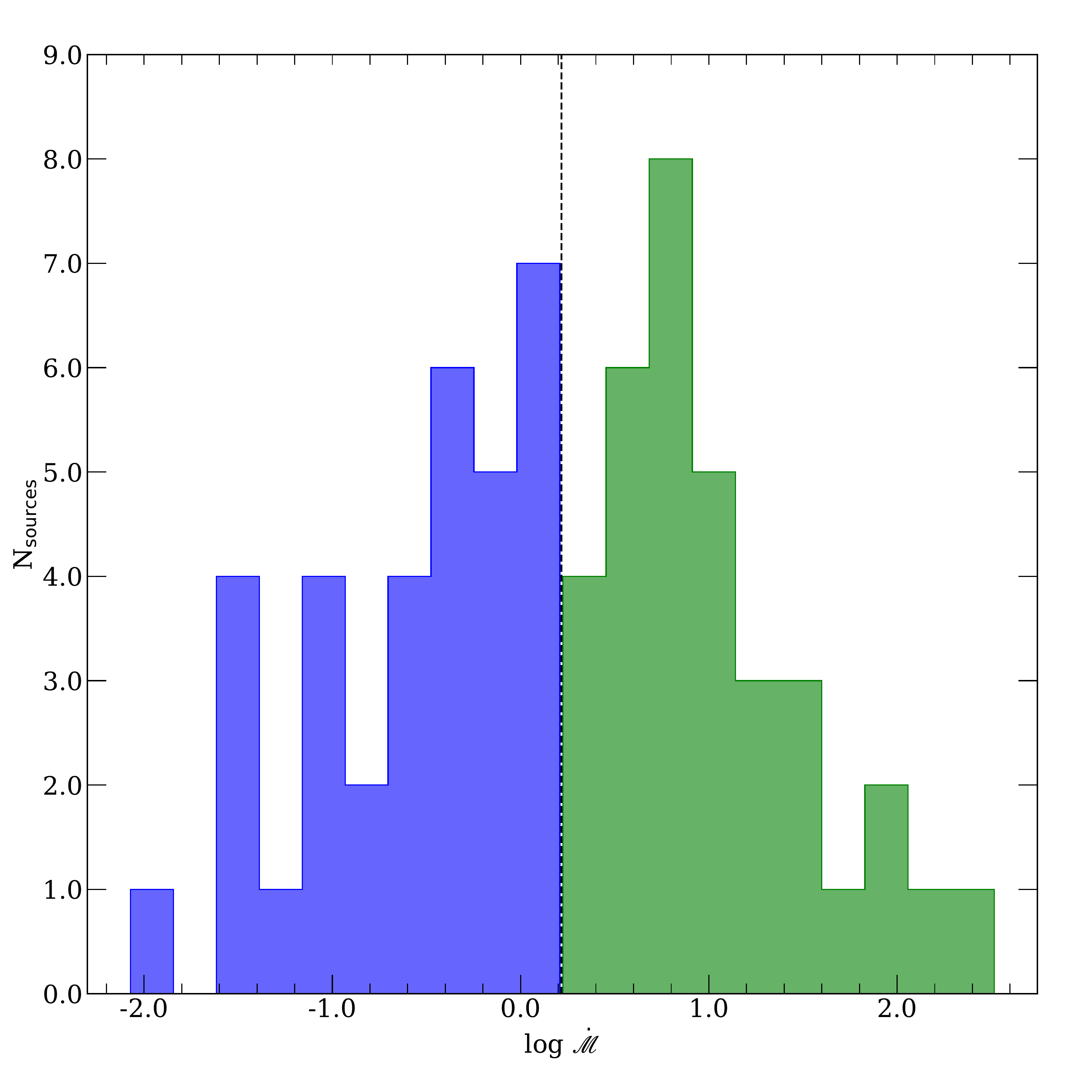}
    \caption{Dimensionless accretion rate distribution for the full sample. Blue and green histogram represent the low and high accretion sub-samples, respectively. Vertical line indicates the median \mdot\ value for the full sample, log \mdot=0.2167}.
    \label{fig:hist_mdot}
\end{figure}

To make use of the strong dependence on \mdot~ in the reduction of the scatter in Sect.~\ref{sect:combi}, we divided the full sample into two sub-samples: low and high accretors. Although limits for low and high accretors have previously been discussed \cite[e.g.][]{marziani2003,du2015}, in this work we consider a division which gives us the sub-samples containing comparable number of objects. Thus, as a reference, we consider the median \mdot\ value (log~\mdot=0.2167) to get an equal number of sources (34 objects) in each sub-sample. In Fig.~\ref{fig:hist_mdot}, the \mdot\ distribution is shown for low and high accretors.  {In Section~\ref{sec:disc_sample_division}, we include a discussion of the accretion rates observed in our sample in a general context considering samples like the DR7 \citep{shen2011} and DR14 \citep{rakshitetal2019}, which also support the \mdot\ division considered in this work.}

\begin{figure*}
    \centering
    \includegraphics[width=\columnwidth]{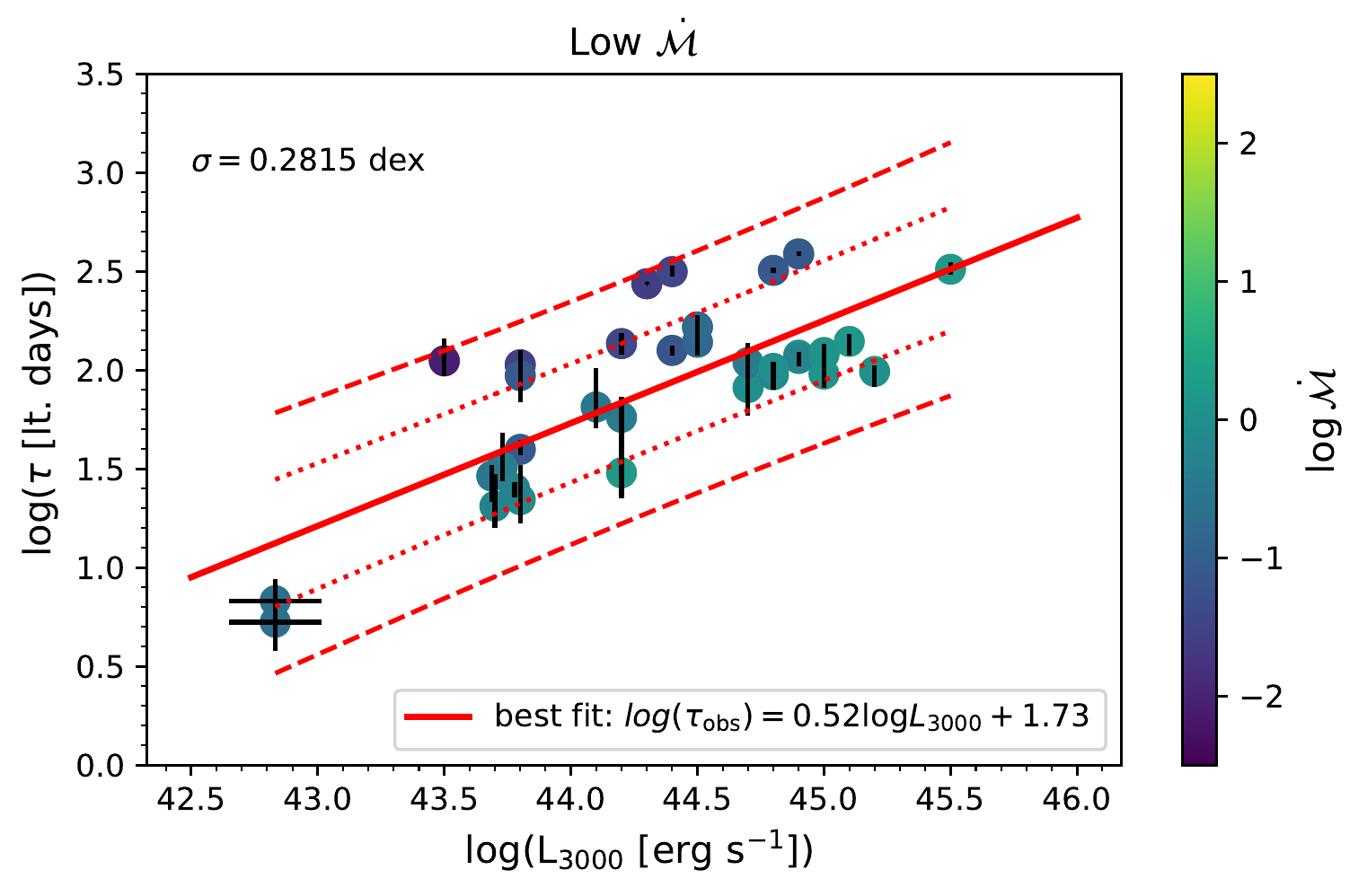}
    \includegraphics[width=\columnwidth]{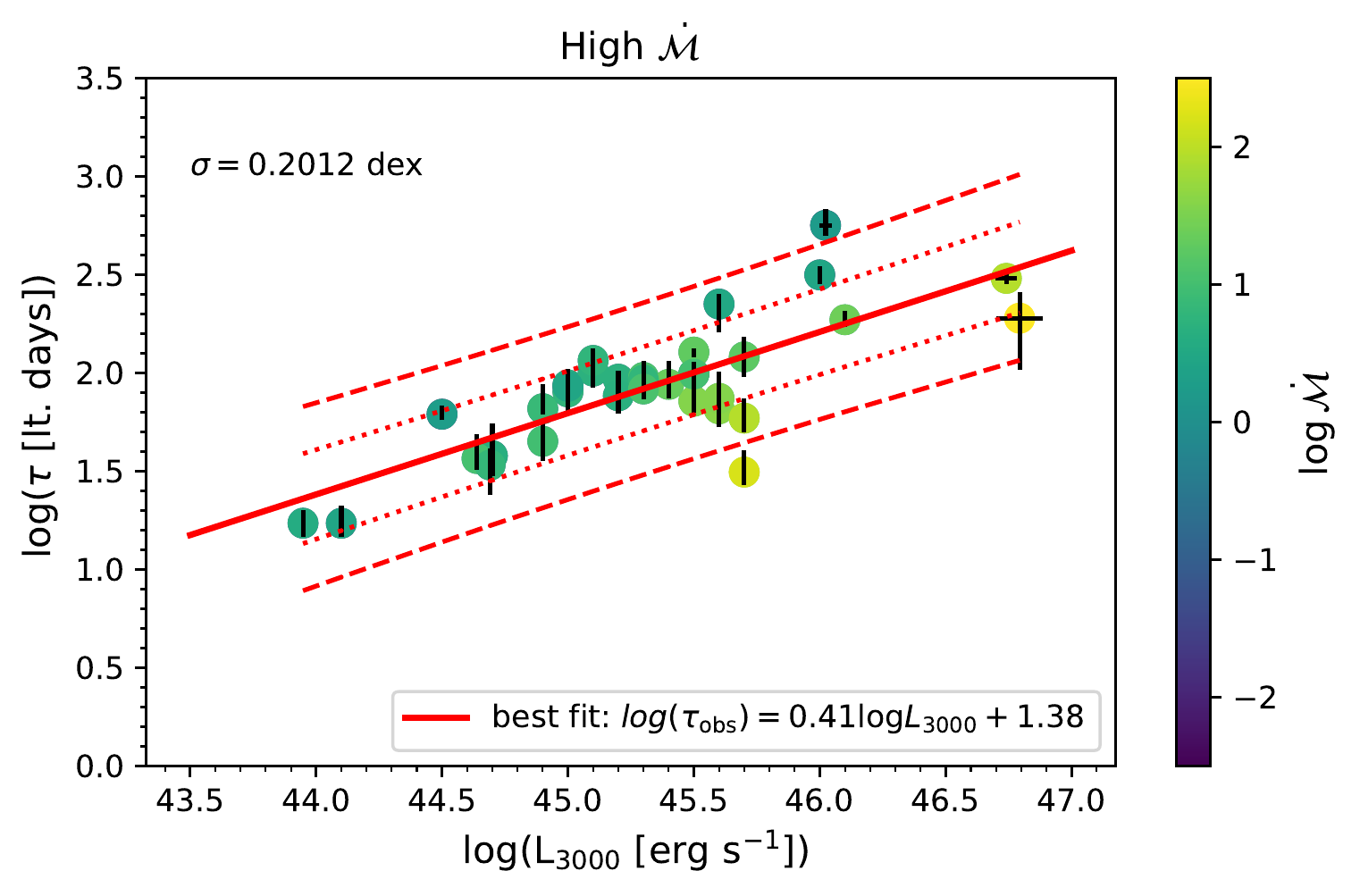}
    \includegraphics[width=\columnwidth]{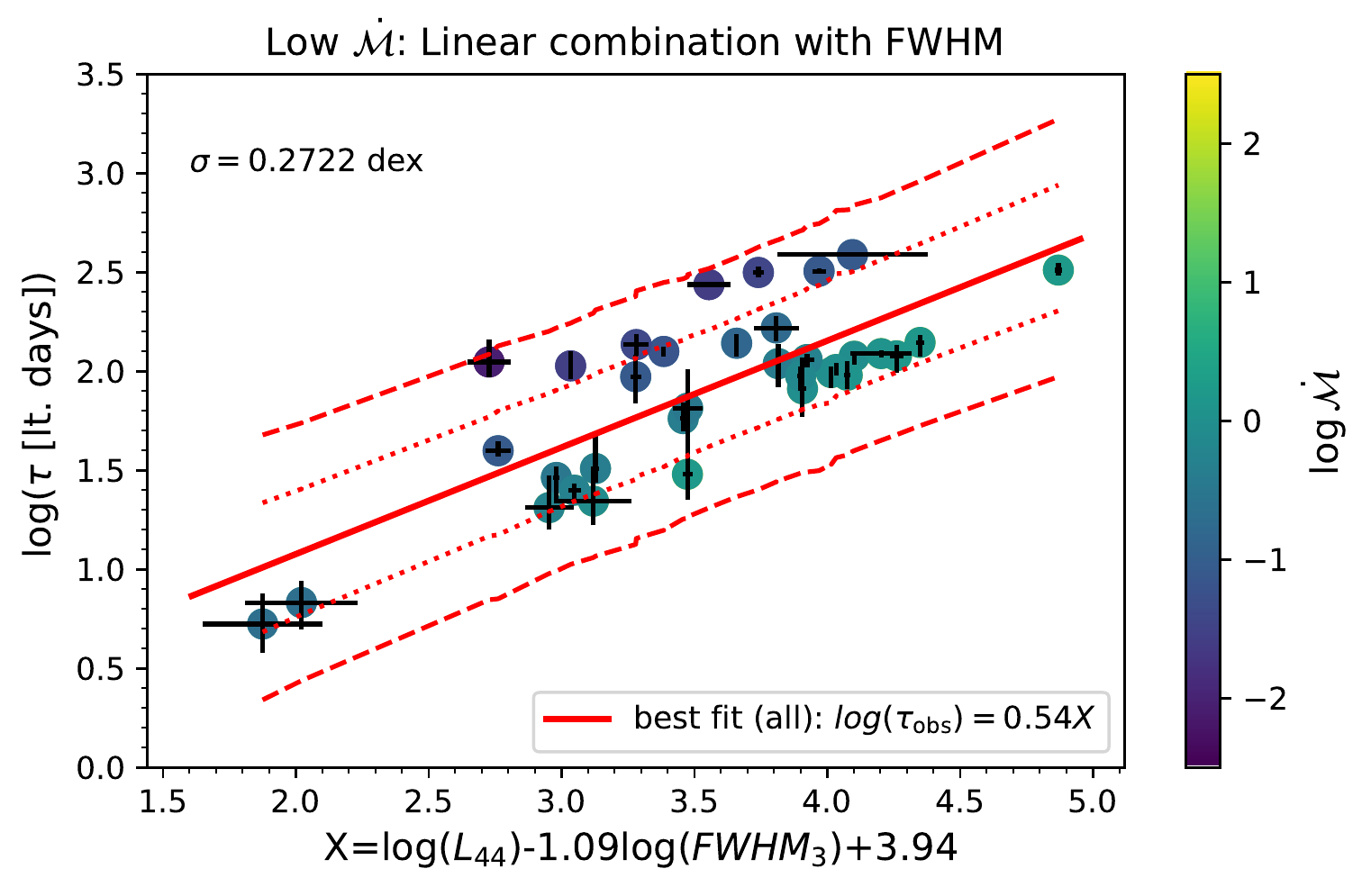}
     \includegraphics[width=\columnwidth]{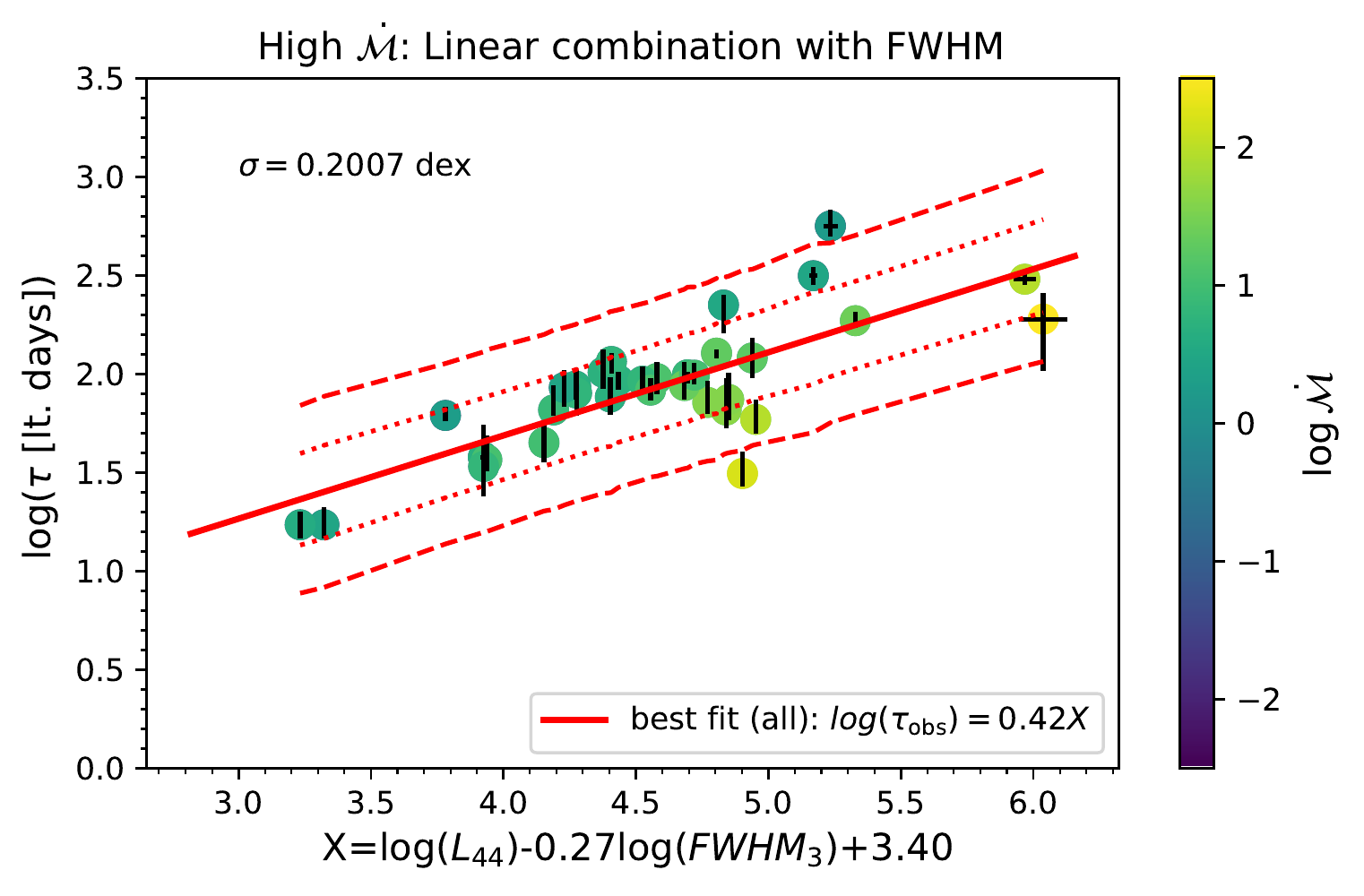}    
    \caption{Observed time-delay ($\tau_{\rm obs}$) expressed as a function of the monochromatic luminosity $L_{3000}$ (top panels) and as a function of the linear combination $\log{L_{44}}+(K_2/K_1)\log{\mathrm{FWHM}_3}+(K_3/K_1$) (bottom panels), {where FWHM$_3$ corresponds to the FWHM is in units of 10$^3$ \kms}. In addition, we divide the Mg II sample into low accretors (left panels) and high accretors (right panels), see Sec.~\ref{sec:sample_division}. {Dotted and dashed lines denote the 68$\%$ and $95\%$ confidence intervals, respectively. Color code represents the intensity of the dimensionless accretion rate, \mdot.} }
    \label{fig_RL_scatter_linear_combination3}
\end{figure*}

The division of the sample into high and low accretors results in significantly reducing the scatter, { particularly for the highly accreting sub-sample} (see Table~\ref{tab_multivariate_regression} and Fig.~\ref{fig_RL_scatter_linear_combination3}). {It is also interesting to note that the slopes in the R-L relation in both sub-samples are steeper than for the whole sample, and much closer to the theoretically expected value of 0.5.} {Comparing both cases, for low accretors, the slope is steeper, closer to the canonical 0.5 value than for the highly accreting sub-sample}, but the difference is within the quoted slope errors. 

The Pearson correlation coefficient increased significantly for both sub-samples supporting the view that mixing sources with different accretion rates spoils the pattern. The dispersion in both sub-samples decreased in comparison with the whole sample, but here the effect in the two sub-samples is clearly different. In low \mdot\ sub-sample, the reduction in the dispersion is not that strong, from $0.3$ dex down to $0.28$ dex. However, the drop in the dispersion for the highly accreting sources is spectacular, from 0.3 dex to 0.2 dex. We should stress here again that this division has been set at \mdot\ = 1.65, and the definition of \mdot\ does not include the accretion efficiency, so for a standard efficiency of 10\%, this corresponds to mild accretion rates above 0.165 in dimensionless units. Hence, a considerable fraction of quasars in the SDSS-RM objects \citep{homayouni2020} belongs to this category (see Section~\ref{sec:disc_sample_division}). 

For low-accretion rate sub-sample, our dispersion in R-L relation for Mg II is still higher than the dispersion of 0.19  dex obtained by \citet{bentz2013} for R-L in H$\beta$, without the removal of outliers. However, our sources are, on average, brighter than those in \citet{bentz2013}, where most of the sources are at $\log L_{5100} \sim 43.5$, and we study a different emission line. For high-accretion rate sources, our dispersion is comparable to \citet{bentz2013}, and we did not remove any outliers in our analysis. Removing outliers (e.g. by 3-$\sigma$ clipping) would clearly tighten the correlation but we use the data from the literature and we do not think we can reliably eliminate some of the available measurements. This shows that for Mg II quasar population, the highly accreting sources are much more attractive for cosmological applications. 

\begin{figure*}
    \centering
    \includegraphics[width=\columnwidth]{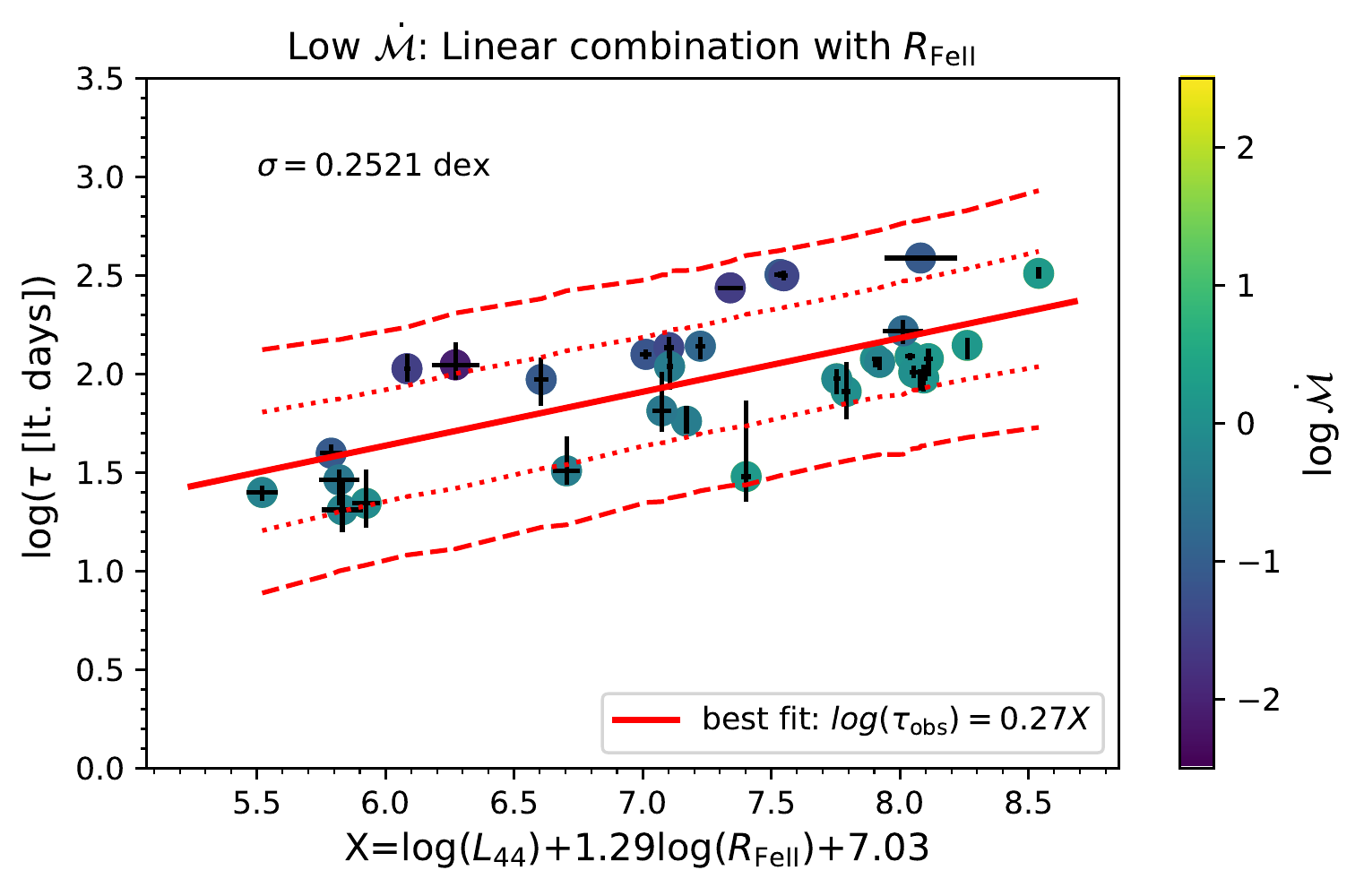}
    \includegraphics[width=\columnwidth]{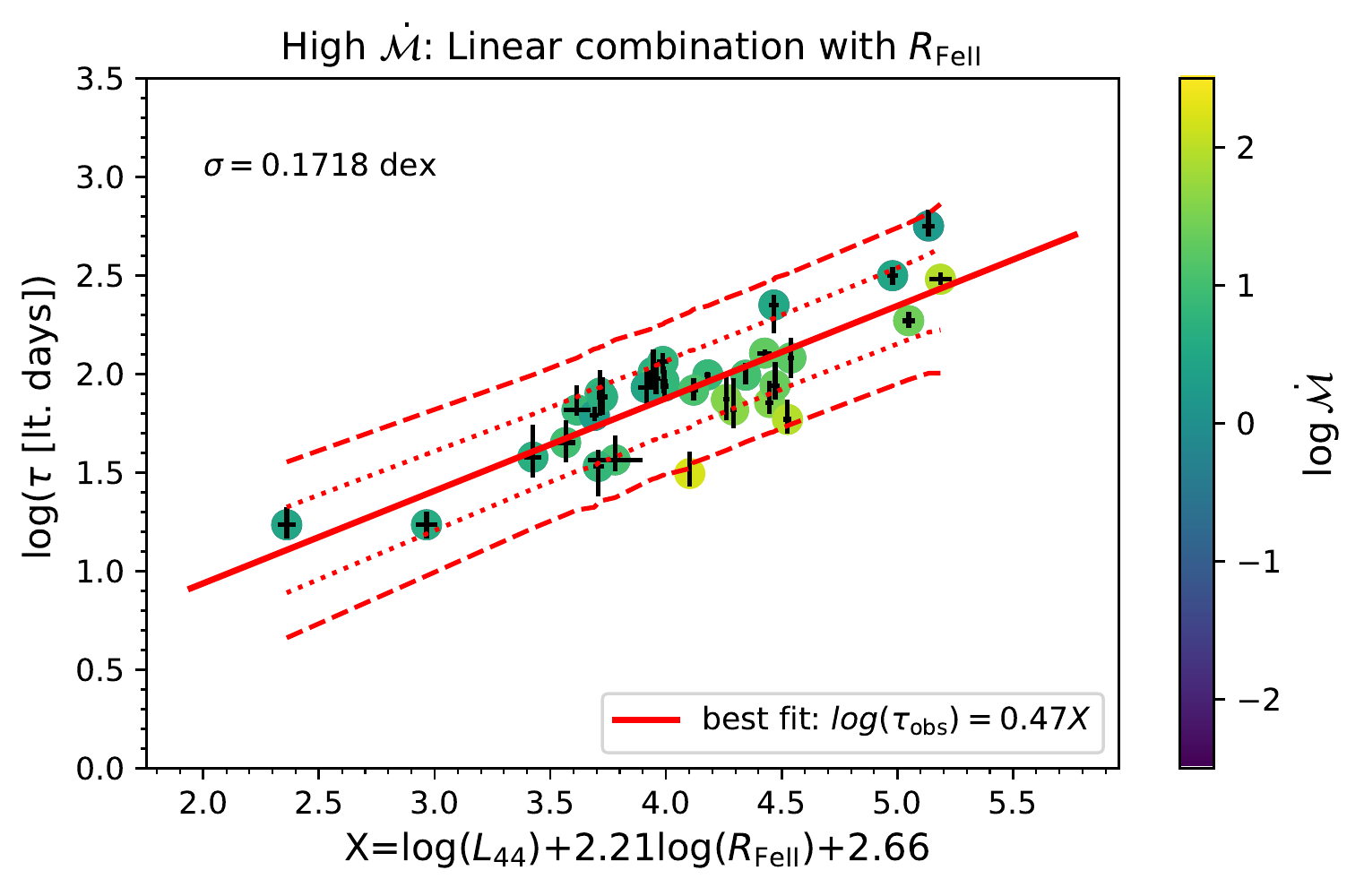}
    \includegraphics[width=\columnwidth]{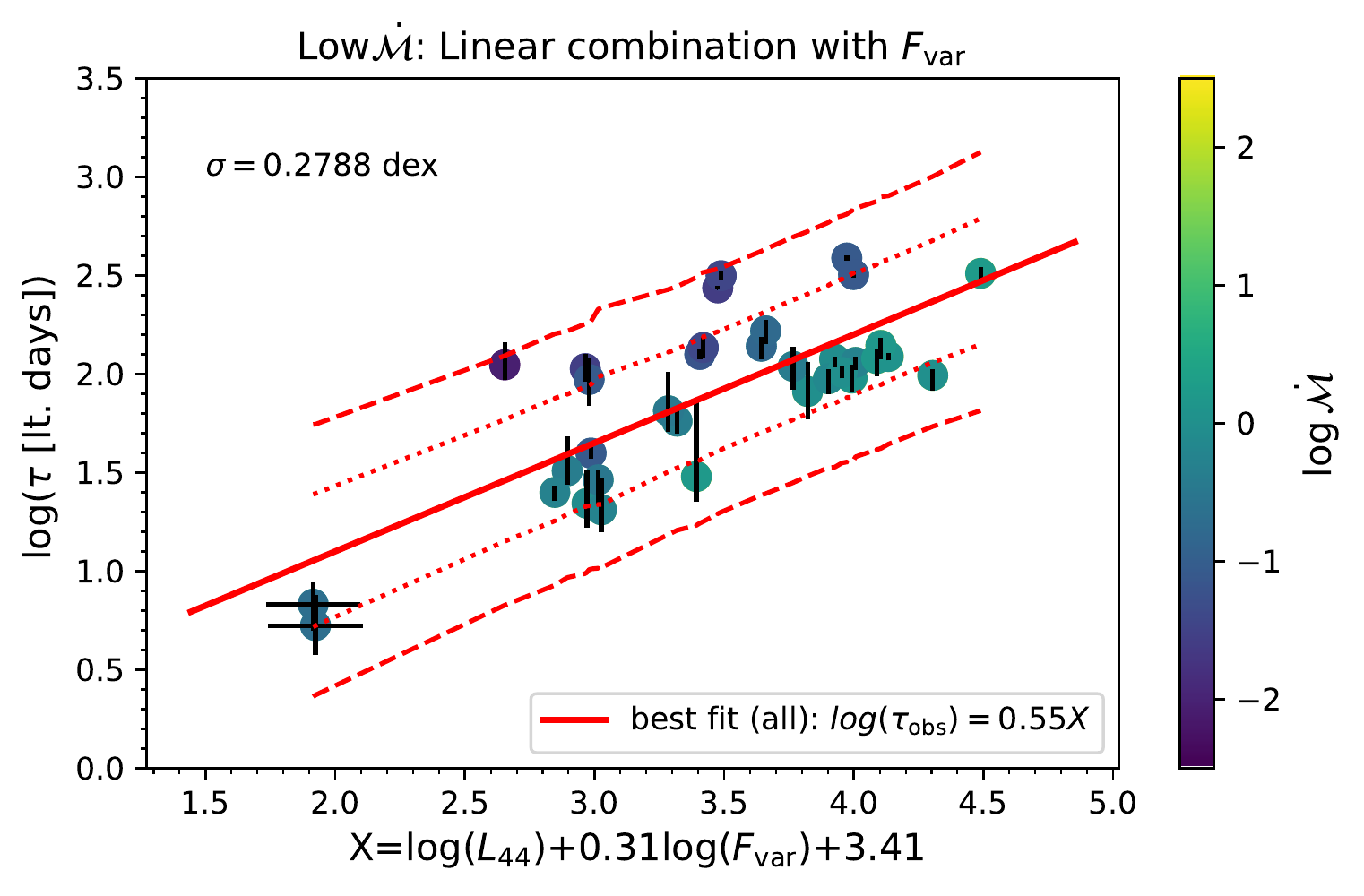}
    \includegraphics[width=\columnwidth]{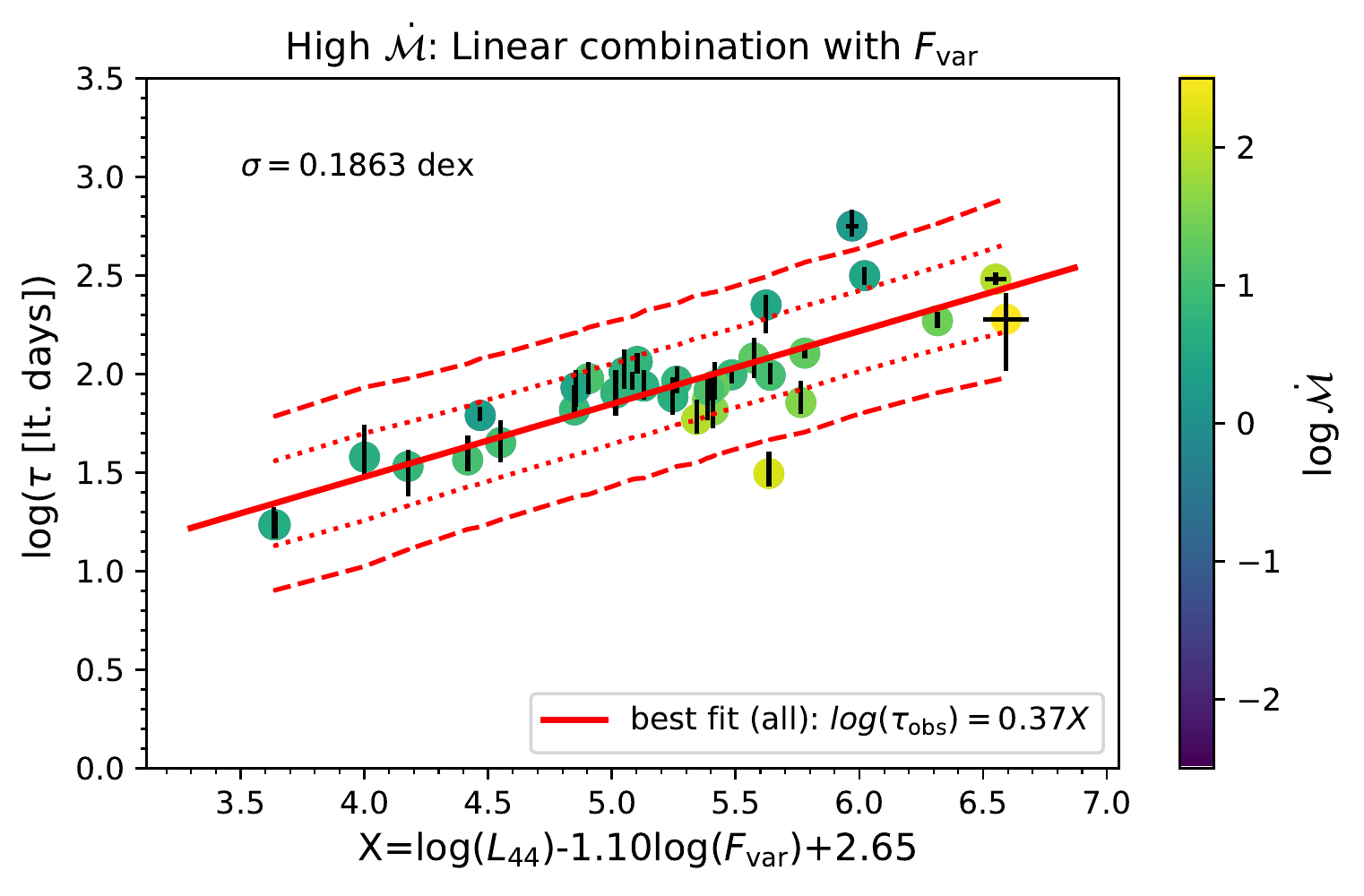}
 \caption{Observed time-delay ($\tau_{\rm obs}$) expressed as a function of the linear combination of $\log{L_{44}}$ and $\log{R_{\rm FeII}}$ or $\log{F_{\rm var}}$. Using the accretion-rate distribution in Fig.~\ref{fig:hist_mdot}, we separately analyze the low accretors (top left panel) and the high accretors (top right panel), for which we obtained a significantly reduced scatter. In comparison with the previous samples, three sources are not included because of the lack of the EW measurements (CTS~252 and two measurements for NGC~4151). In the bottom panels, we show the dependency of the observed time delay on the combination including the fractional variability $F_{\rm var}$ for the {low- (bottom left panel)} and high-\mdot\ sub-samples (bottom right panel). Similarly as for the combination including \rfe\, the scatter is significantly smaller in comparison with the whole sample. {Dotted and dashed lines denote the 68$\%$ and $95\%$ confidence intervals, respectively. Color code represents the intensity of the dimensionless accretion rate, \mdot.} }
    \label{fig_RL_scatter_linear_combination4}
\end{figure*}

We also studied the sub-samples allowing for additional parametric dependencies: FWHM, \rfe\ and \fvar. The inclusion of FWHM in the fit gave some further decrease of the dispersion, but the effect is not strong (see lower panels of Fig.~\ref{fig_RL_scatter_linear_combination3}).  Concerning the fractional variability \fvar, for the low-accretion sources, the improvement was marginal in comparison with the base R-L relation for these sources, see Fig.~\ref{fig_RL_scatter_linear_combination4} (bottom left panel). However, moving towards the highly accreting sub-sample, the reduction in scatter is significant, down to $0.19$ dex (see Fig.~\ref{fig_RL_scatter_linear_combination4}, bottom right panel), and the correlation coefficient also increased. An even larger reduction in the scatter was achieved when we included \rfe\, (see Fig.~\ref{fig_RL_scatter_linear_combination4}, top panels). The effect was clearly visible for both sub-samples, and for the high accretion rate sub-sample, our dispersion reduced to only 0.17 dex, again without the removal of any outliers. This was the smallest scatter achieved {in our study without the use of inter-dependent 
quantities.}

\begin{figure*}
    \centering
    \includegraphics[width=\columnwidth]{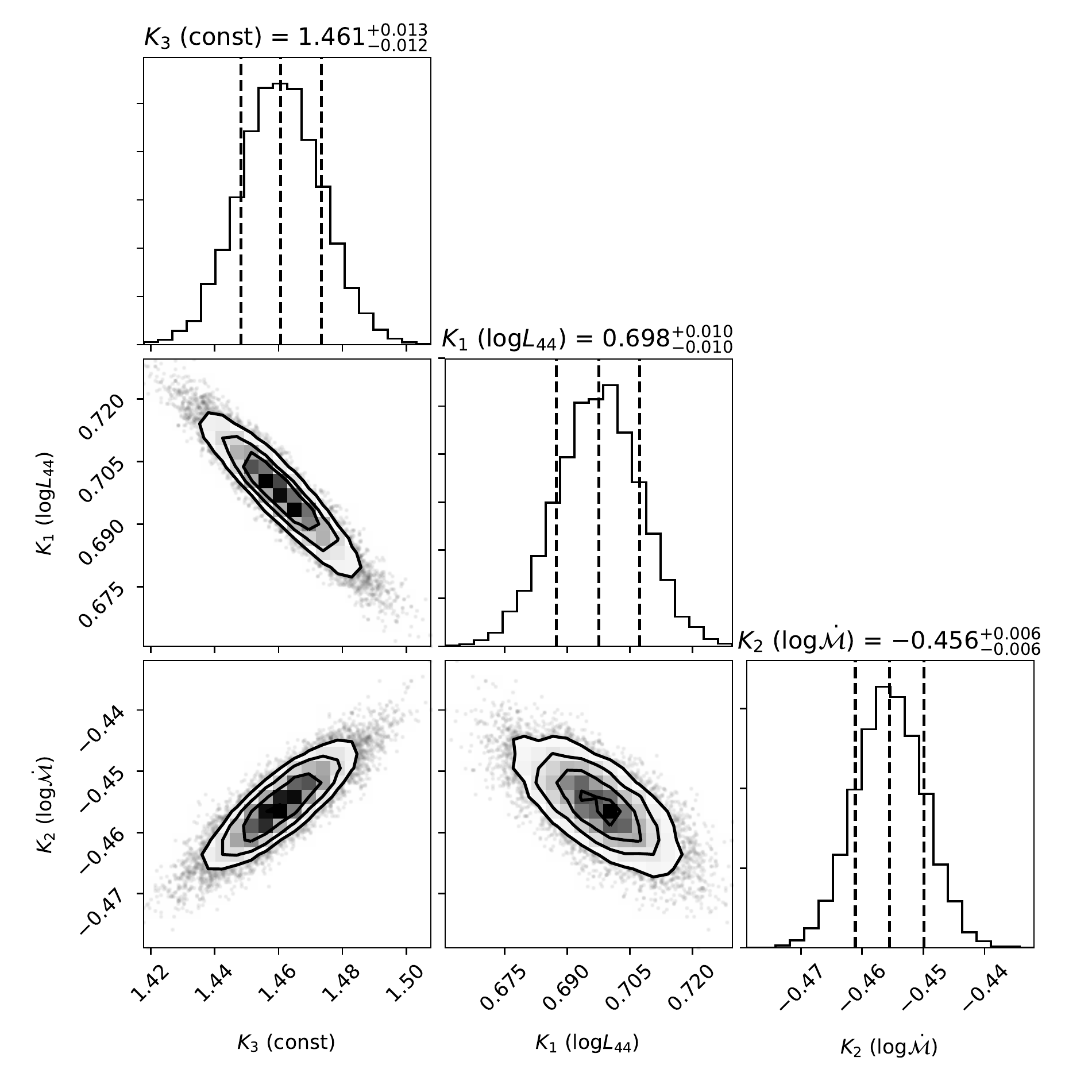}
     \includegraphics[width=\columnwidth]{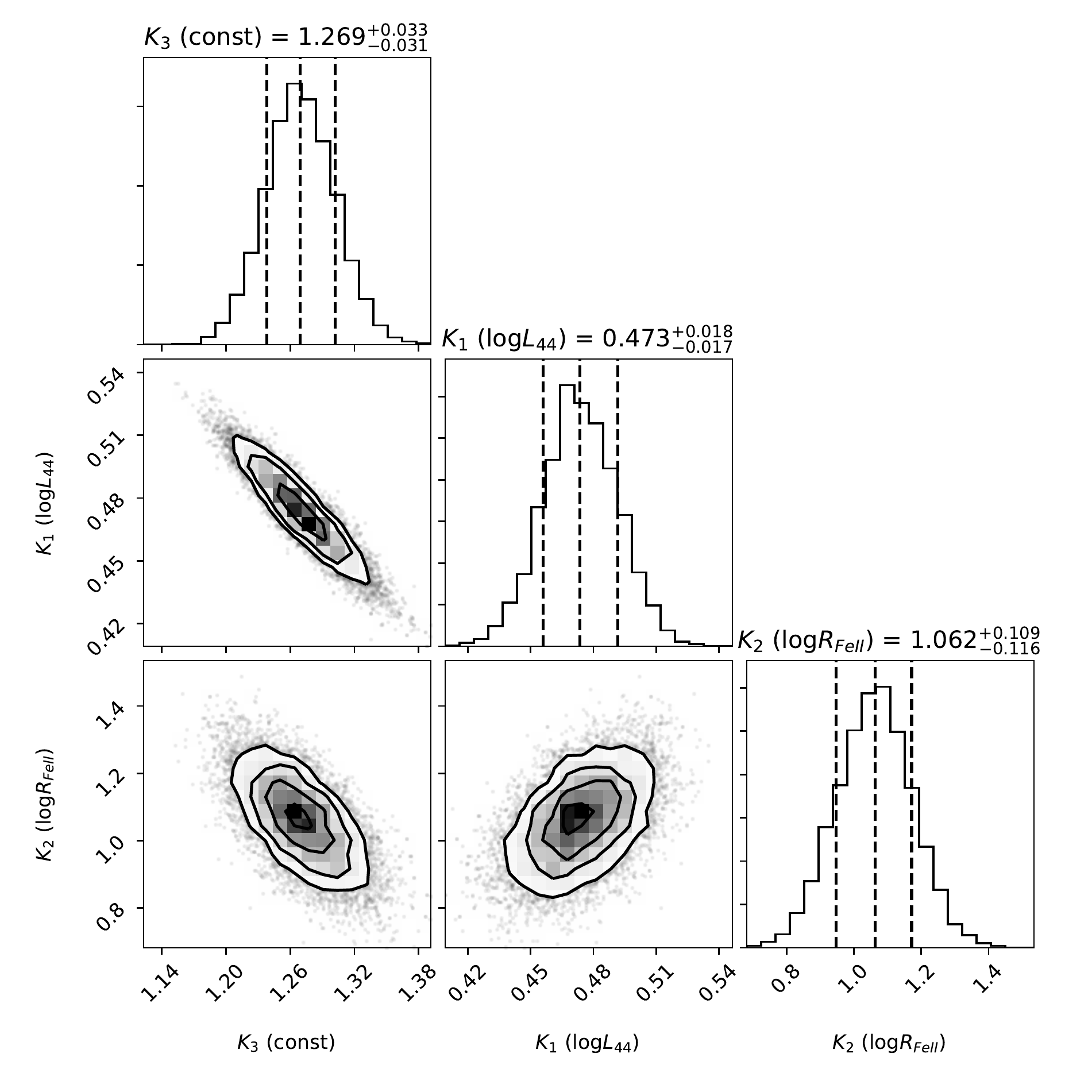}
      \includegraphics[width=\columnwidth]{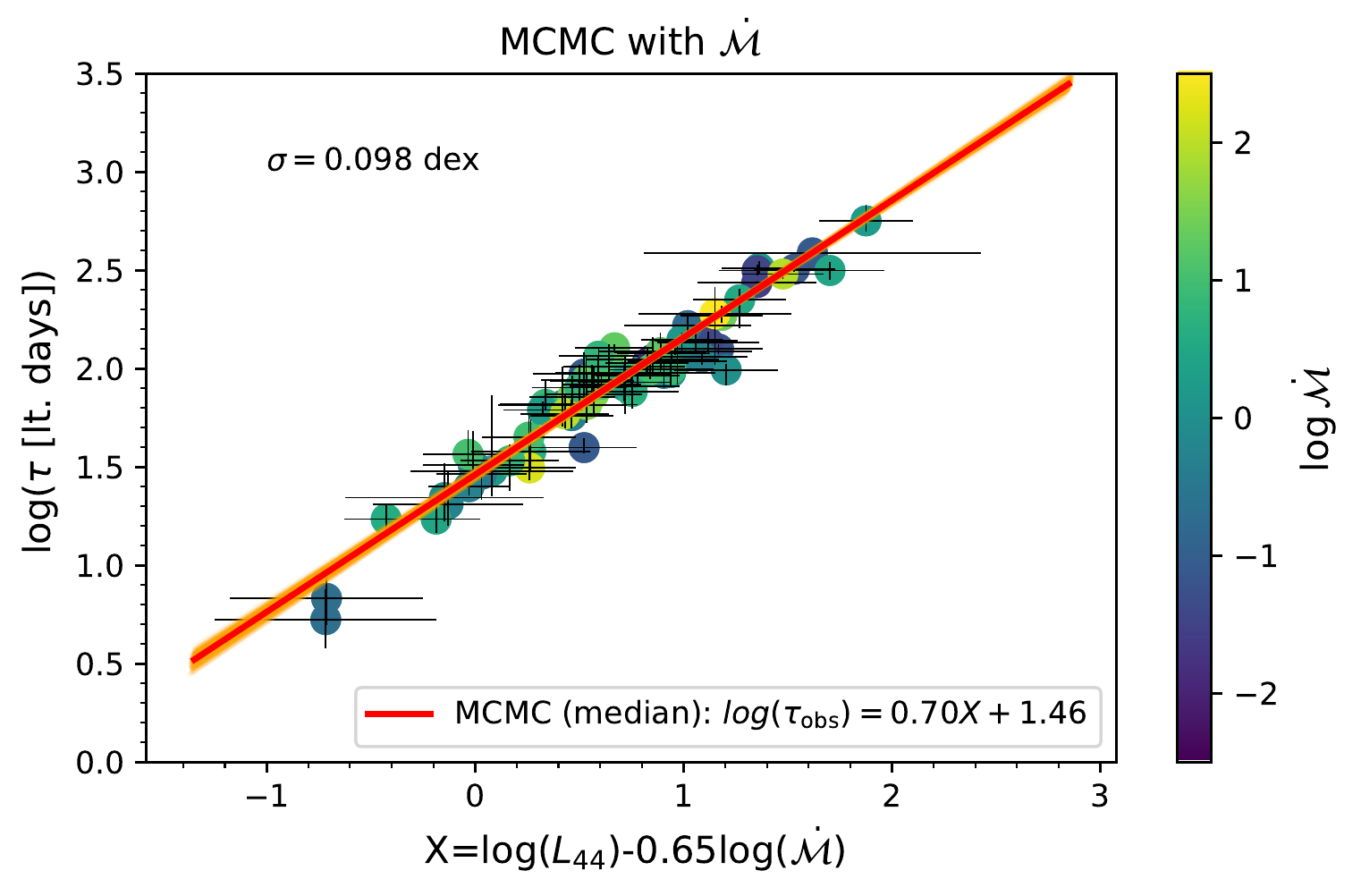}
      \includegraphics[width=\columnwidth]{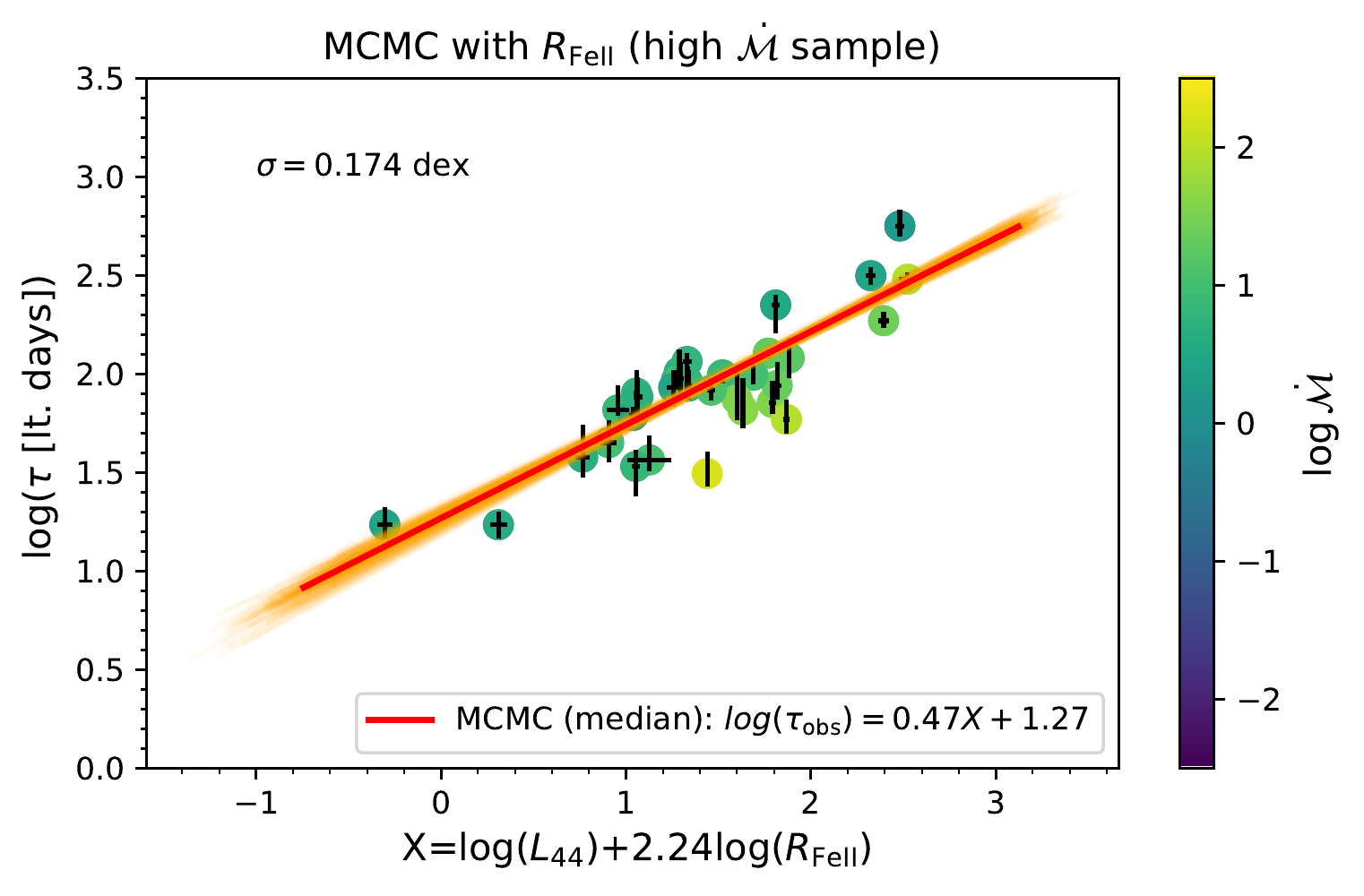}
    \caption{Histograms of the Markov Chain Monte Carlo fitting to the selected combinations of variables with the smallest scatter (top panels). In the left top panel, we show the distributions for $L_{44}$ and $\dot{\mathcal{M}}$ variables. In the right top panel, the distributions for $L_{44}$ and \rfe\ are shown. In the bottom panels, we show the corresponding linear relations between the observed time-delay and the linear combination of variables. In the left bottom panel, the relation between $\log{\tau_{\rm obs}}$ and {$\log{L_{44}}+A\log{\dot{\mathcal{M}}}$} is depicted. The red line represents the maximum likelihood linear relation, while the set of orange lines stands for 300 relations drawn from posterior distributions. In the right bottom panel, we plot the linear relation in analogy to the left panel, but for the observed time-delay expressed as the linear combination of $\log{L_{44}}$ and $\log{R_{\rm FeII}}$ for the high-accretion sub-sample.}
    \label{fig_MCMC}
\end{figure*}

\begin{figure*}
    \centering
    \includegraphics[width=\columnwidth]{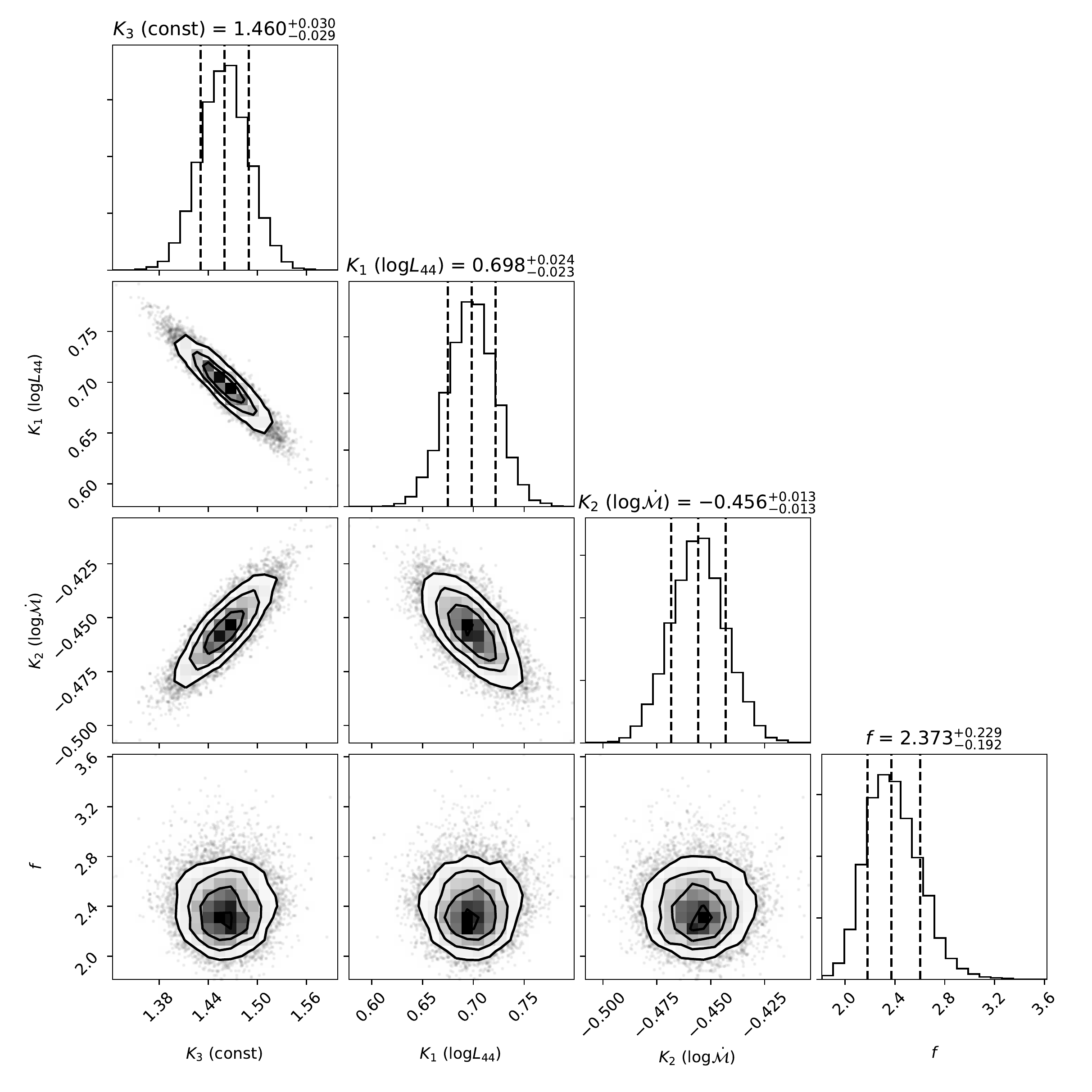}
     \includegraphics[width=\columnwidth]{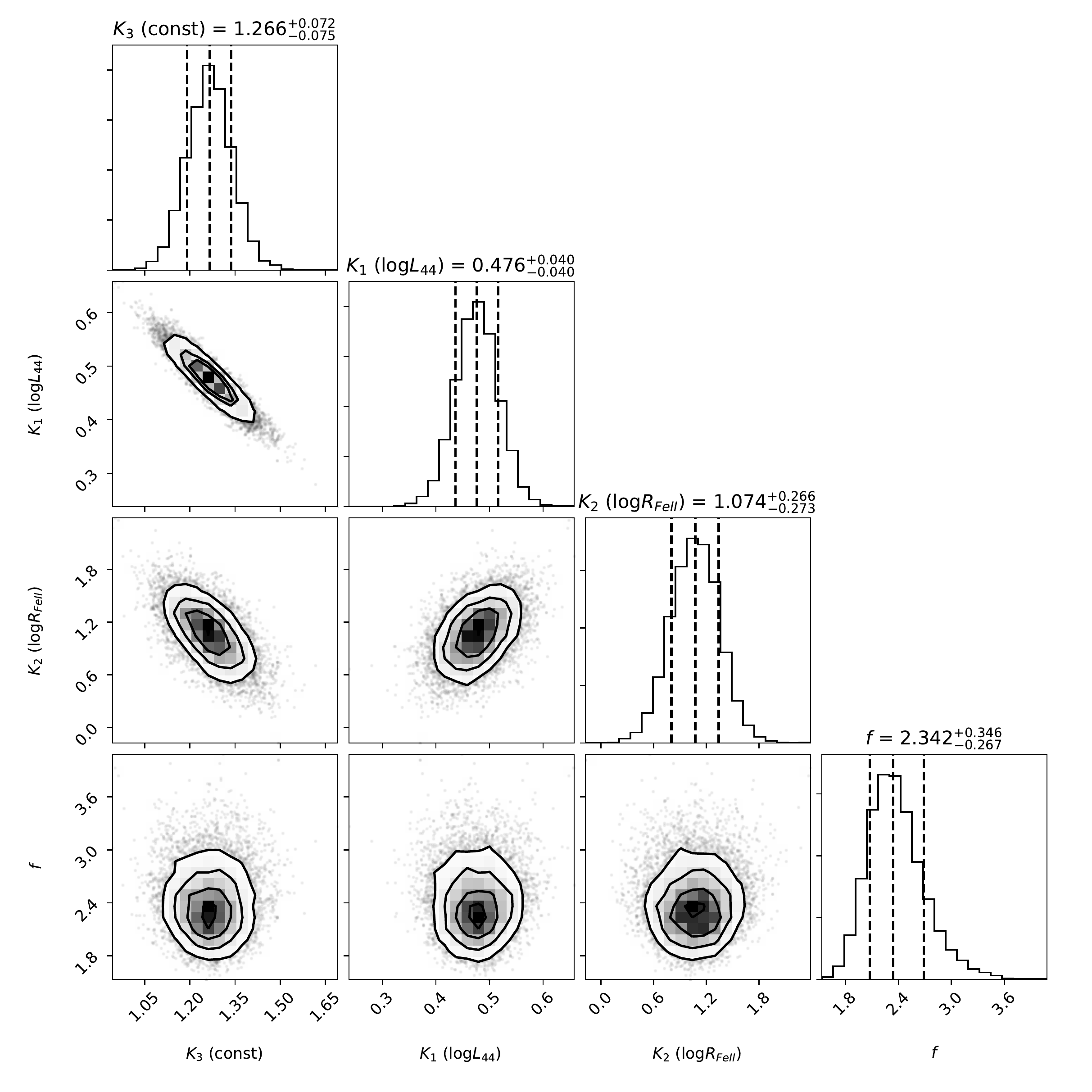}
      \includegraphics[width=\columnwidth]{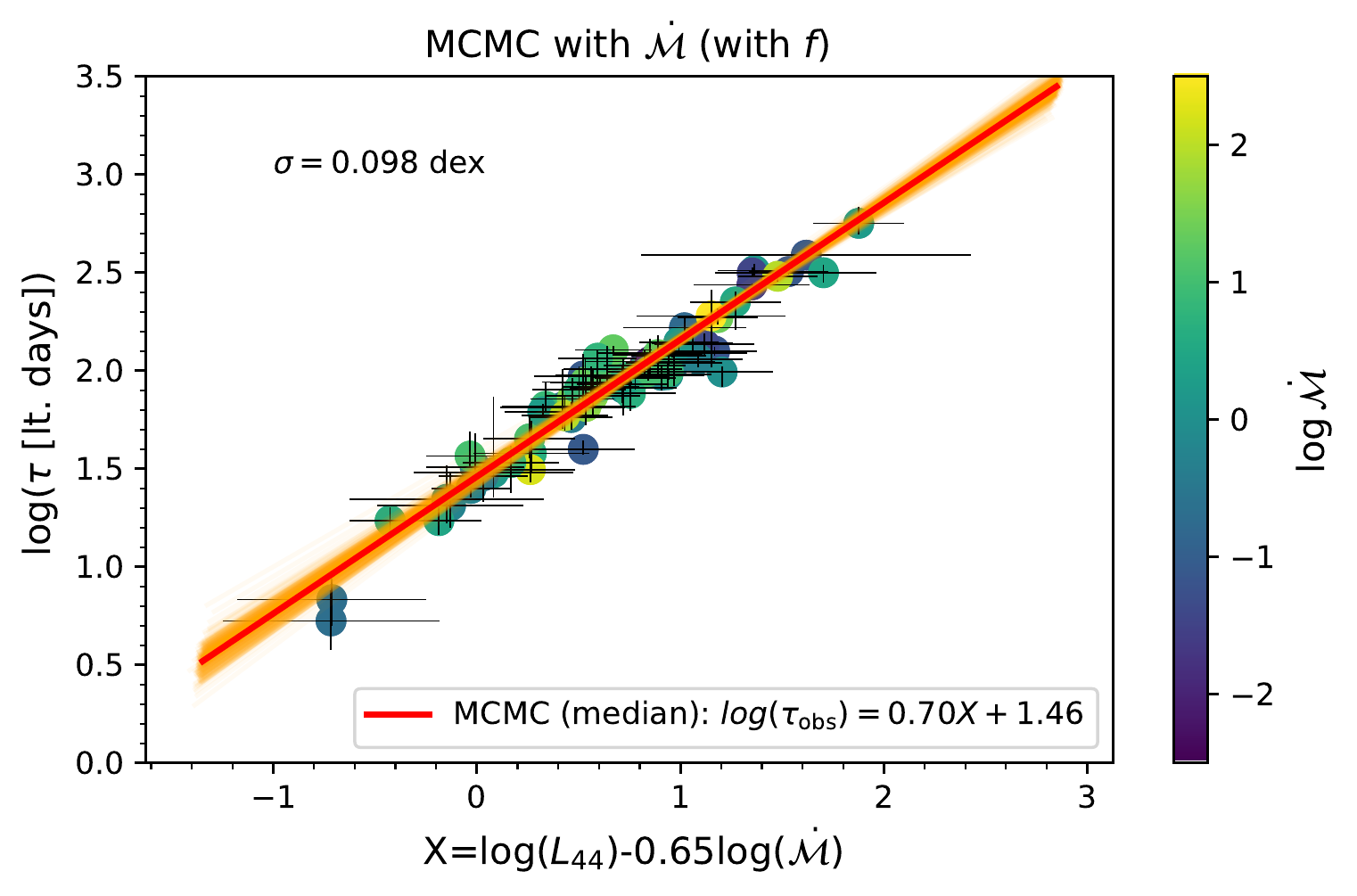}
      \includegraphics[width=\columnwidth]{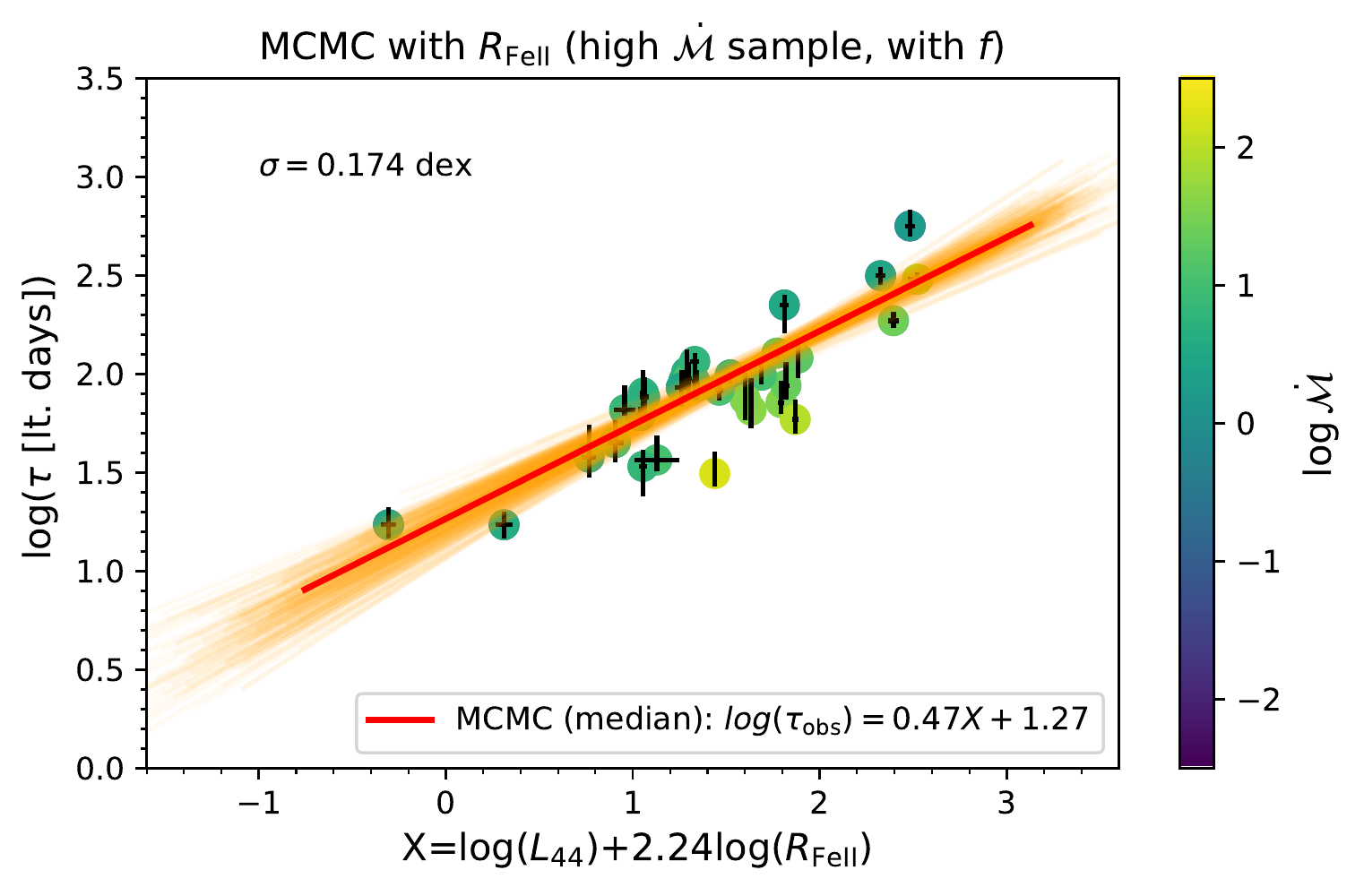}
    \caption{{The same MCMC fitting procedure as in Fig.~\ref{fig_MCMC} but including the underestimation factor $f$ in the likelihood function; see Eq.~\ref{eq_likelihood_withf}.}}
    \label{fig_MCMC_f}
\end{figure*}

\subsection{Ordinary least squares vs. MCMC inference of parameters}
\label{subsec_OLS_vs_MCMC}

For all the studied relations $\log{\tau_{\rm obs}}=K_1\log{L_{44}}+\sum_{i=2}^{n} K_i\log{Q_i}+K_{n+1}$ listed in Table~\ref{tab_multivariate_regression}, we used the higher-dimensional ordinary least square (OLS) method (two- or three-dimensional including the constant factor) as implemented in python packages \textmyfont{sklearn} and \textmyfont{statsmodels}. This allowed us to quickly infer the relevant parameters including standard errors and compare the {RMS} scatter and the correlation coefficient for as many as 15 combinations of relevant quantities. However, hidden correlations between quantities are not apparent when using the multidimensional least squares. Therefore, for the relations with the smallest scatter below 0.2 dex, specifically $\log{\tau_{\rm obs}}=K_1\log{L_{44}}+K_2\log{\dot{\mathcal{M}}}+K_3$ (whole sample) and $\log{\tau_{\rm obs}}=K_1\log{L_{44}}+K_2\log{R_{\rm FeII}}+K_3$ (highly accreting sources), we apply the Markov-Chain Monte Carlo (MCMC) inference of parameters, using the python sampler \textmyfont{emcee}. The two inference techniques -- OLS and MCMC -- are specified in the second column of Table~\ref{tab_multivariate_regression}. The MCMC is a robust Bayesian technique that takes into account measurement errors while inferring the parameters with the maximum likelihood, which in our case is defined as,
\begin{equation}
 \mathcal{L}=-\frac{1}{2}\sum_{i=1}^{N} \left(\frac{\tau_i-\hat{\tau}_i}{\sigma_i^{\tau}}\right)^2\,,
 \label{eq_likelihood_simple}
\end{equation}
where $\tau_i$ are individual time-delay measurements, $\sigma_i^{\tau}$ are the measurement errors of the time-delay, and $\hat{\tau}_i$ are the predicted time-delay values according to the inferred model.  
In addition, MCMC allows us to construct 2D histograms to see potential degeneracies among different parameters.

The results for the two above-mentioned relations are shown in Fig.~\ref{fig_MCMC} for the combination including \mdot\, in the left panels and for the combination including \rfe\, in the right panels. From 2D histograms we can see that the combination with \mdot\, exhibits a degeneracy between the constant $K_3$ coefficient and $K_1$ coefficient as well as between $K_3$ and $K_2$ coefficients. This can be understood in terms of \mdot\, being intrinsically correlated with $\tau_{\rm obs}$. The degeneracy between $K_3$ and $K_2$ parameters is lifted for the combination with \rfe{}\, as \rfe{}\, and $\tau_{\rm obs}$ are intrinsically not correlated.  

As it can be seen from the bottom panels of Fig.~\ref{fig_MCMC}, the parameters with the maximum likelihood are consistent with the best-fit parameters from the least-squares technique. The RMS scatter as well as correlation coefficients are also comparable within uncertainties, see Table~\ref{tab_multivariate_regression}. However, the MCMC uncertainties based on the 16th and the 84th percentiles (1$\sigma$) are by a factor of about two smaller than the 1$\sigma$ errors inferred using the ordinary least squares for the combination involving $\dot{\mathcal{M}}$, see Table~\ref{tab_multivariate_regression}. Similarly, for the combination with \rfe{}, the reduction in uncertainty is by almost a factor of three. {It may seem that the MCMC inference using the maximization of the likelihood function defined by Eq.~\ref{eq_likelihood_simple} is more robust for constraining individual parameters even for general priors of uniformly distributed coefficients in the studied linear combinations. However, the likelihood function defined by Eq.~\ref{eq_likelihood_simple} considers only the measurement errors of $\tau_{i}$. When we take into account that the measurement errors $\sigma_i^{\tau}$ could generally be underestimated by a factor $f$, then the likelihood function takes the following form,
\begin{equation}
 \mathcal{L}_{\rm f}=-\frac{1}{2}\sum_{i=1}^{N} \left[\frac{(\tau_i-\hat{\tau}_i)^2}{s_{i}^2}+\log{s_{i}^2}+\log{(2\pi)}\right]\,,
 \label{eq_likelihood_withf}
\end{equation}
where $s_{i}=f\sigma_i^{\tau}$. Maximizing the functional $\mathcal{L}_{\rm f}$ that depends on the additional underestimation parameter $f$ yields the posterior 2D and 1D marginalized distributions as shown in Fig.~\ref{fig_MCMC_f}. The parameter uncertainties are now comparable to those inferred using the OLS method; see Table~\ref{tab_multivariate_regression} where we list the parameters inferred using the likelihood function $\mathcal{L}_{\rm f}$ under the abbreviation MCMC(f). The difference between uncertainties inferred using either $\mathcal{L}$ or $\mathcal{L}_{\rm f}$ shows the importance of the defined likelihood function for a given problem. In our case, the OLS and MCMC methods are in better agreement in terms of the uncertainties when the factor $f$ is included which suggests that the measurement time-delay uncertainties are underestimated by a factor of about two or three.  

}

Since ordinary least squares and MCMC can be considered as independent inference techniques, their overall consistency confirms the statistical robustness of our results, mainly concerning the low scatter for highly accreting sources.

\section{Discussion} \label{sec:discussion}

\subsection{Mg~II radius-luminosity relation}

{The R-L relation is of key importance for black hole mass measurements in the sources which were not studied through reverberation mapping and thus, relying upon a single epoch spectrum. It is also a promising tool to be applied in cosmology if we have a large enough sample showing a small scatter around the best fit relation. In the present paper, we studied the R-L relation based on Mg II line, using all available data, and we focused on the selection of additional parameters/methods which help to reduce the observed scatter.}

{Considering a virialized and photoionized gas, it is expected that log $\tau\mathrm{_{obs}}\propto \alpha \log L$, where the slope of the luminosity is given by $\alpha=0.5$. In the optical range, the H$\beta$ reverberation-mapping results give a slope for the optical luminosity at 5100~\AA\ of $\alpha=0.533^{+0.035}_{-0.033}$ in the most accepted R-L relation   \citep{bentz2013}. {This slope was nicely consistent with simple predictions of the BLR location based on a fixed ionization parameter \citep[see][and the references therein]{czerny_wang2019} or the Failed Radiatively Accelerated Dusty outflow model \citep[FRADO,][]{czhr2011}.}}

{The slope of R-L relation based on the {Mg~II line}} shows a large diversity \citep{vestergaard-osmer2009, traktenbrot2012, zajacek2020, homayouni2020}. The most recent {Mg~II monitoring from the SDSS-RM project} \citep{homayouni2020}, which increased significantly the number of {Mg~II time lags}, provides a R-L relation with a shallower slope than the slope seen for H$\beta$ line by \citet{bentz2013}.  Based on their time-delay significance criteria, {SDSS-RM sample \citep{homayouni2020} is divided}  into two sub-samples (significant and gold sample), where the slope of the luminosity in both cases is $0.22\pm0.06$ and $0.31^{+0.09}_{-0.10}$, respectively. 

In this work {we combine the data of \citet{homayouni2020} with}  the low- and high-luminosity sources collected by \citet{zajacek2020}. When all {the measurements of Mg II delay} up to now are included, we get a slope of $0.298\pm0.047$, which is in agreement with the one inferred by \citet{homayouni2020}. However, when the sample is divided based on the {median} \mdot\ intensity, the slope becomes much steeper. In the case of low-accretion sub-sample, the slope is very close ($0.520\pm0.078$) to the expected value and the one provided by H$\beta$ results. As we mentioned previously, most of the first H$\beta$ reverberation-mapped objects were selected based on their high variability (high \fvar) {and strong \oiii\ emission}, which indicates a low-accretion rate, and is expected to show an agreement with the standard slope. In the case of the high-\mdot\ sub-sample, the slope is $0.414\pm0.058$, which, within uncertainties, is very close to the expected value. A large sample of highly accreting objects is needed to confirm a real deviation of the slope of the luminosity for this kind of objects.    

When additional independent observational parameters are considered in linear combinations with the luminosity at 3000~\AA, the slope shows different values. Since the typical slope of 0.5 is estimated considering only the luminosity, we cannot claim a deviation from it in the other studied cases where \rfe, \fvar\ and FWHM are used, because new theoretical models taking into account these parameters should be considered.






\subsection{The role of the accretion rate in reducing the scatter}
\label{sec:accretion_rate_effect}

Newer, larger samples for H$\beta$ line brought an additional scatter in comparison with the sample of \citet{bentz2013}, and also the scatter in the Mg II full sample is relatively large, $\sigma_{\rm rms}\sim 0.3$ dex. However, a more advanced approach presented in this paper helped to reduce this scatter considerably. The smallest scatter has been achieved when inter-dependent quantities (\mdot\ and \eddr) are used. These results, thus, have to be treated with care {since they are intrinsically correlated with the time-delay.} On one hand, it is an attractive hypothesis that the scatter in the original R-L relation (i.e. log $\tau\mathrm{_{obs}}$  vs. log $L_{44}$) is due to the spread in the accretion rate intensity. The underlying mechanism for shortening the time delay with an increase in the accretion rate has already been suggested by \citet{wang_shielding2014} who argued that the self-shielding of the accretion disk also leads to the selective shielding of the BLR and its division into two distinctly different regions. In addition, within the FRADO model combined with the shielding effect, such a trend is expected \citep{naddaf2020}. 

The use of the inter-dependent quantities has a drawback, that, despite the small scatter in the final plot, the recovery of the values predicted by the relation comes with a large error. If we want to use {the linear combination with \mdot\ (left panel in Figure~\ref{fig_RL_scatter_linear_combination2}, see also Table~\ref{tab_multivariate_regression}, rows 7 and 8)} to predict the value of the time delay for a given source if we measure the FWHM and the $L_{44}$ assuming a known cosmology, the error of this prediction will become larger than the scatter visible in the plot: the minimum value of the error of $\log \tau\mathrm{_{obs}}$ in this case would be $\delta K_3/0.136 = 0.147$, where $\delta K_3$ is the error of the coefficient $K_3$  in Table~\ref{tab_multivariate_regression}, {row 7}. The same will happen if we use {the linear combination with \mdot} (left panel Figure~\ref{fig_RL_scatter_linear_combination2}) to obtain the absolute luminosity from the measured time delay and FWHM - the minimum error of the predicted $\log L_{44}$ would then be $\delta K_3/0.046 = 0.43$. Thus, for the black hole mass measurements or for the cosmological applications, the use of the relations based on independent quantities still gives much better results. An attractive possibility would be to use the accretion rate or the Eddington ratio as independent parameters. This would require an independent measurement of the black hole mass, e.g. from the broad-band spectral energy distribution (SED) fitting \citep{capellupo2015}.

The use of MCMC fitting method with the inter-dependent \mdot\ parameter reduces the error {for all the coefficients} so the numbers mentioned above will be formally by a factor up to 3 times lower. However, the dispersion around the fit is not affected by the inference method so it is not clear whether the use of MCMC indeed reduces the errors when predictions are made. {As we also showed in Subsection~\ref{subsec_OLS_vs_MCMC}, the uncertainty intervals of the posterior parameter distributions also depend on whether the underestimation parameter is included in the likelihood function or not. If it is included, then the uncertainties are consistent with the OLS method, which suggests that the time-delay uncertainties are generally underestimated by about a factor of two.}

The dominant role of the dimensionless accretion rate or the Eddington ratio is supported by the fact that the division of the sample into two parts representing low and high accretion rates also reduced the scatter in the R-L relation considerably, particularly for the case of the highly accreting sub-sample. In combination with the measurement of \rfe, this reduced the scatter in the R-L relation down to $\sim 0.17$ dex. The scatter for lower accretion-rate sub-sample remained at $\sigma_{\rm rms}\sim 0.25$ dex. This level of scatter is most likely related to the red-noise character of AGN variability in the optical band \citep[e.g.][]{czerny1999,kelly2009,kozlowski2010,kozlowski2016}. A relatively short monitoring allows one to determine the time delay of the lines with respect to the continuum but years of monitoring are needed to determine the mean luminosity level, instead of a part of the lightcurve catching the source in a relatively high or a relatively low state. As was shown by \citet{Ai2010} for SDSS Stripe 82 AGN, higher Eddington-ratio sources vary less in the optical/UV bands. This variability, in particular for a low-accretion sub-sample, may lead to an irreducible scatter in the R-L relation, {as it is clearly seen for our results (Figs.~\ref{fig_RL_scatter_linear_combination3} and \ref{fig_RL_scatter_linear_combination4})}. The same scatter was discussed by \citet{risaliti2019} in the context of the broad-band UV--X-ray relation, where they argue that the variability is relatively unimportant for high-redshift quasars, leading to the scatter of 0.04 dex. However, their selection of predominantly blue quasars contributed to the reduction of this scatter. During the 16-year quasar monitoring, the variability varied from 0.04 to 0.1 dex, depending on the quasar absolute luminosity \citep{hook1994}. In our sample, no pre-selection of objects based on the UV slope has been made. 

{Alternatively, the reduction in the scatter in the sub-sample with high-\mdot\  could be related to the fact that sources radiating close to their Eddington limit saturate toward a limiting value, which leads to the stabilizing of the ratio between the luminosity and black hole mass (which is basically \mdot\ or \eddr), making the sources more steady \citep{marzianisulentic2014}.} Other explanations of the additional scatter include the spin effect and the possibility of a retrograde accretion \citep{wang2014_spin,czerny_wang2019} 

{We stress the fact that \rfe\ and \fvar\ show the smallest scatter (0.17 dex and 0.19 dex, respectively) when the sample is divided into the two considering the \mdot\ intensity. Both observational properties are correlated with  \mdot\ and \eddr\ \citep{marziani2003, wilhite2008, dong2009, dong2011, macleod2010, sanchez-saenz2018, martinez_aldama2019, du2019, Yu2020}, but they are independent. This result suggests that the accretion rate drives the scatter in the R-L relation. }

\subsection{\mdot\ behavior in larger samples }
\label{sec:disc_sample_division}

\begin{figure*}
    \centering
    \includegraphics[width=\textwidth]{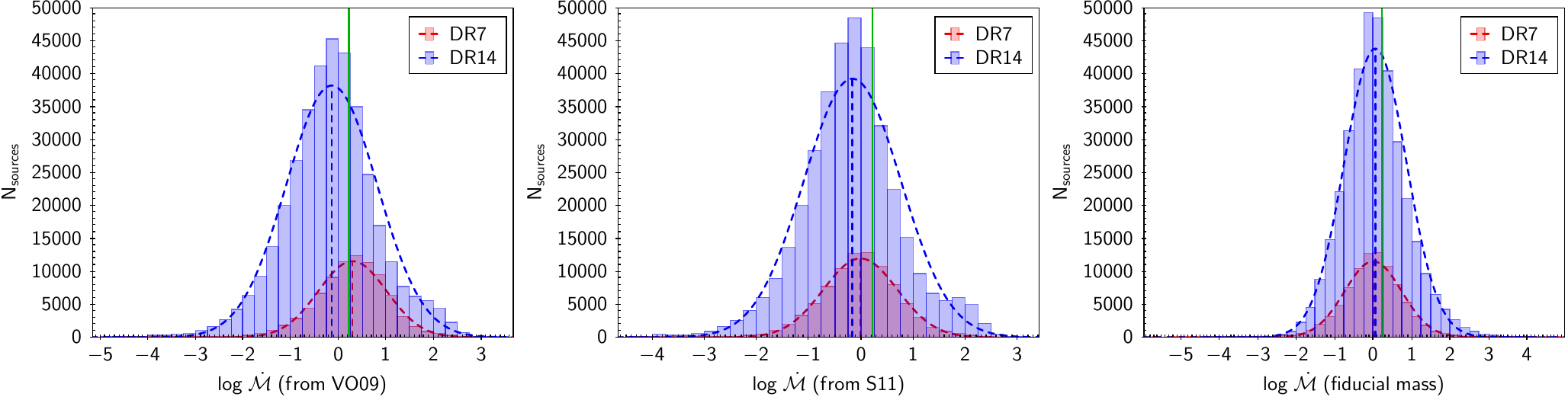}
    \caption{The distribution of \mdot{} (in log-scale) for two representative editions: DR7 \citep{shen2011} and DR14 \citep{rakshitetal2019}. The \mdot{} values are estimated using the Eq.~\eqref{equ:mdot_3000}. The first two panels represent the formalisms for the estimation of the black hole mass using the Mg II line from: (a) \citet{vestergaard-osmer2009}, and (b) \citet{shen2011}. The last panel reports the \mdot{} using the fiducial masses reported in the two catalogues. Gaussian fits (dashed curves) to the histograms are shown with mean values marked (vertical dashed lines). No radiative efficiency has been accounted. The ($\mu\pm\sigma$) values for each panels are: (a) 0.29$\pm$0.72 (DR7) and -0.14$\pm$0.95 (DR14); (b) -0.02$\pm$0.71 (DR7) and -0.17$\pm$0.93 (DR14); and, (c) -0.01$\pm$0.74 (DR7) and 0.05$\pm$0.81 (DR14). The green solid vertical line in each panel indicates the \mdot\ value for our sample, log \mdot{} = 0.2167. }
    \label{fig:dr7-dr14}
\end{figure*}

The division based on the \mdot\ for the current sample could be affected by inclusion of newer sources or reanalyses of existing ones. As a check for completeness, we compare the distribution of \mdot\ estimated for two large SDSS quasar catalogues - for the DR7 release \citep[][, hereafter S11]{shen2011} and for a more recent DR14 release \citep[][, hereafter R20]{rakshitetal2019}, with various spectral parameters estimated for 105,783 and 526,265 sources, respectively. The \mdot\ formalism used in this paper (see Eq.~\eqref{equ:mdot_3000}) is a function of the black hole mass and the monochromatic luminosity at 3000~\AA\ (the associated inclination term, cos $\it{\theta}$ is set to 0.75, which is the mean disk inclination for type 1 AGN).  We filter the catalogues first by limiting to the values that are reported to be positive and non-zero. Additionally, in the latter case (R20), the authors also provide quality flags for selected parameters including the monochromatic luminosity at 3000~\AA\ (L$_{3000}$). For the DR7 QSO catalogue, no such quality flags were provided, thus we use the full sample in this case. For the black holes masses, we use three variants common to the two catalogues: (a) from \citet[][, hereafter VO09]{vestergaard-osmer2009}; (b) from S11; and (c) from the fiducial virial black hole mass values calculated based on (a) H$\beta$ line (for z $<$ 0.8) using the calibration of \citet{vestergaard2006}, (b) Mg II line (for 0.8 $\leq$ z $<$ 1.9) using the calibration provided by VO09, and (c) C IV line (for z $\geq$ 1.9) using \citet{vestergaard2006} calibration. In case of the black hole masses, the DR14 QSO catalogue provides the quality flag only for the fiducial masses and is unavailable for the VO09 and S11 mass estimates. We thus use the quality flags to control the sample for the fiducial mass estimates in this case. Figure \ref{fig:dr7-dr14} demonstrates the \mdot\ distributions computed for the two catalogues. The three panels are synonymous to the three cases of black hole mass estimates incorporated to estimate the \mdot\ values.

Due to the quality control and filtering, the source sample drops to $\sim$ 79\% (DR7) and to $\sim$ 67\% (DR14) of their respective original source counts. The effective number of sources per case of the black hole mass estimates remain almost alike\footnote{for DR7 QSO catalogue, the \mdot\ values were estimated for 83,374 sources with \mbh\ from VO09; for 85,099 sources with \mbh\ from S11; and, for 85,638 sources with fiducial masses. Equivalently, for the DR14 QSO catalogues, these numbers are over 4 times larger - 365,440 sources with \mbh\ each from VO09 and S11; and, for 354,675 sources with fiducial masses.}. To predict the variation with respect to our small sample of Mg II RM-reported sources, we extract the mean ($\mu$) and the standard deviation ($\sigma$) from each \mdot\ distribution (in log-scale) shown in Fig.~\ref{fig:dr7-dr14} using simple Gaussian fits. The ($\mu\pm\sigma$) values for each panels are: (VO09)  0.29$\pm$0.72 (DR7) and -0.14$\pm$0.95 (DR14); (S11) -0.02$\pm$0.71 (DR7) and -0.17$\pm$0.93 (DR14); and, (fiducial) -0.01$\pm$0.74 (DR7) and 0.05$\pm$0.81 (DR14). The median value for our sample, log \mdot\ = 0.2167, is well within 1$\sigma$ limits regardless of the distribution taken from the larger catalogues, and hence, will not have significant effects on the correlations quoted in this paper with the inclusion of more sources in the future.

\section{Conclusions} \label{sec:conclusions}

Using a sample of 68 reverberation-mapped Mg II AGN, we explore the reasons for the scatter along the R-L relation. In addition to the dimensionless accretion-rate parameter \mdot\, and the Eddington ratio, we included independent parameters such as FWMH of Mg II, \fvar\ and \rfe\ in linear combinations with the luminosity ($L_{3000}$ or $L_{44}$) to decrease the scatter. We summarize the important conclusions derived from this analysis as the following:

\begin{itemize}

\item When the whole Mg~II sample is considered, we find the smallest root-mean-square (RMS) scatter of $\sigma_{\rm rms}\sim 0.1$ dex for the combinations that include both the monochromatic luminosity ($L_{3000}$) and either the dimensionless accretion-rate parameter \mdot\, or the Eddington ratio (\eddr). However, for these combinations, there is a caveat that both \mdot\, and \eddr\ are intrinsically correlated with the time-delay. Despite the fact that the scatter decreases significantly, the determination of the time delay or luminosity using the proposed linear combinations provides values with larger errors (both for OLS and MCMC inference techniques).

\item The inclusion of independent parameters such as FWHM, \rfe\ and \fvar\ in the linear combination with the luminosity at 3000~\AA\ leads to a slight decrease of the scatter. In all of the analyzed cases, the scatter ($\sigma_{\rm rms}\sim 0.3$ dex) and the correlation coefficients are similar.

\item For the whole sample, the slope of the luminosity at 3000~\AA\ is less steep ($\alpha=0.298\pm0.047$) than the expected value of $\sim 0.5$. However, after the division of the sample considering the \mdot\ intensity, low-\mdot\ sources follow the expected behavior, while the high \mdot\ sources show a {slightly shallower slope, and the relation is shifted toward shorter time delays}. This manifests the effect of the accretion in the R-L relation.

\item When the sample is divided into low- and high-\mdot\ sub-samples, the scatter decreases significantly and the correlation coefficient increases, in particular for the highly accreting sub-sample. The case with the smallest scatter is the combination including \rfe\ with the scatter of only $\sigma_{\rm rms} \sim 0.17$ dex. Also, the inclusion of \fvar\ results in a low scatter, $\sigma_{\rm rms} \sim 0.19$ dex, which is of interest for future photometric surveys. Since \rfe\ and \fvar\ are independent and at the same time correlated with the accretion rate, our results support the idea that the scatter in the R-L relation is driven by the accretion rate intensity.  In particular, \fvar\ has a potential applicability in the upcoming surveys, such as the {Legacy Survey of Space and Time \citep[LSST, see][]{2019ApJ...873..111I}}, which will provide a large quantity of photometric data. The established relations with physical parameters, such as the accretion rate intensity, could be used as a tool for the classification of sources.

\end{itemize}


\software{\textmyfont{sklearn} \citep{scikit-learn}; \textmyfont{statsmodels} \citep{seabold2010statsmodels}; \textmyfont{emcee} \citep{emcee}; \textmyfont{numpy} \citep{numpy}; \textmyfont{matplotlib} \citep{hunter07}; \textmyfont{TOPCAT} \citep{2005ASPC..347...29T}}

\acknowledgments

The authors would like to acknowledge the anonymous referee for the very helpful comments and suggestions. The project was partially supported by the National Science Centre, Poland, grant No.~2017/\-26/\-A/\-ST9/\-00756 (Maestro 9), and by the Ministry of Science and Higher Education (MNiSW) grant DIR/\-WK/\-2018/\-12.
 Time delay for two quasars reported in this paper were obtained with the Southern African Large Telescope (SALT). Polish participation in SALT is funded by grant No. MNiSW DIR/WK/2016/07. 

%

\appendix

\section{Relation between the UV FeII and the accretion parameters}
\label{sec:fe2uv}

{According to our analysis the linear combinations of the luminosity at 3000\AA\ and the strength of the UV Fe~II (expressed as \rfe) decreases significantly the scatter in the R-L relation. The inclusion of the \rfe\ parameter is justified by the correlation with the accretion parameters \citep{dong2011} and it is also suggested by the similar behavior shown in optical range. Since the optical and the UV Fe~II emission are anti-correlated with the FWHM of \hb\ \citep{kovacevic15, sniegowska}, and in the optical case it is widely shown that this relation is driven by the accretion rate \citep[][e.g. and references therein]{marziani2003, shenandho2014, du2019}, it suggests that for the UV case this should hold as well. }

\begin{figure*}
    \centering
    \includegraphics[width=\textwidth]{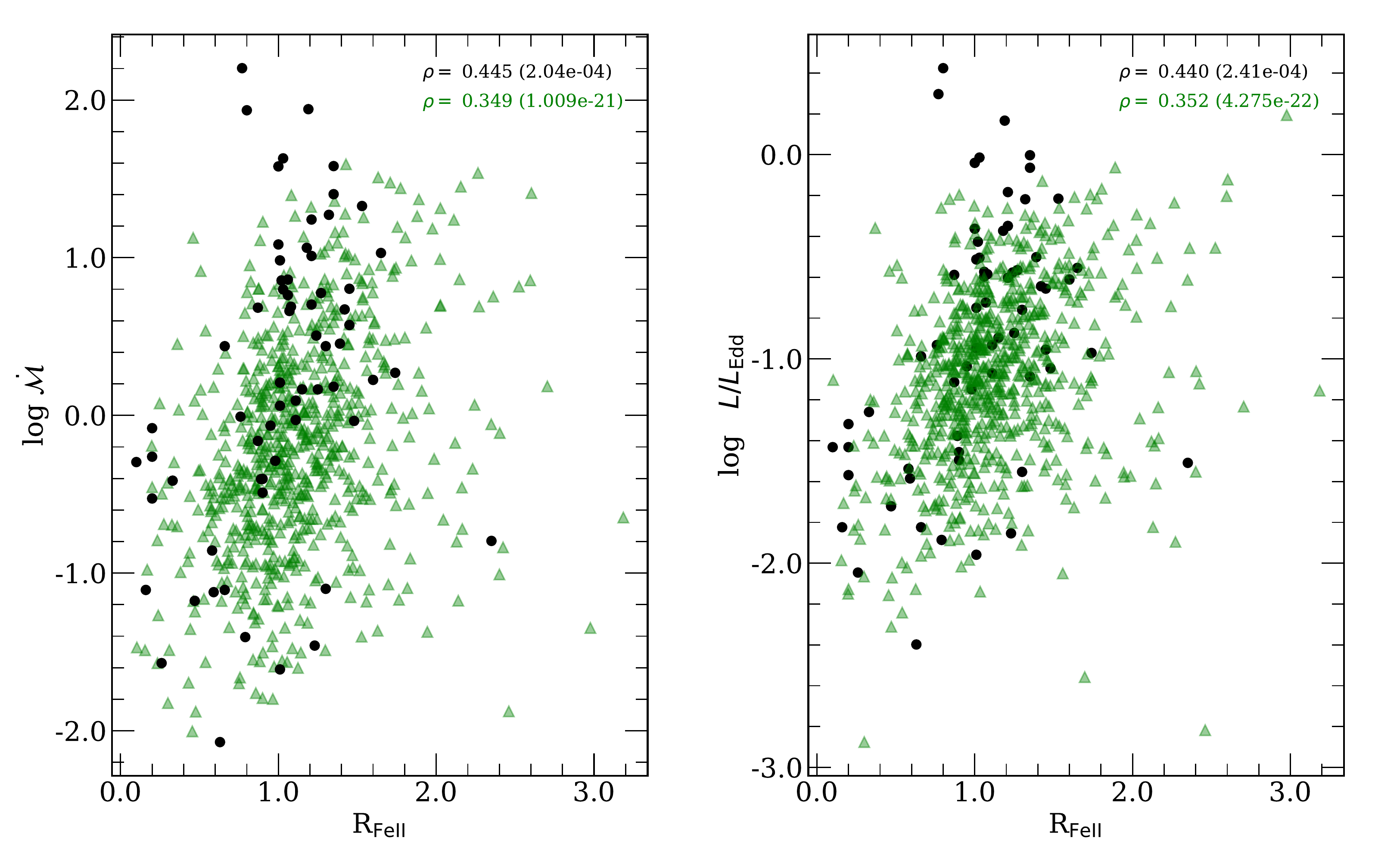}
    \caption{Relation between the \rfe\ parameter, dimensionless accretion rate (left panel) and Eddington ratio (right panel). Black symbols correspond to the presented sample, while green symbols mark the sources from the \citet{shen2019} catalog. In both panels, the Spearman coefficient and $p$-value are included for both samples.}
    \label{fig:fe2uv}
\end{figure*}

{Figure~\ref{fig:fe2uv} shows the relation between the \rfe, dimensionless accretion rate (left panel) and Eddington ratio (right panel). As a reference, we include the Mg~II measurements from the catalog of \citet{shen2019} where \eddr\ was taken from the catalog and \mdot{} was estimated from the single-epoch black hole mass based on Mg~II reported in that paper.  Figure~\ref{fig:fe2uv} also includes the Spearman correlation coefficient ($\rho$) and $p$-values, which indicates a weak correlation between both parameters, but slightly stronger than the correlations found by \citet{dong2009}. Although the correlation is weak, this result opens the possibility to explore this relation in the future and justifies the inclusion of the UV Fe~II in the linear combinations presented in this paper. }

\bibliography{references}

\begin{thebibliography}{}
\expandafter\ifx\csname natexlab\endcsname\relax\def\natexlab#1{#1}\fi
\providecommand{\url}[1]{\href{#1}{#1}}

\bibitem[{{Ai} {et~al.}(2010){Ai}, {Yuan}, {Zhou}, {Wang}, {Dong}, {Wang}, \&
  {Lu}}]{Ai2010}
{Ai}, Y.~L., {Yuan}, W., {Zhou}, H.~Y., {et~al.} 2010, \apjl, 716, L31

\bibitem[{{Alexander}(1997)}]{1997ASSL..218..163A}
{Alexander}, T. 1997, in Astrophysics and Space Science Library, Vol. 218,
  Astronomical Time Series, ed. D.~{Maoz}, A.~{Sternberg}, \& E.~M.
  {Leibowitz}, 163

\bibitem[{{Bentz} {et~al.}(2013){Bentz}, {Denney}, {Grier}, {Barth},
  {Peterson}, {Vestergaard}, {Bennert}, {Canalizo}, {De Rosa}, {Filippenko},
  {Gates}, {Greene}, {Li}, {Malkan}, {Pogge}, {Stern}, {Treu}, \&
  {Woo}}]{bentz2013}
{Bentz}, M.~C., {Denney}, K.~D., {Grier}, C.~J., {et~al.} 2013, \apj, 767, 149

\bibitem[{{Blandford} \& {McKee}(1982)}]{1982ApJ...255..419B}
{Blandford}, R.~D., \& {McKee}, C.~F. 1982, \apj, 255, 419

\bibitem[{{Boroson}(2002)}]{boroson2002}
{Boroson}, T.~A. 2002, \apj, 565, 78

\bibitem[{{Capellupo} {et~al.}(2015){Capellupo}, {Netzer}, {Lira},
  {Trakhtenbrot}, \& {Mej{\'\i}a-Restrepo}}]{capellupo2015}
{Capellupo}, D.~M., {Netzer}, H., {Lira}, P., {Trakhtenbrot}, B., \&
  {Mej{\'\i}a-Restrepo}, J. 2015, \mnras, 446, 3427

\bibitem[{{Chelouche} {et~al.}(2017){Chelouche}, {Pozo-Nu{\~n}ez}, \&
  {Zucker}}]{2017ApJ...844..146C}
{Chelouche}, D., {Pozo-Nu{\~n}ez}, F., \& {Zucker}, S. 2017, \apj, 844, 146

\bibitem[{{Collin} {et~al.}(2006){Collin}, {Kawaguchi}, {Peterson}, \&
  {Vestergaard}}]{collin2006}
{Collin}, S., {Kawaguchi}, T., {Peterson}, B.~M., \& {Vestergaard}, M. 2006,
  \aap, 456, 75

\bibitem[{{Czerny}(2019)}]{2019OAst...28..200C}
{Czerny}, B. 2019, Open Astronomy, 28, 200

\bibitem[{{Czerny} \& {Hryniewicz}(2011)}]{czhr2011}
{Czerny}, B., \& {Hryniewicz}, K. 2011, \aap, 525, L8

\bibitem[{{Czerny} {et~al.}(2013){Czerny}, {Hryniewicz}, {Maity},
  {Schwarzenberg-Czerny}, {{\.Z}ycki}, \& {Bilicki}}]{czerny2013}
{Czerny}, B., {Hryniewicz}, K., {Maity}, I., {et~al.} 2013, \aap, 556, A97

\bibitem[{{Czerny} {et~al.}(1999){Czerny}, {Schwarzenberg-Czerny}, \&
  {Loska}}]{czerny1999}
{Czerny}, B., {Schwarzenberg-Czerny}, A., \& {Loska}, Z. 1999, \mnras, 303, 148

\bibitem[{{Czerny} {et~al.}(2017){Czerny}, {Li}, {Hryniewicz}, {Panda},
  {Wildy}, {Sniegowska}, {Wang}, {Sredzinska}, \& {Karas}}]{czerny2017}
{Czerny}, B., {Li}, Y.-R., {Hryniewicz}, K., {et~al.} 2017, \apj, 846, 154

\bibitem[{{Czerny} {et~al.}(2019{\natexlab{a}}){Czerny}, {Olejak},
  {Ra{\l}owski}, {Koz{\l}owski}, {Loli Martinez Aldama}, {Zajacek}, {Pych},
  {Hryniewicz}, {Pietrzy{\'n}ski}, {Sobrino Figaredo}, {Haas},
  {{\'S}redzi{\'n}ska}, {Krupa}, {Kurcz}, {Udalski}, {Gorski}, {Karas},
  {Panda}, {Sniegowska}, {Naddaf}, {Bilicki}, \& {Sarna}}]{czerny2019}
{Czerny}, B., {Olejak}, A., {Ra{\l}owski}, M., {et~al.} 2019{\natexlab{a}},
  \apj, 880, 46

\bibitem[{{Czerny} {et~al.}(2019{\natexlab{b}}){Czerny}, {Wang}, {Du},
  {Hryniewicz}, {Karas}, {Li}, {Pand a}, {Sniegowska}, {Wildy}, \&
  {Yuan}}]{czerny_wang2019}
{Czerny}, B., {Wang}, J.-M., {Du}, P., {et~al.} 2019{\natexlab{b}}, \apj, 870,
  84

\bibitem[{{Dalla Bont\`a} {et~al.}(2020){Dalla Bont\`a}, {Peterson}, {Bentz},
  {Brandt}, {Ciroi}, {De Rosa}, {Fonseca Alvarez}, {Grier}, {Hall}, {Hernandez
  Santisteban}, {Ho}, {Homayouni}, {Horne}, {Kochanek}, {Li}, {Morelli},
  {Pizzella}, {Pogge}, {Schneider}, {Shen}, {Trump}, \&
  {Vestergaard}}]{dallabonta2020}
{Dalla Bont\`a}, E., {Peterson}, B.~M., {Bentz}, M.~C., {et~al.} 2020, arXiv
  e-prints, arXiv:2007.02963

\bibitem[{{Denney} {et~al.}(2013){Denney}, {Pogge}, {Assef}, {Kochanek},
  {Peterson}, \& {Vestergaard}}]{denney2013}
{Denney}, K.~D., {Pogge}, R.~W., {Assef}, R.~J., {et~al.} 2013, \apj, 775, 60

\bibitem[{{Dong} {et~al.}(2011){Dong}, {Wang}, {Ho}, {Wang}, {Fan}, {Wang},
  {Zhou}, \& {Yuan}}]{dong2011}
{Dong}, X.-B., {Wang}, J.-G., {Ho}, L.~C., {et~al.} 2011, \apj, 736, 86

\bibitem[{{Dong} {et~al.}(2009){Dong}, {Wang}, {Wang}, {Fan}, {Wang}, {Zhou},
  \& {Yuan}}]{dong2009}
{Dong}, X.-B., {Wang}, T.-G., {Wang}, J.-G., {et~al.} 2009, \apjl, 703, L1

\bibitem[{{Du} \& {Wang}(2019)}]{du2019}
{Du}, P., \& {Wang}, J.-M. 2019, \apj, 886, 42

\bibitem[{{Du} {et~al.}(2014){Du}, {Hu}, {Lu}, {Wang}, {Qiu}, {Li}, {Bai},
  {Kaspi}, {Netzer}, {Wang}, \& {SEAMBH Collaboration}}]{dupu2014}
{Du}, P., {Hu}, C., {Lu}, K.-X., {et~al.} 2014, \apj, 782, 45

\bibitem[{{Du} {et~al.}(2015){Du}, {Hu}, {Lu}, {Huang}, {Cheng}, {Qiu}, {Li},
  {Zhang}, {Fan}, {Bai}, {Bian}, {Yuan}, {Kaspi}, {Ho}, {Netzer}, {Wang}, \&
  {SEAMBH Collaboration}}]{du2015}
---. 2015, \apj, 806, 22

\bibitem[{{Du} {et~al.}(2016){Du}, {Lu}, {Zhang}, {Huang}, {Wang}, {Hu}, {Qiu},
  {Li}, {Fan}, {Fang}, {Bai}, {Bian}, {Yuan}, {Ho}, {Wang}, \& {SEAMBH
  Collaboration}}]{du2016}
{Du}, P., {Lu}, K.-X., {Zhang}, Z.-X., {et~al.} 2016, \apj, 825, 126

\bibitem[{{Du} {et~al.}(2018){Du}, {Zhang}, {Wang}, {Huang}, {Zhang}, {Lu},
  {Hu}, {Li}, {Bai}, {Bian}, {Yuan}, {Ho}, {Wang}, \& {SEAMBH
  Collaboration}}]{dupu2018}
{Du}, P., {Zhang}, Z.-X., {Wang}, K., {et~al.} 2018, \apj, 856, 6

\bibitem[{{Edelson} \& {Krolik}(1988)}]{1988ApJ...333..646E}
{Edelson}, R.~A., \& {Krolik}, J.~H. 1988, \apj, 333, 646

\bibitem[{{Fonseca Alvarez} {et~al.}(2019){Fonseca Alvarez}, {Trump},
  {Homayouni}, {Grier}, {Shen}, {Horne}, {I-Hsiu Li}, {Brandt}, {Ho},
  {Peterson}, \& {Schneider}}]{fonseca-alvarez2019}
{Fonseca Alvarez}, G., {Trump}, J.~R., {Homayouni}, Y., {et~al.} 2019, arXiv
  e-prints, arXiv:1910.10719

\bibitem[{{Foreman-Mackey} {et~al.}(2013){Foreman-Mackey}, {Hogg}, {Lang}, \&
  {Goodman}}]{emcee}
{Foreman-Mackey}, D., {Hogg}, D.~W., {Lang}, D., \& {Goodman}, J. 2013, \pasp,
  125, 306

\bibitem[{{Gaskell}(2009)}]{2009NewAR..53..140G}
{Gaskell}, C.~M. 2009, \nar, 53, 140

\bibitem[{{Gaskell} \& {Peterson}(1987)}]{1987ApJS...65....1G}
{Gaskell}, C.~M., \& {Peterson}, B.~M. 1987, \apjs, 65, 1

\bibitem[{{Gravity Collaboration} {et~al.}(2018){Gravity Collaboration},
  {Sturm}, {Dexter}, {Pfuhl}, {Stock}, {Davies}, {Lutz}, {Cl{\'e}net},
  {Eckart}, {Eisenhauer}, {Genzel}, {Gratadour}, {H{\"o}nig}, {Kishimoto},
  {Lacour}, {Millour}, {Netzer}, {Perrin}, {Peterson}, {Petrucci}, {Rouan},
  {Waisberg}, {Woillez}, {Amorim}, {Brandner}, {F{\"o}rster Schreiber},
  {Garcia}, {Gillessen}, {Ott}, {Paumard}, {Perraut}, {Scheithauer},
  {Straubmeier}, {Tacconi}, \& {Widmann}}]{GRAVITY2018}
{Gravity Collaboration}, {Sturm}, E., {Dexter}, J., {et~al.} 2018, \nat, 563,
  657

\bibitem[{{Grier} {et~al.}(2017){Grier}, {Trump}, {Shen}, {Horne}, {Kinemuchi},
  {McGreer}, {Starkey}, {Brandt}, {Hall}, {Kochanek}, {Chen}, {Denney},
  {Greene}, {Ho}, {Homayouni}, {I-Hsiu Li}, {Pei}, {Peterson}, {Petitjean},
  {Schneider}, {Sun}, {AlSayyad}, {Bizyaev}, {Brinkmann}, {Brownstein},
  {Bundy}, {Dawson}, {Eftekharzadeh}, {Fernandez-Trincado}, {Gao},
  {Hutchinson}, {Jia}, {Jiang}, {Oravetz}, {Pan}, {Paris}, {Ponder}, {Peters},
  {Rogerson}, {Simmons}, {Smith}, \& {Wang}}]{grier2017}
{Grier}, C.~J., {Trump}, J.~R., {Shen}, Y., {et~al.} 2017, \apj, 851, 21

\bibitem[{{Haas} {et~al.}(2011){Haas}, {Chini}, {Ramolla}, {Pozo Nu{\~n}ez},
  {Westhues}, {Watermann}, {Hoffmeister}, \& {Murphy}}]{haas2011}
{Haas}, M., {Chini}, R., {Ramolla}, M., {et~al.} 2011, \aap, 535, A73

\bibitem[{{Homayouni} {et~al.}(2020){Homayouni}, {Trump}, {Grier}, {Horne},
  {Shen}, {Brandt}, {Dawson}, {Fonseca Alvarez}, {Green}, {Hall}, {Hernandez
  Santisteban}, {Ho}, {Kinemuchi}, {Kochanek}, {I-Hsiu Li}, {Peterson},
  {Schneider}, {Starkey}, {Bizyaev}, {Pan}, {Oravetz}, \&
  {Simmons}}]{homayouni2020}
{Homayouni}, Y., {Trump}, J.~R., {Grier}, C.~J., {et~al.} 2020, arXiv e-prints,
  arXiv:2005.03663

\bibitem[{{Hook} {et~al.}(1994){Hook}, {McMahon}, {Boyle}, \&
  {Irwin}}]{hook1994}
{Hook}, I.~M., {McMahon}, R.~G., {Boyle}, B.~J., \& {Irwin}, M.~J. 1994,
  \mnras, 268, 305

\bibitem[{{Hunter}(2007)}]{hunter07}
{Hunter}, J.~D. 2007, Computing in Science and Engineering, 9, 90

\bibitem[{{Ichikawa} {et~al.}(2015){Ichikawa}, {Packham}, {Ramos Almeida},
  {Asensio Ramos}, {Alonso-Herrero}, {Gonz{\'a}lez-Mart{\'\i}n},
  {Lopez-Rodriguez}, {Ueda}, {D{\'\i}az-Santos}, {Elitzur}, {H{\"o}nig},
  {Imanishi}, {Levenson}, {Mason}, {Perlman}, \& {Alsip}}]{ichikawa2015}
{Ichikawa}, K., {Packham}, C., {Ramos Almeida}, C., {et~al.} 2015, \apj, 803,
  57

\bibitem[{{Ivezi{\'c}} {et~al.}(2019){Ivezi{\'c}}, {Kahn}, {Tyson}, {Abel},
  {Acosta}, {Allsman}, {Alonso}, {AlSayyad}, {Anderson}, {Andrew}, {Angel},
  {Angeli}, {Ansari}, {Antilogus}, {Araujo}, {Armstrong}, {Arndt}, {Astier},
  {Aubourg}, {Auza}, {Axelrod}, {Bard}, {Barr}, {Barrau}, {Bartlett}, {Bauer},
  {Bauman}, {Baumont}, {Bechtol}, {Bechtol}, {Becker}, {Becla}, {Beldica},
  {Bellavia}, {Bianco}, {Biswas}, {Blanc}, {Blazek}, {Bland ford}, {Bloom},
  {Bogart}, {Bond}, {Booth}, {Borgland}, {Borne}, {Bosch}, {Boutigny},
  {Brackett}, {Bradshaw}, {Brand t}, {Brown}, {Bullock}, {Burchat}, {Burke},
  {Cagnoli}, {Calabrese}, {Callahan}, {Callen}, {Carlin}, {Carlson}, {Chand
  rasekharan}, {Charles-Emerson}, {Chesley}, {Cheu}, {Chiang}, {Chiang},
  {Chirino}, {Chow}, {Ciardi}, {Claver}, {Cohen-Tanugi}, {Cockrum}, {Coles},
  {Connolly}, {Cook}, {Cooray}, {Covey}, {Cribbs}, {Cui}, {Cutri}, {Daly},
  {Daniel}, {Daruich}, {Daubard}, {Daues}, {Dawson}, {Delgado}, {Dellapenna},
  {de Peyster}, {de Val-Borro}, {Digel}, {Doherty}, {Dubois},
  {Dubois-Felsmann}, {Durech}, {Economou}, {Eifler}, {Eracleous}, {Emmons},
  {Fausti Neto}, {Ferguson}, {Figueroa}, {Fisher-Levine}, {Focke}, {Foss},
  {Frank}, {Freemon}, {Gangler}, {Gawiser}, {Geary}, {Gee}, {Geha}, {Gessner},
  {Gibson}, {Gilmore}, {Glanzman}, {Glick}, {Goldina}, {Goldstein}, {Goodenow},
  {Graham}, {Gressler}, {Gris}, {Guy}, {Guyonnet}, {Haller}, {Harris},
  {Hascall}, {Haupt}, {Hernand ez}, {Herrmann}, {Hileman}, {Hoblitt},
  {Hodgson}, {Hogan}, {Howard}, {Huang}, {Huffer}, {Ingraham}, {Innes},
  {Jacoby}, {Jain}, {Jammes}, {Jee}, {Jenness}, {Jernigan}, {Jevremovi{\'c}},
  {Johns}, {Johnson}, {Johnson}, {Jones}, {Juramy-Gilles}, {Juri{\'c}},
  {Kalirai}, {Kallivayalil}, {Kalmbach}, {Kantor}, {Karst}, {Kasliwal},
  {Kelly}, {Kessler}, {Kinnison}, {Kirkby}, {Knox}, {Kotov}, {Krabbendam},
  {Krughoff}, {Kub{\'a}nek}, {Kuczewski}, {Kulkarni}, {Ku}, {Kurita}, {Lage},
  {Lambert}, {Lange}, {Langton}, {Le Guillou}, {Levine}, {Liang}, {Lim},
  {Lintott}, {Long}, {Lopez}, {Lotz}, {Lupton}, {Lust}, {MacArthur}, {Mahabal},
  {Mand elbaum}, {Markiewicz}, {Marsh}, {Marshall}, {Marshall}, {May},
  {McKercher}, {McQueen}, {Meyers}, {Migliore}, {Miller}, {Mills}, {Miraval},
  {Moeyens}, {Moolekamp}, {Monet}, {Moniez}, {Monkewitz}, {Montgomery},
  {Morrison}, {Mueller}, {Muller}, {Mu{\~n}oz Arancibia}, {Neill}, {Newbry},
  {Nief}, {Nomerotski}, {Nordby}, {O'Connor}, {Oliver}, {Olivier}, {Olsen},
  {O'Mullane}, {Ortiz}, {Osier}, {Owen}, {Pain}, {Palecek}, {Parejko},
  {Parsons}, {Pease}, {Peterson}, {Peterson}, {Petravick}, {Libby Petrick},
  {Petry}, {Pierfederici}, {Pietrowicz}, {Pike}, {Pinto}, {Plante}, {Plate},
  {Plutchak}, {Price}, {Prouza}, {Radeka}, {Rajagopal}, {Rasmussen},
  {Regnault}, {Reil}, {Reiss}, {Reuter}, {Ridgway}, {Riot}, {Ritz}, {Robinson},
  {Roby}, {Roodman}, {Rosing}, {Roucelle}, {Rumore}, {Russo}, {Saha},
  {Sassolas}, {Schalk}, {Schellart}, {Schindler}, {Schmidt}, {Schneider},
  {Schneider}, {Schoening}, {Schumacher}, {Schwamb}, {Sebag}, {Selvy},
  {Sembroski}, {Seppala}, {Serio}, {Serrano}, {Shaw}, {Shipsey}, {Sick},
  {Silvestri}, {Slater}, {Smith}, {Smith}, {Sobhani}, {Soldahl},
  {Storrie-Lombardi}, {Stover}, {Strauss}, {Street}, {Stubbs}, {Sullivan},
  {Sweeney}, {Swinbank}, {Szalay}, {Takacs}, {Tether}, {Thaler}, {Thayer},
  {Thomas}, {Thornton}, {Thukral}, {Tice}, {Trilling}, {Turri}, {Van Berg},
  {Vanden Berk}, {Vetter}, {Virieux}, {Vucina}, {Wahl}, {Walkowicz}, {Walsh},
  {Walter}, {Wang}, {Wang}, {Warner}, {Wiecha}, {Willman}, {Winters},
  {Wittman}, {Wolff}, {Wood-Vasey}, {Wu}, {Xin}, {Yoachim}, \&
  {Zhan}}]{2019ApJ...873..111I}
{Ivezi{\'c}}, {\v{Z}}., {Kahn}, S.~M., {Tyson}, J.~A., {et~al.} 2019, \apj,
  873, 111

\bibitem[{{Kaspi} {et~al.}(2000){Kaspi}, {Smith}, {Netzer}, {Maoz}, {Jannuzi},
  \& {Giveon}}]{kaspi2000}
{Kaspi}, S., {Smith}, P.~S., {Netzer}, H., {et~al.} 2000, \apj, 533, 631

\bibitem[{{Kelly} {et~al.}(2009){Kelly}, {Bechtold}, \&
  {Siemiginowska}}]{kelly2009}
{Kelly}, B.~C., {Bechtold}, J., \& {Siemiginowska}, A. 2009, \apj, 698, 895

\bibitem[{{Kova{\v c}evi{\'c}-Doj{\v c}inovi{\'c}} \&
  {Popovi{\'c}}(2015)}]{kovacevic15}
{Kova{\v c}evi{\'c}-Doj{\v c}inovi{\'c}}, J., \& {Popovi{\'c}}, L.~{\v C}.
  2015, \apjs, 221, 35

\bibitem[{{Koz{\l}owski}(2016)}]{kozlowski2016}
{Koz{\l}owski}, S. 2016, \apj, 826, 118

\bibitem[{{Koz{\l}owski} {et~al.}(2010){Koz{\l}owski}, {Kochanek}, {Udalski},
  {Wyrzykowski}, {Soszy{\'n}ski}, {Szyma{\'n}ski}, {Kubiak}, {Pietrzy{\'n}ski},
  {Szewczyk}, {Ulaczyk}, {Poleski}, \& {OGLE Collaboration}}]{kozlowski2010}
{Koz{\l}owski}, S., {Kochanek}, C.~S., {Udalski}, A., {et~al.} 2010, \apj, 708,
  927

\bibitem[{{Lawrence} \& {Elvis}(2010)}]{lawrenceandelvis2010}
{Lawrence}, A., \& {Elvis}, M. 2010, \apj, 714, 561

\bibitem[{{Li} {et~al.}(2019){Li}, {Shen}, {Brandt}, {Grier}, {Hall}, {Ho},
  {Homayouni}, {Horne}, {Schneider}, {Trump}, \&
  {Starkey}}]{2019ApJ...884..119L}
{Li}, I-Hsiu, J., {Shen}, Y., {Brandt}, W.~N., {et~al.} 2019, \apj, 884, 119

\bibitem[{{Lira} {et~al.}(2018){Lira}, {Kaspi}, {Netzer}, {Botti}, {Morrell},
  {Mej{\'{\i}}a-Restrepo}, {S{\'a}nchez-S{\'a}ez}, {Mart{\'{\i}}nez-Palomera},
  \& {L{\'o}pez}}]{lira2018}
{Lira}, P., {Kaspi}, S., {Netzer}, H., {et~al.} 2018, \apj, 865, 56

\bibitem[{{Lusso} \& {Risaliti}(2017)}]{lusso2017}
{Lusso}, E., \& {Risaliti}, G. 2017, \aap, 602, A79

\bibitem[{{MacLeod} {et~al.}(2010){MacLeod}, {Ivezi{\'c}}, {Kochanek},
  {Koz{\l}owski}, {Kelly}, {Bullock}, {Kimball}, {Sesar}, {Westman}, {Brooks},
  {Gibson}, {Becker}, \& {de Vries}}]{macleod2010}
{MacLeod}, C.~L., {Ivezi{\'c}}, {\v{Z}}., {Kochanek}, C.~S., {et~al.} 2010,
  \apj, 721, 1014

\bibitem[{{Mart{\'\i}nez-Aldama} {et~al.}(2019){Mart{\'\i}nez-Aldama},
  {Czerny}, {Kawka}, {Karas}, {Panda}, {Zaja{\v{c}}ek}, \&
  {{\.Z}ycki}}]{martinez_aldama2019}
{Mart{\'\i}nez-Aldama}, M.~L., {Czerny}, B., {Kawka}, D., {et~al.} 2019, \apj,
  883, 170

\bibitem[{{Marziani} \& {Sulentic}(2014)}]{marzianisulentic2014}
{Marziani}, P., \& {Sulentic}, J.~W. 2014, \mnras, 442, 1211

\bibitem[{{Marziani} {et~al.}(2003){Marziani}, {Zamanov}, {Sulentic}, \&
  {Calvani}}]{marziani2003}
{Marziani}, P., {Zamanov}, R.~K., {Sulentic}, J.~W., \& {Calvani}, M. 2003,
  \mnras, 345, 1133

\bibitem[{{McLure} \& {Jarvis}(2002)}]{mclure2002}
{McLure}, R.~J., \& {Jarvis}, M.~J. 2002, \mnras, 337, 109

\bibitem[{{Mej{\'\i}a-Restrepo} {et~al.}(2018){Mej{\'\i}a-Restrepo}, {Lira},
  {Netzer}, {Trakhtenbrot}, \& {Capellupo}}]{mejia2018}
{Mej{\'\i}a-Restrepo}, J.~E., {Lira}, P., {Netzer}, H., {Trakhtenbrot}, B., \&
  {Capellupo}, D.~M. 2018, Nature Astronomy, 2, 63

\bibitem[{{Metzroth} {et~al.}(2006){Metzroth}, {Onken}, \&
  {Peterson}}]{metzroth2006}
{Metzroth}, K.~G., {Onken}, C.~A., \& {Peterson}, B.~M. 2006, \apj, 647, 901

\bibitem[{Naddaf {et~al.}(2020)Naddaf, Czerny, \& Szczerba}]{naddaf2020}
Naddaf, M.-H., Czerny, B., \& Szczerba, R. 2020, Frontiers in Astronomy and
  Space Sciences, 7, 15.
\newblock \url{https://www.frontiersin.org/article/10.3389/fspas.2020.00015}

\bibitem[{{Negrete} {et~al.}(2018){Negrete}, {Dultzin}, {Marziani}, {Esparza},
  {Sulentic}, {del Olmo}, {Mart{\'\i}nez-Aldama}, {Garc{\'\i}a L{\'o}pez},
  {D'Onofrio}, {Bon}, \& {Bon}}]{negrete2018}
{Negrete}, C.~A., {Dultzin}, D., {Marziani}, P., {et~al.} 2018, \aap, 620, A118

\bibitem[{{Netzer}(2013)}]{netzer2013book}
{Netzer}, H. 2013, {The Physics and Evolution of Active Galactic Nuclei}
  (Cambridge University Press; 1st edition)

\bibitem[{{Netzer}(2020)}]{netzer2020}
---. 2020, \mnras, 494, 1611

\bibitem[{Oliphant(2015)}]{numpy}
Oliphant, T. 2015, {NumPy}: A guide to {NumPy}, 2nd edn., USA: CreateSpace
  Independent Publishing Platform, , , [Online; accessed <today>].
\newblock \url{http://www.numpy.org/}

\bibitem[{{Panda} {et~al.}(2019{\natexlab{a}}){Panda}, {Mart{\'\i}nez-Aldama},
  \& {Zaja{\v{c}}ek}}]{2019arXiv190905572P}
{Panda}, S., {Mart{\'\i}nez-Aldama}, M.~L., \& {Zaja{\v{c}}ek}, M.
  2019{\natexlab{a}}, Frontiers in Astronomy and Space Sciences, 6, 75

\bibitem[{{Panda} {et~al.}(2019{\natexlab{b}}){Panda}, {Marziani}, \&
  {Czerny}}]{2019ApJ...882...79P}
{Panda}, S., {Marziani}, P., \& {Czerny}, B. 2019{\natexlab{b}}, \apj, 882, 79

\bibitem[{{Pedregosa} {et~al.}(2011){Pedregosa}, {Varoquaux}, {Gramfort},
  {Michel}, {Thirion}, {Grisel}, {Blondel}, {Prettenhofer}, {Weiss}, {Dubourg},
  {Vanderplas}, {Passos}, {Cournapeau}, {Brucher}, {Perrot}, \&
  {Duchesnay}}]{scikit-learn}
{Pedregosa}, F., {Varoquaux}, G., {Gramfort}, A., {et~al.} 2011, Journal of
  Machine Learning Research, 12, 2825

\bibitem[{{Peterson} \& {Horne}(2004)}]{2004AN....325..248P}
{Peterson}, B.~M., \& {Horne}, K. 2004, Astronomische Nachrichten, 325, 248

\bibitem[{{Peterson} {et~al.}(1998{\natexlab{a}}){Peterson}, {Wanders},
  {Bertram}, {Hunley}, {Pogge}, \& {Wagner}}]{peterson1998}
{Peterson}, B.~M., {Wanders}, I., {Bertram}, R., {et~al.} 1998{\natexlab{a}},
  \apj, 501, 82

\bibitem[{{Peterson} {et~al.}(1998{\natexlab{b}}){Peterson}, {Wanders},
  {Horne}, {Collier}, {Alexander}, {Kaspi}, \& {Maoz}}]{1998PASP..110..660P}
{Peterson}, B.~M., {Wanders}, I., {Horne}, K., {et~al.} 1998{\natexlab{b}},
  \pasp, 110, 660

\bibitem[{{Peterson} {et~al.}(2004){Peterson}, {Ferrarese}, {Gilbert}, {Kaspi},
  {Malkan}, {Maoz}, {Merritt}, {Netzer}, {Onken}, {Pogge}, {Vestergaard}, \&
  {Wandel}}]{peterson2004}
{Peterson}, B.~M., {Ferrarese}, L., {Gilbert}, K.~M., {et~al.} 2004, \apj, 613,
  682

\bibitem[{{Rakshit} {et~al.}(2019){Rakshit}, {Stalin}, \&
  {Kotilainen}}]{rakshitetal2019}
{Rakshit}, S., {Stalin}, C.~S., \& {Kotilainen}, J. 2019, arXiv e-prints,
  arXiv:1910.10395

\bibitem[{{Richards} {et~al.}(2006){Richards}, {Lacy}, {Storrie-Lombardi},
  {Hall}, {Gallagher}, {Hines}, {Fan}, {Papovich}, {Vanden Berk}, {Trammell},
  {Schneider}, {Vestergaard}, {York}, {Jester}, {Anderson}, {Budav{\'a}ri}, \&
  {Szalay}}]{richards2006}
{Richards}, G.~T., {Lacy}, M., {Storrie-Lombardi}, L.~J., {et~al.} 2006, \apjs,
  166, 470

\bibitem[{{Risaliti} \& {Lusso}(2019)}]{risaliti2019}
{Risaliti}, G., \& {Lusso}, E. 2019, Nature Astronomy, 3, 272

\bibitem[{{Rodr{\'{\i}}guez-Pascual} {et~al.}(1997){Rodr{\'{\i}}guez-Pascual},
  {Alloin}, {Clavel}, {Crenshaw}, {Horne}, {Kriss}, {Krolik}, {Malkan},
  {Netzer}, {O'Brien}, {Peterson}, {Reichert}, {Wamsteker}, {Alexander},
  {Barr}, {Blandford}, {Bregman}, {Carone}, {Clements}, {Courvoisier}, {De
  Robertis}, {Dietrich}, {Dottori}, {Edelson}, {Filippenko}, {Gaskell},
  {Huchra}, {Hutchings}, {Kollatschny}, {Koratkar}, {Korista}, {Laor},
  {MacAlpine}, {Martin}, {Maoz}, {McCollum}, {Morris}, {Perola}, {Pogge},
  {Ptak}, {Recondo-Gonz{\'a}lez}, {Rodr{\'{\i}}guez-Espinoza}, {Rokaki},
  {Santos-Lle{\'o}}, {Sekiguchi}, {Shull}, {Snijders}, {Sparke}, {Stirpe},
  {Stoner}, {Sun}, {Wagner}, {Wanders}, {Wilkes}, {Winge}, \&
  {Zheng}}]{rodriguez-pascual1997}
{Rodr{\'{\i}}guez-Pascual}, P.~M., {Alloin}, D., {Clavel}, J., {et~al.} 1997,
  \apjs, 110, 9

\bibitem[{{S{\'a}nchez-S{\'a}ez} {et~al.}(2018){S{\'a}nchez-S{\'a}ez}, {Lira},
  {Mej{\'\i}a-Restrepo}, {Ho}, {Ar{\'e}valo}, {Kim}, {Cartier}, \&
  {Coppi}}]{sanchez-saenz2018}
{S{\'a}nchez-S{\'a}ez}, P., {Lira}, P., {Mej{\'\i}a-Restrepo}, J., {et~al.}
  2018, \apj, 864, 87

\bibitem[{{Seabold} \& {Perktold}(2010)}]{seabold2010statsmodels}
{Seabold}, S., \& {Perktold}, J. 2010, in {P}roceedings of the 9th {P}ython in
  {S}cience {C}onference, ed. {S}t\'efan van~der {W}alt \& {J}arrod {M}illman,
  92 -- 96

\bibitem[{{Shakura} \& {Sunyaev}(1973)}]{SS1973}
{Shakura}, N.~I., \& {Sunyaev}, R.~A. 1973, \aap, 500, 33

\bibitem[{{Shen} \& {Ho}(2014)}]{shenandho2014}
{Shen}, Y., \& {Ho}, L.~C. 2014, \nat, 513, 210

\bibitem[{{Shen} {et~al.}(2011){Shen}, {Richards}, {Strauss}, {Hall},
  {Schneider}, {Snedden}, {Bizyaev}, {Brewington}, {Malanushenko},
  {Malanushenko}, {Oravetz}, {Pan}, \& {Simmons}}]{shen2011}
{Shen}, Y., {Richards}, G.~T., {Strauss}, M.~A., {et~al.} 2011, \apjs, 194, 45

\bibitem[{{Shen} {et~al.}(2016){Shen}, {Horne}, {Grier}, {Peterson}, {Denney},
  {Trump}, {Sun}, {Brandt}, {Kochanek}, {Dawson}, {Green}, {Greene}, {Hall},
  {Ho}, {Jiang}, {Kinemuchi}, {McGreer}, {Petitjean}, {Richards}, {Schneider},
  {Strauss}, {Tao}, {Wood-Vasey}, {Zu}, {Pan}, {Bizyaev}, {Ge}, {Oravetz}, \&
  {Simmons}}]{shen2006}
{Shen}, Y., {Horne}, K., {Grier}, C.~J., {et~al.} 2016, \apj, 818, 30

\bibitem[{{Shen} {et~al.}(2019){Shen}, {Hall}, {Horne}, {Zhu}, {McGreer},
  {Simm}, {Trump}, {Kinemuchi}, {Brandt}, {Green}, {Grier}, {Guo}, {Ho},
  {Homayouni}, {Jiang}, {I-Hsiu Li}, {Morganson}, {Petitjean}, {Richards},
  {Schneider}, {Starkey}, {Wang}, {Chambers}, {Kaiser}, {Kudritzki}, {Magnier},
  \& {Waters}}]{shen2019}
{Shen}, Y., {Hall}, P.~B., {Horne}, K., {et~al.} 2019, \apjs, 241, 34

\bibitem[{{{\'S}niegowska} {et~al.}(2018){{\'S}niegowska}, {Koz{\l}owski},
  {Czerny}, \& {Panda}}]{sniegowska}
{{\'S}niegowska}, M., {Koz{\l}owski}, S., {Czerny}, B., \& {Panda}, S. 2018,
  arXiv e-prints, arXiv:1810.09363

\bibitem[{{Taylor}(2005)}]{2005ASPC..347...29T}
{Taylor}, M.~B. 2005, in Astronomical Data Analysis Software and Systems XIV,
  Vol. 347, 29

\bibitem[{{Trakhtenbrot} \& {Netzer}(2012)}]{traktenbrot2012}
{Trakhtenbrot}, B., \& {Netzer}, H. 2012, \mnras, 427, 3081

\bibitem[{{Vestergaard} \& {Osmer}(2009)}]{vestergaard-osmer2009}
{Vestergaard}, M., \& {Osmer}, P.~S. 2009, \apj, 699, 800

\bibitem[{{Vestergaard} \& {Peterson}(2006)}]{vestergaard2006}
{Vestergaard}, M., \& {Peterson}, B.~M. 2006, \apj, 641, 689

\bibitem[{{Vestergaard} \& {Wilkes}(2001)}]{vestergaard2001}
{Vestergaard}, M., \& {Wilkes}, B.~J. 2001, \apjs, 134, 1

\bibitem[{{Wang} {et~al.}(2014{\natexlab{a}}){Wang}, {Du}, {Li}, {Ho}, {Hu}, \&
  {Bai}}]{wang2014_spin}
{Wang}, J.-M., {Du}, P., {Li}, Y.-R., {et~al.} 2014{\natexlab{a}}, \apjl, 792,
  L13

\bibitem[{{Wang} {et~al.}(2014{\natexlab{b}}){Wang}, {Qiu}, {Du}, \&
  {Ho}}]{wang_shielding2014}
{Wang}, J.-M., {Qiu}, J., {Du}, P., \& {Ho}, L.~C. 2014{\natexlab{b}}, \apj,
  797, 65

\bibitem[{{Wang} {et~al.}(2014{\natexlab{c}}){Wang}, {Du}, {Hu}, {Netzer},
  {Bai}, {Lu}, {Kaspi}, {Qiu}, {Li}, {Wang}, \& {SEAMBH
  Collaboration}}]{wang2014_mdot}
{Wang}, J.-M., {Du}, P., {Hu}, C., {et~al.} 2014{\natexlab{c}}, \apj, 793, 108

\bibitem[{{Watson} {et~al.}(2011){Watson}, {Denney}, {Vestergaard}, \&
  {Davis}}]{watson2011}
{Watson}, D., {Denney}, K.~D., {Vestergaard}, M., \& {Davis}, T.~M. 2011,
  \apjl, 740, L49

\bibitem[{{Wilhite} {et~al.}(2008){Wilhite}, {Brunner}, {Grier}, {Schneider},
  \& {vanden Berk}}]{wilhite2008}
{Wilhite}, B.~C., {Brunner}, R.~J., {Grier}, C.~J., {Schneider}, D.~P., \&
  {vanden Berk}, D.~E. 2008, \mnras, 383, 1232

\bibitem[{{Wilhite} {et~al.}(2005){Wilhite}, {Vanden Berk}, {Kron},
  {Schneider}, {Pereyra}, {Brunner}, {Richards}, \& {Brinkmann}}]{wilhite2005}
{Wilhite}, B.~C., {Vanden Berk}, D.~E., {Kron}, R.~G., {et~al.} 2005, \apj,
  633, 638

\bibitem[{{Yu} {et~al.}(2020{\natexlab{a}}){Yu}, {Bian}, {Zhang}, {Zhao},
  {Wang}, {Ge}, {Zhu}, \& {Chen}}]{yu2020_sigmaline}
{Yu}, L.-M., {Bian}, W.-H., {Zhang}, X.-G., {et~al.} 2020{\natexlab{a}}, arXiv
  e-prints, arXiv:2008.06623

\bibitem[{{Yu} {et~al.}(2020{\natexlab{b}}){Yu}, {Zhao}, {Bian}, {Wang}, \&
  {Ge}}]{Yu2020}
{Yu}, L.-M., {Zhao}, B.-X., {Bian}, W.-H., {Wang}, C., \& {Ge}, X.
  2020{\natexlab{b}}, \mnras, 491, 5881

\bibitem[{{{Yu}, Z.} {et~al.}(2020){{Yu}, Z.}, {Kochanek}, {Peterson}, {Zu},
  {Brandt}, {Cackett}, {Fausnaugh}, \& {McHardy}}]{2020MNRAS.491.6045Y}
{{Yu}, Z.}, {Kochanek}, C.~S., {Peterson}, B.~M., {et~al.} 2020, \mnras, 491,
  6045

\bibitem[{{Zaja{\v{c}}ek} {et~al.}(2019){Zaja{\v{c}}ek}, {Czerny},
  {Mart{\'\i}nez-Aldama}, \& {Karas}}]{2019arXiv190703910Z}
{Zaja{\v{c}}ek}, M., {Czerny}, B., {Mart{\'\i}nez-Aldama}, M.~L., \& {Karas},
  V. 2019, Astronomische Nachrichten, 340, 577

\bibitem[{{Zaja{\v{c}}ek} {et~al.}(2020){Zaja{\v{c}}ek}, {Czerny},
  {Martinez-Aldama}, {Ra{\l}owski}, {Olejak}, {Panda}, {Hryniewicz},
  {{\'S}niegowska}, {Naddaf}, {Pych}, {Pietrzy{\'n}ski}, {Figaredo}, {Haas},
  {{\'S}redzi{\'n}ska}, {Krupa}, {Kurcz}, {Udalski}, {Gorski}, \&
  {Sarna}}]{zajacek2020}
{Zaja{\v{c}}ek}, M., {Czerny}, B., {Martinez-Aldama}, M.~L., {et~al.} 2020,
  \apj, 896, 146

\bibitem[{{Zu} {et~al.}(2016){Zu}, {Kochanek}, {Koz{\l}owski}, \&
  {Peterson}}]{2016ApJ...819..122Z}
{Zu}, Y., {Kochanek}, C.~S., {Koz{\l}owski}, S., \& {Peterson}, B.~M. 2016,
  \apj, 819, 122

\bibitem[{{Zu} {et~al.}(2013){Zu}, {Kochanek}, {Koz{\l}owski}, \&
  {Udalski}}]{2013ApJ...765..106Z}
{Zu}, Y., {Kochanek}, C.~S., {Koz{\l}owski}, S., \& {Udalski}, A. 2013, \apj,
  765, 106

\bibitem[{{Zu} {et~al.}(2011){Zu}, {Kochanek}, \&
  {Peterson}}]{2011ApJ...735...80Z}
{Zu}, Y., {Kochanek}, C.~S., \& {Peterson}, B.~M. 2011, \apj, 735, 80

\end{thebibliography}
\bibliographystyle{aasjournal}





\appendix

\setcounter{table}{0}
\renewcommand\thetable{\Alph{section}.\arabic{table}}

\section{Observational properties for the full sample}
\label{appendix}

\begin{center}
\scriptsize
\small
\begin{longtable}[c]{ccccccccc}
\caption{Observational properties for the full sample}
\label{tab:measurements}\\
 \hline\hline\noalign{\vskip 0.1cm}
\multirow{2}{*}{Object} & log $L_{3000}$ & $\tau_\mathrm{{obs}}$ & FWHM Mg~II & \multirow{2}{*}{log~\mdot} & \multirow{2}{*}{\eddr} & \multirow{2}{*}{R${_\mathrm{FeII}}$} & \multirow{2}{*}{\fvar}  & \multirow{2}{*}{Class}\\
\endfirsthead
\endhead
&  [\ergs] & [1 lt-day] & [km s$^{-1}$] & & & & \\
(1) & (2) & (3) & (4) & (5) & (6) & (7) & (8)  & (9)\\ 

 \hline\hline\noalign{\vskip 0.1cm}
\noalign{\vskip 0.1cm} 																										
\multicolumn{9}{c}{SDSS-RM sample \citep{homayouni2020}} \\ 				

\hline \noalign{\vskip 0.1cm}

18	&	44.4	$\pm$	0.0009	&	125.9	$^{+	6.8	}_{-	7.0	}$  &	7416	$\pm$	123	&	-1.176	$^{+	0.321	}_{-	0.321	}$  &	0.019	$^{+	0.008	}_{-	0.008	}$  &	0.470	$\pm$	0.019	&	0.050	&	1	\\
28	&	45.6	$\pm$	0.0004	&	65.7	$^{+	24.8	}_{-	14.2	}$  &	3899	$\pm$	75	&	1.630	$^{+	0.328	}_{-	0.424	}$  &	0.968	$^{+	0.511	}_{-	0.415	}$  &	1.030	$\pm$	0.013	&	0.088	&	2	\\
38	&	45.7	$\pm$	0.0003	&	120.7	$^{+	27.9	}_{-	28.7	}$  &	3954	$\pm$	87	&	1.242	$^{+	0.340	}_{-	0.336	}$  &	0.656	$^{+	0.287	}_{-	0.289	}$  &	1.210	$\pm$	0.015	&	0.077	&	2	\\
44	&	44.9	$\pm$	0.0013	&	65.8	$^{+	18.8	}_{-	4.8	}$  &	2583	$\pm$	114	&	0.861	$^{+	0.288	}_{-	0.375	}$  &	0.267	$^{+	0.127	}_{-	0.104	}$  &	1.060	$\pm$	0.064	&	0.066	&	2	\\
102	&	45.0	$\pm$	0.0005	&	86.9	$^{+	16.2	}_{-	13.3	}$  &	2977	$\pm$	78	&	0.672	$^{+	0.299	}_{-	0.313	}$  &	0.227	$^{+	0.094	}_{-	0.091	}$  &	1.420	$\pm$	0.026	&	0.045	&	2	\\
114	&	46.1	$\pm$	0.0003	&	186.6	$^{+	20.3	}_{-	15.4	}$  &	4318	$\pm$	226	&	1.403	$^{+	0.299	}_{-	0.305	}$  &	0.994	$^{+	0.403	}_{-	0.397	}$  &	1.350	$\pm$	0.040	&	0.038	&	2	\\
118	&	45.1	$\pm$	0.0006	&	102.2	$^{+	27.0	}_{-	19.5	}$  &	2885	$\pm$	64	&	0.703	$^{+	0.314	}_{-	0.352	}$  &	0.250	$^{+	0.113	}_{-	0.104	}$  &	1.210	$\pm$	0.027	&	0.066	&	2	\\
123	&	44.7	$\pm$	0.0009	&	81.6	$^{+	28.0	}_{-	26.6	}$  &	4647	$\pm$	126	&	-0.029	$^{+	0.398	}_{-	0.408	}$  &	0.085	$^{+	0.044	}_{-	0.043	}$  &	1.110	$\pm$	0.032	&	0.117	&	1	\\
135	&	45.2	$\pm$	0.0005	&	93.0	$^{+	9.6	}_{-	9.8	}$  &	4128	$\pm$	64	&	0.689	$^{+	0.285	}_{-	0.284	}$  &	0.260	$^{+	0.100	}_{-	0.100	}$  &	1.080	$\pm$	0.010	&	0.075	&	2	\\
158	&	44.9	$\pm$	0.0012	&	119.1	$^{+	4.0	}_{-	11.8	}$  &	4699	$\pm$	69	&	-0.065	$^{+	0.289	}_{-	0.278	}$  &	0.092	$^{+	0.035	}_{-	0.036	}$  &	0.950	$\pm$	0.021	&	0.058	&	1	\\
159	&	45.5	$\pm$	0.0006	&	324.2	$^{+	25.3	}_{-	19.4	}$  &	3298	$\pm$	87	&	0.208	$^{+	0.273	}_{-	0.277	}$  &	0.178	$^{+	0.067	}_{-	0.067	}$  &	1.010	$\pm$	0.018	&	0.044	&	1	\\
160	&	43.8	$\pm$	0.0013	&	106.5	$^{+	18.2	}_{-	16.6	}$  &	4386	$\pm$	56	&	-1.571	$^{+	0.303	}_{-	0.310	}$  &	0.009	$^{+	0.004	}_{-	0.004	}$  &	0.260	$\pm$	0.005	&	0.160	&	1	\\
170	&	45.2	$\pm$	0.0005	&	98.5	$^{+	6.7	}_{-	17.7	}$  &	10594	$\pm$	121	&	-0.008	$^{+	0.395	}_{-	0.368	}$  &	0.117	$^{+	0.055	}_{-	0.058	}$  &	0.760	$\pm$	0.011	&	0.103	&	1	\\
185	&	44.9	$\pm$	0.0094	&	387.9	$^{+	3.3	}_{-	3.0	}$  &	4765	$\pm$	2835	&	-1.100	$^{+	1.239	}_{-	1.239	}$  &	0.028	$^{+	0.040	}_{-	0.040	}$  &	1.300	$\pm$	0.326	&	0.082	&	1	\\
191	&	43.8	$\pm$	0.0012	&	93.9	$^{+	24.3	}_{-	29.1	}$  &	2619	$\pm$	109	&	-1.107	$^{+	0.388	}_{-	0.358	}$  &	0.015	$^{+	0.007	}_{-	0.007	}$  &	0.660	$\pm$	0.034	&	0.178	&	1	\\
228	&	44.7	$\pm$	0.0011	&	37.9	$^{+	14.4	}_{-	9.1	}$  &	4481	$\pm$	460	&	0.662	$^{+	0.401	}_{-	0.475	}$  &	0.189	$^{+	0.111	}_{-	0.095	}$  &	1.070	$\pm$	0.042	&	0.256	&	2	\\
232	&	44.3	$\pm$	0.0014	&	273.8	$^{+	5.1	}_{-	4.1	}$  &	4202	$\pm$	713	&	-1.611	$^{+	0.436	}_{-	0.436	}$  &	0.011	$^{+	0.006	}_{-	0.006	}$  &	1.010	$\pm$	0.090	&	0.173	&	1	\\
240	&	44.1	$\pm$	0.0021	&	17.2	$^{+	3.5	}_{-	2.8	}$  &	4547	$\pm$	126	&	0.439	$^{+	0.312	}_{-	0.330	}$  &	0.103	$^{+	0.045	}_{-	0.043	}$  &	0.660	$\pm$	0.028	&	0.157	&	2	\\
260	&	45.3	$\pm$	0.0004	&	94.9	$^{+	18.7	}_{-	17.2	}$  &	2814	$\pm$	90	&	1.084	$^{+	0.314	}_{-	0.321	}$  &	0.434	$^{+	0.183	}_{-	0.180	}$  &	1.000	$\pm$	0.018	&	0.135	&	2	\\
280	&	45.5	$\pm$	0.0003	&	99.1	$^{+	3.3	}_{-	9.5	}$  &	5751	$\pm$	256	&	0.856	$^{+	0.315	}_{-	0.305	}$  &	0.375	$^{+	0.152	}_{-	0.156	}$  &	1.020	$\pm$	0.014	&	0.061	&	2	\\
285	&	44.5	$\pm$	0.0020	&	138.5	$^{+	15.2	}_{-	21.1	}$  &	5139	$\pm$	65	&	-0.857	$^{+	0.311	}_{-	0.297	}$  &	0.029	$^{+	0.012	}_{-	0.012	}$  &	0.580	$\pm$	0.019	&	0.137	&	1	\\
291	&	43.8	$\pm$	0.0016	&	39.7	$^{+	4.2	}_{-	2.6	}$  &	7788	$\pm$	761	&	-1.107	$^{+	0.382	}_{-	0.389	}$  &	0.015	$^{+	0.007	}_{-	0.007	}$  &	0.160	$\pm$	0.012	&	0.188	&	1	\\
294	&	45.5	$\pm$	0.0004	&	71.8	$^{+	17.8	}_{-	9.5	}$  &	3008	$\pm$	52	&	1.581	$^{+	0.289	}_{-	0.342	}$  &	0.863	$^{+	0.382	}_{-	0.337	}$  &	1.350	$\pm$	0.025	&	0.034	&	2	\\
301	&	44.2	$\pm$	0.0011	&	136.3	$^{+	17.0	}_{-	16.9	}$  &	6052	$\pm$	599	&	-1.406	$^{+	0.373	}_{-	0.373	}$  &	0.013	$^{+	0.006	}_{-	0.006	}$  &	0.790	$\pm$	0.027	&	0.239	&	1	\\
303	&	44.2	$\pm$	0.0013	&	57.7	$^{+	10.5	}_{-	8.3	}$  &	4173	$\pm$	95	&	-0.404	$^{+	0.300	}_{-	0.315	}$  &	0.042	$^{+	0.017	}_{-	0.017	}$  &	0.890	$\pm$	0.017	&	0.114	&	1	\\
329	&	45.4	$\pm$	0.0007	&	87.5	$^{+	23.8	}_{-	14.0	}$  &	2720	$\pm$	29	&	1.328	$^{+	0.300	}_{-	0.355	}$  &	0.609	$^{+	0.278	}_{-	0.244	}$  &	1.530	$\pm$	0.033	&	0.057	&	2	\\
338	&	43.8	$\pm$	0.0013	&	22.1	$^{+	8.8	}_{-	6.2	}$  &	3662	$\pm$	1102	&	-0.081	$^{+	0.708	}_{-	0.750	}$  &	0.048	$^{+	0.043	}_{-	0.040	}$  &	0.200	$\pm$	0.019	&	0.168	&	1	\\
419	&	45.0	$\pm$	0.0011	&	95.5	$^{+	15.2	}_{-	15.5	}$  &	6132	$\pm$	135	&	0.094	$^{+	0.330	}_{-	0.329	}$  &	0.117	$^{+	0.050	}_{-	0.050	}$  &	1.110	$\pm$	0.026	&	0.046	&	1	\\
422	&	44.7	$\pm$	0.0011	&	109.3	$^{+	25.4	}_{-	29.6	}$  &	5628	$\pm$	94	&	-0.414	$^{+	0.373	}_{-	0.353	}$  &	0.055	$^{+	0.025	}_{-	0.026	}$  &	0.330	$\pm$	0.007	&	0.078	&	1	\\
440	&	44.9	$\pm$	0.0004	&	114.6	$^{+	7.4	}_{-	10.8	}$  &	6825	$\pm$	403	&	-0.288	$^{+	0.339	}_{-	0.334	}$  &	0.071	$^{+	0.031	}_{-	0.031	}$  &	0.980	$\pm$	0.014	&	0.105	&	1	\\
441	&	45.5	$\pm$	0.0004	&	127.7	$^{+	5.7	}_{-	7.3	}$  &	2276	$\pm$	91	&	1.272	$^{+	0.288	}_{-	0.286	}$  &	0.605	$^{+	0.234	}_{-	0.235	}$  &	1.320	$\pm$	0.042	&	0.033	&	2	\\
449	&	45.0	$\pm$	0.0013	&	119.8	$^{+	14.7	}_{-	24.4	}$  &	4149	$\pm$	216	&	0.165	$^{+	0.338	}_{-	0.307	}$  &	0.127	$^{+	0.052	}_{-	0.056	}$  &	1.150	$\pm$	0.034	&	0.091	&	1	\\
457	&	43.7	$\pm$	0.0029	&	20.5	$^{+	7.7	}_{-	5.3	}$  &	4213	$\pm$	810	&	-0.262	$^{+	0.524	}_{-	0.575	}$  &	0.037	$^{+	0.025	}_{-	0.023	}$  &	0.200	$\pm$	0.028	&	0.524	&	1	\\
459	&	45.0	$\pm$	0.0011	&	122.8	$^{+	5.1	}_{-	5.7	}$  &	4686	$\pm$	1134	&	0.061	$^{+	0.564	}_{-	0.564	}$  &	0.113	$^{+	0.076	}_{-	0.076	}$  &	1.010	$\pm$	0.035	&	0.127	&	1	\\
469	&	45.6	$\pm$	0.0002	&	224.1	$^{+	27.9	}_{-	74.3	}$  &	4246	$\pm$	57	&	0.506	$^{+	0.395	}_{-	0.291	}$  &	0.265	$^{+	0.104	}_{-	0.132	}$  &	1.240	$\pm$	0.029	&	0.056	&	2	\\
492	&	45.3	$\pm$	0.0004	&	92.0	$^{+	16.3	}_{-	12.7	}$  &	4436	$\pm$	103	&	0.799	$^{+	0.300	}_{-	0.315	}$  &	0.313	$^{+	0.130	}_{-	0.125	}$  &	1.030	$\pm$	0.017	&	0.064	&	2	\\
493	&	46.0	$\pm$	0.0004	&	315.6	$^{+	30.7	}_{-	35.7	}$  &	7102	$\pm$	823	&	0.455	$^{+	0.402	}_{-	0.399	}$  &	0.315	$^{+	0.158	}_{-	0.159	}$  &	1.390	$\pm$	0.036	&	0.056	&	2	\\
501	&	44.9	$\pm$	0.0009	&	44.9	$^{+	11.7	}_{-	10.4	}$  &	3511	$\pm$	110	&	0.983	$^{+	0.337	}_{-	0.353	}$  &	0.307	$^{+	0.139	}_{-	0.135	}$  &	1.010	$\pm$	0.044	&	0.123	&	2	\\
505	&	44.8	$\pm$	0.0011	&	94.7	$^{+	10.8	}_{-	16.7	}$  &	5819	$\pm$	160	&	-0.162	$^{+	0.333	}_{-	0.312	}$  &	0.077	$^{+	0.032	}_{-	0.034	}$  &	0.870	$\pm$	0.021	&	0.101	&	1	\\
522	&	45.1	$\pm$	0.0006	&	115.8	$^{+	11.3	}_{-	16.0	}$  &	2214	$\pm$	33	&	0.776	$^{+	0.300	}_{-	0.288	}$  &	0.272	$^{+	0.106	}_{-	0.109	}$  &	1.270	$\pm$	0.032	&	0.059	&	2	\\
556	&	45.5	$\pm$	0.0005	&	98.7	$^{+	13.9	}_{-	10.8	}$  &	4616	$\pm$	90	&	1.011	$^{+	0.292	}_{-	0.302	}$  &	0.448	$^{+	0.180	}_{-	0.176	}$  &	1.210	$\pm$	0.016	&	0.044	&	2	\\
588	&	45.6	$\pm$	0.0002	&	74.3	$^{+	23.0	}_{-	18.2	}$  &	3596	$\pm$	42	&	1.579	$^{+	0.340	}_{-	0.377	}$  &	0.912	$^{+	0.438	}_{-	0.402	}$  &	1.000	$\pm$	0.013	&	0.092	&	2	\\
593	&	45.0	$\pm$	0.0006	&	80.1	$^{+	21.4	}_{-	20.8	}$  &	2890	$\pm$	41	&	0.763	$^{+	0.348	}_{-	0.352	}$  &	0.253	$^{+	0.115	}_{-	0.114	}$  &	1.060	$\pm$	0.018	&	0.057	&	2	\\
622	&	44.5	$\pm$	0.0005	&	61.7	$^{+	6.0	}_{-	4.3	}$  &	2768	$\pm$	128	&	0.270	$^{+	0.287	}_{-	0.293	}$  &	0.107	$^{+	0.042	}_{-	0.042	}$  &	1.740	$\pm$	0.037	&	0.063	&	2	\\
645	&	44.2	$\pm$	0.0009	&	30.2	$^{+	26.8	}_{-	8.9	}$  &	4035	$\pm$	158	&	0.182	$^{+	0.378	}_{-	0.819	}$  &	0.082	$^{+	0.079	}_{-	0.039	}$  &	1.350	$\pm$	0.044	&	0.197	&	1	\\
649	&	44.5	$\pm$	0.0013	&	165.5	$^{+	22.2	}_{-	25.1	}$  &	3753	$\pm$	666	&	-0.796	$^{+	0.466	}_{-	0.462	}$  &	0.031	$^{+	0.018	}_{-	0.018	}$  &	2.350	$\pm$	0.330	&	0.155	&	1	\\
651	&	45.2	$\pm$	0.0011	&	76.5	$^{+	18.0	}_{-	15.6	}$  &	5331	$\pm$	85	&	0.683	$^{+	0.336	}_{-	0.351	}$  &	0.258	$^{+	0.117	}_{-	0.113	}$  &	0.870	$\pm$	0.022	&	0.054	&	2	\\
675	&	45.1	$\pm$	0.0005	&	139.8	$^{+	12.0	}_{-	22.6	}$  &	4250	$\pm$	132	&	0.165	$^{+	0.310	}_{-	0.286	}$  &	0.134	$^{+	0.052	}_{-	0.055	}$  &	1.250	$\pm$	0.015	&	0.049	&	1	\\
678	&	45.3	$\pm$	0.0007	&	82.9	$^{+	11.9	}_{-	10.2	}$  &	3446	$\pm$	67	&	1.063	$^{+	0.287	}_{-	0.294	}$  &	0.424	$^{+	0.167	}_{-	0.164	}$  &	1.180	$\pm$	0.025	&	0.049	&	2	\\
709	&	45.0	$\pm$	0.0010	&	85.4	$^{+	17.7	}_{-	19.3	}$  &	4277	$\pm$	197	&	0.439	$^{+	0.346	}_{-	0.337	}$  &	0.174	$^{+	0.076	}_{-	0.078	}$  &	1.300	$\pm$	0.047	&	0.080	&	2	\\
714	&	44.8	$\pm$	0.0012	&	320.1	$^{+	11.3	}_{-	11.2	}$  &	5031	$\pm$	266	&	-1.121	$^{+	0.300	}_{-	0.300	}$  &	0.026	$^{+	0.010	}_{-	0.010	}$  &	0.590	$\pm$	0.022	&	0.206	&	1	\\
756	&	44.4	$\pm$	0.0023	&	315.3	$^{+	20.5	}_{-	16.4	}$  &	3505	$\pm$	151	&	-1.460	$^{+	0.281	}_{-	0.284	}$  &	0.014	$^{+	0.005	}_{-	0.005	}$  &	1.230	$\pm$	0.034	&	0.092	&	1	\\
761	&	44.8	$\pm$	0.0024	&	102.1	$^{+	8.2	}_{-	7.4	}$  &	4393	$\pm$	79	&	-0.035	$^{+	0.280	}_{-	0.282	}$  &	0.090	$^{+	0.034	}_{-	0.034	}$  &	1.480	$\pm$	0.042	&	0.149	&	1	\\
771	&	45.7	$\pm$	0.0004	&	31.3	$^{+	8.1	}_{-	4.6	}$  &	5391	$\pm$	57	&	2.202	$^{+	0.312	}_{-	0.363	}$  &	1.980	$^{+	0.920	}_{-	0.818	}$  &	0.770	$\pm$	0.008	&	0.068	&	2	\\
774	&	45.7	$\pm$	0.0004	&	58.9	$^{+	13.7	}_{-	10.1	}$  &	3537	$\pm$	125	&	1.942	$^{+	0.311	}_{-	0.340	}$  &	1.468	$^{+	0.647	}_{-	0.604	}$  &	1.190	$\pm$	0.021	&	0.125	&	2	\\
792	&	43.5	$\pm$	0.0030	&	111.4	$^{+	29.5	}_{-	20.0	}$  &	4451	$\pm$	772	&	-2.072	$^{+	0.474	}_{-	0.501	}$  &	0.004	$^{+	0.003	}_{-	0.002	}$  &	0.630	$\pm$	0.103	&	0.148	&	1	\\
848	&	44.1	$\pm$	0.0015	&	65.1	$^{+	29.4	}_{-	16.3	}$  &	3264	$\pm$	378	&	-0.490	$^{+	0.414	}_{-	0.527	}$  &	0.035	$^{+	0.023	}_{-	0.018	}$  &	0.900	$\pm$	0.058	&	0.186	&	1	\\

\hline\noalign{\vskip   0.1cm}								\noalign{\vskip 0.1cm}																			
\multicolumn{9}{c}{\citet{zajacek2020} sample} \\
\hline \noalign{\vskip 0.1cm}	

J141214	&	44.6	$\pm$	0.0004	&	36.7	$^{+	10.4	}_{-	4.8	}$  &	2391	$\pm$	46	&	1.030	$^{+	0.367	}_{-	0.295	}$  &	0.279	$^{+	0.131	}_{-	0.110	}$  &	1.650	$\pm$	0.201	&	0.094	&	2	\\
J141018	&	43.7	$\pm$	0.0051	&	32.3	$^{+	12.9	}_{-	5.3	}$  &	3101	$\pm$	76	&	-0.403	$^{+	0.438	}_{-	0.303	}$  &	0.032	$^{+	0.017	}_{-	0.013	}$  &	0.900	$\pm$	0.085	&	0.162	&	1	\\
J141417	&	43.7	$\pm$	0.0029	&	29.1	$^{+	3.6	}_{-	8.8	}$  &	3874	$\pm$	86	&	-0.527	$^{+	0.290	}_{-	0.376	}$  &	0.027	$^{+	0.011	}_{-	0.013	}$  &	0.200	$\pm$	0.028	&	0.524	&	1	\\
J142049	&	44.7	$\pm$	0.0009	&	34.0	$^{+	6.7	}_{-	12.0	}$  &	4108	$\pm$	39	&	0.803	$^{+	0.318	}_{-	0.407	}$  &	0.221	$^{+	0.093	}_{-	0.113	}$  &	1.450	$\pm$	0.034	&	0.174	&	2	\\
J141650	&	43.8	$\pm$	0.0020	&	25.1	$^{+	2.0	}_{-	2.6	}$  &	4066	$\pm$	202	&	-0.296	$^{+	0.294	}_{-	0.299	}$  &	0.037	$^{+	0.015	}_{-	0.015	}$  &	0.100	$\pm$	0.011	&	0.080	&	1	\\
J141644	&	43.9	$\pm$	0.0010	&	17.2	$^{+	2.7	}_{-	2.7	}$  &	2681	$\pm$	96	&	0.573	$^{+	0.307	}_{-	0.307	}$  &	0.111	$^{+	0.045	}_{-	0.045	}$  &	1.450	$\pm$	0.068	&	0.113	&	2	\\
CTS~252	&	46.8	$\pm$	0.0914	&	190.0	$^{+	59.0	}_{-	114.0	}$  &	3800	$\pm$	380	&	2.516	$^{+	0.450	}_{-	0.634	}$  &	5.334	$^{+	3.063	}_{-	4.109	}$  &	--			&	0.090	&	2	\\
NGC~4151	&	42.8	$\pm$	0.1821	&	6.8	$^{+	1.7	}_{-	2.1	}$  &	4823	$\pm$	1105	&	-0.698	$^{+	0.643	}_{-	0.662	}$  &	0.013	$^{+	0.011	}_{-	0.011	}$  &	--			&	0.088	&	1	\\
NGC~4151	&	42.8	$\pm$	0.1821	&	5.3	$^{+	1.9	}_{-	1.8	}$  &	6558	$\pm$	1850	&	-0.692	$^{+	0.768	}_{-	0.762	}$  &	0.014	$^{+	0.013	}_{-	0.013	}$  &	--			&	0.094	&	1	\\
CTS~C30.10	&	46.0	$\pm$	0.0260	&	564.0	$^{+	109.0	}_{-	71.0	}$  &	5009	$\pm$	325	&	0.225	$^{+	0.353	}_{-	0.329	}$  &	0.245	$^{+	0.112	}_{-	0.106	}$  &	1.600 	$^{+0.005}_{-0.003}$	&	0.066	&	2	\\
HE~0413-4013	&	46.7	$\pm$	0.0434	&	302.9	$^{+	23.7	}_{-	19.1	}$  &	4380	$\pm$	14	&	1.935	$^{+	0.290	}_{-	0.294	}$  &	2.652	$^{+	1.052	}_{-	1.062	}$  &	0.800	$\pm$	0.020	&	0.088	&	2	\\

\hline \noalign{\vskip 0.1cm}

\end{longtable}
\end{center}
\footnotesize{{\sc Notes.} Columns are as follows: (1) Object identification. For SDSS-RM sample, the object identification corresponds to the number (RMID) in the original catalog \citep{homayouni2020}. (2) Logarithm of  continuum luminosity at 3000\AA. (3) Time delay in units of light--day. SDSS-RM time delay reported correspond to ones obtained with the JAVELIN method.  (4) Full width at half maximum of Mg II. (5) Dimensionless accretion rate. (6) Eddington ratio. (7) R$_\mathrm{Fe}$ parameter. (8) \fvar\ parameter.  (9) Classification based on the \mdot\ intensity, numbers 1 and 2 correspond to the low and high accretion rate sub-sample, respectively. See Sec.~\ref{sec:sample_division}. }

\end{document}